\documentclass[hello,arxiv]{informs3_modified}

\OneAndAHalfSpacedXI %

\usepackage[dvipsnames]{xcolor}
\usepackage{endnotes}
\let\footnote=\endnote
\let\enotesize=\normalsize
\def\notesname{Endnotes}%
\def\makeenmark{\hbox to1.275em{\theenmark.\enskip\hss}}
\def\enoteformat{\rightskip0pt\leftskip0pt\parindent=1.275em
  \leavevmode\llap{\makeenmark}}

\newcommand{\bvx}{X} %
\newcommand{\vx}{x} %
\newcommand{\vy}{y} %
\newcommand{\vz}{z} %

\newcommand{\E}{\mathbb{E}}
\newcommand{\cF}{\mathcal{F}}
\newcommand{\bbP}{\mathbb{P}}
\newcommand{\OPT}{\mathrm{OPT}}
\usepackage{algorithm,algpseudocode}
\usepackage[colorlinks=true,citecolor=blue]{hyperref}

\usepackage{natbib}
 \bibpunct[, ]{(}{)}{,}{a}{}{,}%
 \def\bibfont{\small}%

\TheoremsNumberedThrough     %
\ECRepeatTheorems

\EquationsNumberedThrough    %

\UseRawInputEncoding

\begin{document}

\RUNAUTHOR{Lee, Vojnovic, and Yun}

\RUNTITLE{Test Score Algorithms for Budgeted Stochastic Utility Maximization}

\TITLE{Test Score Algorithms for Budgeted Stochastic Utility Maximization}

\ARTICLEAUTHORS{%
	\AUTHOR{Dabeen Lee}
	\AFF{Discrete Mathematics Group, Institute for Basic Science (IBS), Daejeon, South Korea, \EMAIL{dabeenl@ibs.re.kr}} %
	\AUTHOR{Milan Vojnovic}
	\AFF{Department of Statistics, London School of Economics (LSE), London, UK, \EMAIL{m.vojnovic@lse.ac.uk}}
	\AUTHOR{Se-Young Yun}
	\AFF{Graduate School of AI, KAIST, Daejeon, South Korea, \EMAIL{yunseyoung@kaist.ac.kr}}
} %

\ABSTRACT{
Motivated by recent developments in designing algorithms based on individual item scores for solving utility maximization problems, we study the framework of using test scores, defined as a statistic of observed individual item performance data, for solving the budgeted stochastic utility maximization problem. We extend an existing scoring mechanism, namely the replication test scores, to incorporate heterogeneous item costs as well as item values. We show that a natural greedy algorithm that selects items solely based on their replication test scores outputs solutions within a constant factor of the optimum for 
the class of functions satisfying an extended diminishing returns property.
Our algorithms and approximation guarantees assume that test scores are noisy estimates of certain expected values with respect to marginal distributions of individual item values, thus making our algorithms practical and extending previous work that assumes noiseless estimates. Moreover, we show how our algorithm can be adapted to the setting where items arrive in a streaming fashion while maintaining the same approximation guarantee. We present numerical results, using synthetic data and data sets from the Academia.StackExchange Q\&A forum, which show that our test score algorithm can achieve competitiveness, and in some cases better performance than a benchmark algorithm that requires access to a value oracle to evaluate function values.
}


\maketitle

\section{Introduction}\label{sec:intro}

The problem of selecting the most profitable group of items from a large pool of items arises in a wide range of applications. A common approach for solving such problems is to rank items according to certain criteria and then selecting a top set of items based on this ranking. In online shopping websites, an important task is to display the most relevant items in a limited space in order to encourage customers to purchase and thereby maximize the revenue, e.g., assortment optimization~\citep{assortment,combinatorial-assortment}. For sports teams, it is crucial to recruit high-performance players, and so is for team selection in online gaming~\citep{skill-rating}. In online freelancing platforms, such as Fiverr, PeoplePerHour, and Topcoder, the goal is to hire freelancers to perform a service for a client based on their expertise and hourly pay rates. Content recommendation in online platforms~\citep{recommend} is also a notable application. In the context of information retrieval, the goal is to return a set of relevant items to a user search query, which often amounts to selecting items based on their individual relevance scores, an area referred to as learning to rank \citep{Li11}.

As a concrete motivating practical application scenario, consider an online platform in which users contribute solutions to tasks posted by other users. For example, tasks may require providing consultancy or problem solving by experts in crowdsourcing or labour market platforms, and different experts may require different payments depending on their expertise levels. Having received several solutions for a task, the value of the set of solutions to the task owner may be according to different valuation functions. For example, the task owner may only extract a value from the best solution, or a few best solutions, and in some applications, the task owner may also extract value from overall user engagement. The quality of submitted solutions would positively correlate with the expertise level of experts, and would have some intrinsic randomness.

When implementing the ranking based approach for group selection, there are several practical challenges. Item values are often random quantities, and the underlying uncertainty and variability therein make it difficult to evaluate them. Besides this uncertainty in item values, it is not straightforward how to rank the items. Sometimes, adding ``high-risk high-return" items to the selected set is a good strategy as their low average performance can be mitigated by other items. Like this example, the group value of a set of items is often not linear in individual item values, sometimes due to diminishing marginal utilities or substitutional and complementary interaction among items. Hence, it is important to come up with a ranking metric that reflects not only individual item values but also their contribution to the objective of maximizing the group value.

Motivated by these challenges, the focus of this paper is on solving the group selection problem, and other more general combinatorial optimization problems, by ranking items based on only some statistics of item values and costs. Throughout this paper, any quantities defined to evaluate items based on individual item values and costs will be referred to as test scores. As test score of an item depends only on its value distribution, test scores can be computed in a decentralized or distributed computing fashion. Hence, algorithms based on test scores are more efficient in their nature than algorithms that require access to multiple items' information at the same time. However, it may be difficult to infer the true value of a group from the scores of its individual members, due to complicated interactions between them. With this in mind, the goal of this paper is to understand the performance of test score-based algorithms.

\cite{kleinberg15} (see~\cite{Kleinberg18} for the journal version) first introduced the notion of test scores in the context of the team formation problem. Each player takes multiple trials, and only the best performance out of them is kept as the final score of the player. As the performance at each trial is a random quantity, the test score of a player is defined as the expectation of the best performance from a given number of trials. They used the test scores to solve the team selection problem where the team value is measured by the sum of \emph{the best few performance values} amongst the team members, i.e., the objective is the so-called \emph{top-$r$} function (defined formally in Section~\ref{sec:problem}%
). The important idea is that each individual item is evaluated based on a score metric which depends only on the distribution of a single item. \cite{kleinberg15} proved that selection of top players according to the test scores achieves a constant factor approximation with the constant factor of value roughly $1/30$. They provided a negative result as well; note that the top-$r$ function is submodular, but there exists a submodular set function for which no test score-based algorithm can guarantee a constant factor approximation; we discuss this further in Section~\ref{sec:examples}.

\cite{sekar2019test-score} studied the concept of using test scores for more general objective functions, and introduced a framework that allows to establish constant-factor approximation guarantees for more general classes of submodular objectives. They devised the notion of \emph{replication test scores} that generalizes Kleinberg and Raghu's test scores. The replication test score of an item measures how well the item performs when assigned to a group defined as the value of a virtual group of multiple independent copies of the same item, which is similar to the interpretation of performing multiple trials. They used the replication test scores to solve \emph{stochastic utility function maximization subject to a cardinality constraint} problem where the utility function comes from the class of functions satisfying a diminishing returns property (defined in Section~\ref{sec:problem}), thus generalizing previous results holding for maximum or best performance functions. \cite{sekar2019test-score} showed that greedy selection of items according to replication test scores guarantees a constant-factor approximation with the constant factor of value roughly 1/7.3.

The previous work on using individual item scores focused on combinatorial optimization subject to a cardinality constraint; basically, the problem of selecting at most a certain number of items maximizing a given objective. The item score essentially quantifies the expected marginal contribution of the item to a set when a given number of other items are already in the set.
Although the cardinality constraint allows us to accommodate important applications, it is limited to settings with equal item costs, %
while the case when items have heterogeneous costs often arise in applications, like in the aforementioned examples. For the heterogeneous costs, score metrics, based merely on the value distribution of an item, may no longer be the right measure for the item's performance. To take into account item costs, we should consider how many more items can be selected together with a given item when evaluating the marginal contribution. Even with a score metric that considers both the cost of an item and its value distribution, the algorithm of simply selecting a few items with the highest score values, studied by~\cite{kleinberg15} and~\cite{sekar2019test-score}, may not work. An algorithm should consider the costs of items explicitly, even if the chosen metric already has some dependency on the item costs.

The goal of this paper is to show that the combinatorial optimization problem of selecting a set of items that have random values and arbitrary deterministic  costs can be (approximately) solved by an efficient and scalable test score-based algorithm. The problem we consider is the \emph{budgeted stochastic utility function maximization}: given a ground set of items $\Omega = \{1,\ldots, n\}$, the problem is to find a set that solves
\begin{equation}
\hbox{ maximize  } u(S)=\E[f(\bvx_S)] \hbox{ over } S\subseteq \Omega \hbox{ subject to } \sum_{i\in S}c_i \leq B
\label{equ:probdef}
\end{equation}
where $f$ is a valuation function, $\bvx_S$ are independent random values of items in set $S$, $c_1,\ldots,c_n$ are postive-valued item costs, and $B$ is the budget limit. The budget constraint is often referred to as a \emph{knapsack constraint}, and the \emph{knapsack problem} is a special case of (\ref{equ:probdef}) where the valuation function $f$ computes the sum of the performance values of items in a given set $S$ (amounting to a modular utility set function). We consider problem (\ref{equ:probdef}) for a subset of monotone submodular utility functions, defined for group valuation functions that are monotone and satisfy the extended diminishing returns property. 

The framework of test score algorithms requires to first represent each item in the ground set $\Omega$ with a deterministic scalar score by using a value oracle access to distributions of item values (or samples from these  distributions) and item cost values, and then approximately solving problem (\ref{equ:probdef}) by using only the item score values and the budget constraint parameters. Note that the framework rules out algorithms that require value oracle access for the utility function $u$ value for arbitrary subsets of the ground set of items, which is commonly used by algorithms for solving submodular function maximization problems. We consider a test score measure to evaluate individual items based on their costs and values. Given the value distribution $P$ and the price $c$ of an item, a test score of the item is defined as $a(P, c)$ for some score function $a$. Here, our goal is to provide a proper score function $a$ so that an algorithm selecting items solely based on the corresponding test scores can attain a constant-factor approximation. The choice of the test score function $a$ depends on the group valuation function $f$.

A related problem setting is when items arrive in a \emph{streaming fashion} and the decision-maker can access or keep only a small number of items at any given point. In practice, there are typically a massive number of items stored across a large number of servers in a distributed computing system, and it may be reasonable to assume that the list of all items is not available. 
This is problematic for algorithms that rely on value oracle access for function evaluations, as it is often necessary for the algorithms to evaluate combinations of items. Estimating the values of items and sets of items is another hurdle especially when we need to process incoming items in a timely manner.
From this perspective, test score-based algorithms are advantageous as such algorithms use only some statistics of item values and can be implemented in a distributed computing setting. This paper demonstrates that a test score-based algorithm developed for the budgeted stochastic utility function maximization problem can be adapted to the streaming setting while keeping the same approximation guarantee. 

Another practical concern is that item value distributions are often unknown to the decision-maker, which makes it difficult to compute the exact value of a test score. Therefore, one might be interested in understanding how the performance of an algorithm is affected by noisy estimation, and at the same time, we look for an algorithm that is robust to estimation noise. The previous works by~\cite{kleinberg15} and~\cite{sekar2019test-score} on test scores assume that the value distribution of each item is given so that the exact score value of each item can be computed, but in practice, a test score-based algorithm would choose items based on the estimated values of their scores using a dataset containing observations of item values or by actively testing individual items. In this paper, we unravel how the presence of estimation noise affects a test score-based algorithm's approximation guarantee. %

\subsection{Summary of our contributions}\label{sec:contributions}

Our contributions can be summarized in the following points.
\begin{itemize}
    \item When ranking items that have different costs, the item costs as well as their values need to be taken into consideration. To account for this, we design a test score method that takes item costs as well as value distributions into account when evaluating individual items with heterogeneous costs. We generalize the definition of replication test scores such that individual item costs are also part of the definition; we define replication test scores, for each item $i\in \Omega$, as the expected value of a set of items that consists of $k_i$ independent copies of this item, where $k_i = \lfloor B/c_i\rfloor$. This definition combines the effect of item values and their costs, and the formal definition is in Section~\ref{sec:score}. Intuitively, the smaller the cost of an item, the larger its replication test score.
    \item We develop a simple greedy algorithm that selects items solely based on estimated cost-dependent replication test scores. Despite our test score algorithm uses only some statistic for representing each individual items, we show that our algorithm achieves a constant-factor approximation for a wide range of utility functions (Section~\ref{sec:algorithm}). Our analysis may be of independent interest; it is a novel framework based on lower and upper bounding the utility of a given set by the maximum, minimum, and average scores of items in the set. Moreover, unlike previous results on test score algorithms, we allow noisy evaluations and estimation errors in individual test score computations. We give a refined analysis that delineates how the performance of our algorithm depends on given estimation errors.
    \item We also find out that our test score-based algorithm as well as its analysis can be simply translated to the \emph{streaming utility function maximization}. %
    In Section~\ref{sec:streaming}, we show that we can adapt the test score greedy algorithm for the streaming setting, thereby achieving the same approximation guarantee as for the non-streaming setting. To our knowledge, our greedy algorithm is the most memory efficient single pass algorithm that can guarantee a constant-factor approximation. 
    \item In Section~\ref{sec:sampling}, we derive sufficient sample sizes for estimation of replication test scores within a prescribed statistical estimation error, which is specially tailored for the underlying optimization problem. These results specify sufficient sample sizes in terms of key parameters such as the budget value, individual item cost values, and curvature properties of the underlying group valuation function. %
    These bounds provide insights into how the sample sizes depend on the underlying group valuation function. 
    \item In Section~\ref{sec:num}, we report the results of a numerical evaluation whose goal is to evaluate the efficiency of our algorithm for the budgeted stochastic utility maximization problem under various assumptions about item value distributions and costs of items, and by varying various factors such as the choice of the group valuation function and the test score estimation sample size%
    . We have studied this by carefully designed experiments using synthetic simulated data for generating item values and costs, and real-world datasets. For the latter, we used historical data for the online Q\&A service StackExchange, where items correspond to experts providing answers to questions, and the group value corresponds to value derived from answers submitted for a question. We compared the performance with a benchmark algorithm that uses value oracle calls. Our results demonstrate the efficiency of the proposed test score algorithm across a wide range of settings. 
\end{itemize}

\subsection{Related Work}

Submodular maximization is a common framework for modeling situations where the decision-maker needs to select an assortment of items that has the maximum group utility. To name a few applications, submodular maximization is used for sensor placement~\citep{sensor-placement}, influence maximization~\citep{influence}, risk-averse utility maximization~\citep{MNL}, document summarization~\citep{HLT10,HLT11}, image summarization~\citep{image1,streaming-matchoid2,streaming-matchoid3},  feature and variable selection~\citep{variable-selection}, and active set selection in non-parametric learning~\citep{active-set}. 

Near-optimal algorithms have been proposed for variants of the problem, e.g., algorithms for the cardinality constraint~\citep{cardinality1,cardinality2,cardinality3} and for the knapsack constraint~\citep{kdd-knapsack,Sviridenko,Yoshida,Ene}. \cite{Feldman} studied algorithms for monotone continuous submodular function maximization subject to a linear constraint on a compact set of positive real numbers. \cite{Niazadeh} studied algorithms for non-monotone continuous submodular functions over a hypercube. All these papers studied algorithms that use value oracle queries for the objective function or its gradient, which is not possible in our test score framework.

\cite{asadpour2016submod} studied the problem of maximizing a stochastic monotone submodular function subject to a matroid constraint and showed that the adaptivity gap---defined as the ratio between the values of optimal adaptive and optimal non-adaptive policies---is bounded and is equal to $1/(1-1/e)$. \cite{golovin} further developed the framework for more general partially observable stochastic optimization problems, and introduced the property of adaptive submodularity under which a simple adaptive greedy algorithm is guaranteed to achieve near-optimal solutions. In these papers, an adaptive policy specifies which item to select next conditional on having already selected a set of items and having observed realizations of their values. The greedy adaptive policies studied in these papers use value oracle queries for the conditional expected marginal utility of adding an item to a set given observed values of items in this set. \cite{stosub1} and \cite{stosub2} studied stochastic variants of the continuous greedy algorithm for submodular function maximization, which also rely on value oracle queries.

Maximization of submodular set functions, defined as the expected value of a function of random variables, was studied by \cite{influence}, for the problem of maximizing the expected size of an information cascade, and by \cite{coverage} for more general stochastic submodular maximization problems whose objective can be expressed as a coverage function. Both these papers require value oracle calls and cover different classes of submodular functions than in our paper. Another related line work is on noisy submodular maximization~\citep{noisy1,noisy2,hassidim2017submodular,Qian}, where randomness is incorporated in noise of value oracle outputs (e.g. additive or multiplicative noise). Our algorithm may be seen as using access to an approximate value oracle, which outputs marginal valuations of items based solely on some test score representation of items. 

The problem we study is different from the aforementioned previous works as, following the framework of test scores, all items in the ground set need first to be represented with deterministic scalar score values computed by using a value oracle access to item value distributions (or samples from item value distributions) and item cost values, and then items must be selected by using only these score values and budget constraint parameters. In our case, while selecting items, the algorithm does not have access to realized values of selected items.

The most closely related work to our paper is that on test score algorithms. As discussed in the introduction, the framework of test score algorithms was introduced by \cite{kleinberg15}, and greedy selection of items according to some test scores was shown to guarantee a constant-factor approximation, for group valuation function defined as the sum of a fixed number of highest values from the set of values of selected items; defined as top-$r$ function in Section~\ref{sec:problem}. \cite{sekar2019test-score} showed that a constant-factor guarantee can be achieved by test score algorithms for a broader class of group valuation functions satisfying the extended diminishing returns property. %
\cite{MNPR20} studied the problem of selecting a set of items of given cardinality that maximizes the expected highest item value or the expected second highest item value. They provided a PTAS for the former problem that runs in polynomial time in the total number of items and the cardinality parameter $k$, and a constant-factor approximation algorithm for the latter problem. For maximizing the expected highest item value, their algorithm uses discretized distributions of item values, with $\Theta(k\log(k))$ support sizes, which are computed by a joint computation using value oracle access to all item value distribution.

We next discuss related work on streaming utility maximization problems, focused on designing algorithms that require a small number of passes over a stream of items and a small size memory. Various algorithms have been proposed for streaming submodular maximization subject to cardinality constraints~\citep{examplar,streaming-cardinality,streaming-subsampling},  knapsack constraints~\citep{streaming-knapsack1,streaming-knapsack2,streaming-knapsack3}, intersection of matroid constraints~\citep{streaming-matroid},  $p$-matchoid constraints~\citep{streaming-matchoid1,streaming-matchoid2,streaming-matchoid3}, and $p$-extendible and $p$-set systems~\citep{streaming-set}. We remark that cardinality constraints are nothing but matroid constraints that correspond to uniform matroids. $p$-matchoid constraints generalize matroid constraints, in that the intersection of $p$ matroids is a $p$-matchoid, and $p$-extendible systems and $p$-set systems further generalize $p$-matchoids.
Knapsack constraints are neither matroid nor matchoid constraints. A knapsack constraint is a $\lceil c_{\max}/c_{\min}\rceil$-extendible system where $c_{\max}=\max_{i\in\Omega} c_i$ and $c_{\min}=\min_{i\in\Omega} c_i$, but here, $\lceil c_{\max}/c_{\min}\rceil$ can be arbitrarily large. The challenge in the streaming setting is that, due to the restriction on the memory size and the number of passes of streams, the list of combinations of items that are accessible at a given time is limited and it also depends on the order in which items are input to the algorithm. Our algorithm is advantageous for this setting because the requirement of selecting the items with the highest replication test scores can be accomplished by iterative replacements in a single pass over input stream of items.
	
\section{Problem formulation}
\label{sec:problem}

In this section, we formulate the budgeted stochastic submodular maximization problem that we study and the class of algorithms based on using test scores.

\subsection{Budgeted stochastic utility maximization problem} Let $u:2^{\Omega}\rightarrow \mathbb{R}_+$ be a set function, where $\Omega=\{1,\ldots, n\}$ is a ground set of items. Our focus is on set functions that admit the following form 
\begin{equation}
u(S) = \E[f(X_S)], \hbox{ for } S\subseteq \Omega
\label{eq:utility}
\end{equation}
where $f:\mathcal{X} \rightarrow \mathbb{R}_+$ is some given function on domain $\mathcal{X} = \mathcal{X}_1\times \cdots \times \mathcal{X}_n\subseteq \mathbb{R}_+^n$, and $X_S$ denotes the $n$-dimensional vector with each coordinate $i\in S$ equal to $X_i$ and, otherwise, equal to $0$. We refer to $f$ as a \emph{group valuation} function, or a \emph{valuation} function, interchangably. The expectation in (\ref{eq:utility}) is with respect to a product-form distribution of $X_1, \ldots, X_n$ with respective marginal distributions $P_1, \ldots, P_n$. We may interpret $u(S)$ as the expected value of a set of items $S$, where the value of a set of items is defined to be according to function $f$ of independent random item values.

We consider the \emph{budgeted stochastic utility maximization} problem defined as: find a set $S^*$ that solves
\begin{equation}\label{eq:problem-def}
\hbox{maximize } u(S) \hbox{ over } S\in \cF=\left\{S'\subseteq \Omega:\ c(S')=\sum_{i\in S'}c_i\leq B\right\}
\end{equation}
where $c_1,\ldots,c_n$ are given positive-valued item costs, $B$ is the budget limit, and $u$ admits the form in (\ref{eq:utility}) for some valuation function $f$. A special case is when there is a cardinality constraint, in which case, $c(S)\leq B$ represents $|S|\leq k$ for some given positive integer $k$.

We assume that $f$ satisfies the following conditions:
\begin{description}
\item (C1) $f$ is a permutation invariant function;
\item (C2) $f$ is a monotone submodular function;
\item (C3) $f$ satisfies extended diminishing returns property;
\item (C4) $f$ is a continuous function. 
\end{description}
These conditions accommodate various group valuation functions. In the following section, we provide definitions of the function properties stated in the above conditions, except for continuity which is a basic function property, and discuss how they relate to various other definitions. We then discuss various examples of group valuation functions that satisfy conditions (C1)-(C4).

\subsection{Function classes} 

A function $f$ is said to be \emph{permutation invariant} if for every $x = (x_1, \ldots, x_n)\in \mathcal{X}$ and every permutation $\pi(x)$ of elements $x_1, \ldots, x_n$, $f(x) = f(\pi(x))$. Intuitively, permutation invariance means that the value of a group depends only on individual values and not on their identities. Note that under the permutation invariance condition, $f$ defines a set function that maps any given set of points $\{x_i: i\in S\}$, with $x_i\in \mathcal{X}_i$, for $i\in S$, and $S\subseteq \Omega$, to a real value $f(x_S)$. Under permutation invariance condition, we can write $f(\{x_i: i\in S\})$ in lieu of $f(x_S)$ because the value of the function is invariant to permutation of its input variables.

A function $f$ is said to be \emph{monotone} if $f(x)\leq f(y)$ for every $x,y\in \mathcal{X}$ such that $x\leq y$, where the last inequality holds coordinate-wise. A function $f$ is said to be \emph{submodular} if $f(x\wedge y) + f(x\vee y)\leq f(x)+f(y)$ for every $x,y\in \mathcal{X}$, where $\wedge$ and $\vee$ denote coordinate-wise minimum and maximum operators, respectively. By \cite{bian},  a function $f$ is submodular if, and only if, it satisfies the following \emph{weak diminishing returns condition}: for every $x,y\in \mathcal{X}$ such that $x\leq y$ and $x_i = y_i$ for some $i\in \Omega$, and all $z\in \mathbb{R}_+$ such that $x+ze_i\in \mathcal{X}$ and $y+ze_i \in \mathcal{X}$, $f(y+ze_i)-f(y)\leq f(x+ze_i)-f(x)$. By \cite{topkis}, if $f$ is twice-differentiable on its domain, then $f$ is submodular if, and only if, all off-diagonal elements of the Hessian
matrix of $f$ are non-positive, i.e. $\partial^2 f(x)/ \partial x_i \partial x_j\leq 0$, for all $i,j\in \Omega$ such that $i\neq j$ and $x\in \mathcal{X}$.

A set function $u$ on domain $2^{\Omega}$ is said to be submodular if $u(S\cap T) + u(S\cup T)\leq u(S)+u(T)$, for every $S,T\subseteq \Omega$. Equivalently, $u$ is submodular if it satisfies the following \emph{diminishing returns property}: $u(T\cup \{i\}) - u(T) \leq u(S\cup \{i\})-u(S)$, for all $i\in \Omega$ and $S,T\subseteq \Omega$ such that $S\subseteq T$. By \cite{asadpour2016submod}, if $u$ is a set function of the form (\ref{eq:utility}) where $f$ is a monotone submodular function, then $u$ is a monotone submodular set function.

Following \cite{sekar2019test-score}, a function $f$ is said to satisfy \emph{the extended diminishing returns property} if for any $v\geq 0$ that has a non-empty preimage under $f$ and $i\in \Omega$, there exists $y\in \mathcal{X}$ with $y_i = 0$ such that (i) $f(y)=v$ and (ii) $f(y+ze_i)-f(y)\leq f(x+ze_i)-f(x)$ for any $z\in \mathbb{R}_+$ and any $x\in \mathcal{X}$ such that $f(x)\leq f(y)=v$ and $x_i = 0$. A simple but different property is that $f(y+ze_i)-f(y)\leq f(x+ze_i)-f(x)$ holds for every $z\in \mathbb{R}_+$ and $x,y\in \mathcal{X}$ such that $f(x)\leq f(y)$, that is, the marginal return is smaller for a vector with a larger value according to function $f$. In fact, this is strictly stronger than the extended diminishing returns property, so we admit the definition of the extended diminishing returns property to cover a larger set of functions.

It is worthwhile to discuss how the extended diminishing returns property compares to the known notion of diminishing returns (DR) submodular functions \citep{bian,Soma}. A function $f$ is said to be \emph{DR-submodular} if for all $x,y\in \mathcal{X}$ such that $x\leq y$ and any standard basis vector $e_i$ and $z\in \mathbb{R}_+$ such that $x+ze_i \in \mathcal{X}$ and $y+ze_i \in \mathcal{X}$, $f(y+ze_i)-f(y)\leq f(x+ze_i)-f(x)$. Clearly, DR-submodular functions are a proper subset of submodular functions. If $f$ is twice-differentiable, then $f$ is DR-submodular if, and only if, all elements of the Hessian
matrix of $f$ are non-positive. There are functions that satisfy the extended diminishing returns conditions and are not DR-submodular. For example, any CES function with $r > 1$ is convex, and thus coordinate-wise convex, and, hence, is not DR-submodular, while it satisfies the extended diminishing returns property. We define CES and some other functions that satisfy the extended diminishing returns property in the following section.

\subsection{Examples of functions satisfying the extended diminishing returns property}\label{sec:examples} We present examples of monotone submodular functions that satisfy the extended diminishing returns property, which were used for modelling a group value in various previous works.

\paragraph{(1) Total production:} $f(x)=g(\sum_{i=1}^n x_i)$ where $g:\mathbb{R}_+ \rightarrow \mathbb{R}_+$ is a non-decreasing concave function. The \emph{stochastic knapsack problem with a concave objective}~\citep{s-knapsack,nonlinear-knapsack} can be formulated by using the total production function:
\begin{equation}\label{eq:knapsack-problem}
\max\bigg\{g\bigg(\sum_{i\in S}p_i\bigg):\ c(S)\leq B,\ S\subseteq \Omega \bigg\}
\end{equation}
where $p_i$ is the random profit of item $i$ for $i\in \Omega$. The constrained assortment optimization problem under the multi nomial logit (MNL) choice model with unit prices studied by~\cite{MNL,MNL-submodular} can also be formulated as the optimization problem of the form~\eqref{eq:knapsack-problem}. To be specific, the objective is ${p\sum_{i\in \Omega}v_iz_i}/({v_0+\sum_{i\in \Omega}v_iz_i})$ where $p$ is the fixed price of items, $v_i$ is the probability of item $i$ being purchased when displayed in an assorment, and $v_0$ is the probability of no purchase. Here, $g(z)={pz}/({v_0+z})$ is a concave function. Moreover, \emph{client selection} within \emph{federated learning} framework can be formulated as~\eqref{eq:knapsack-problem}~\citep{client-selection}. Another application is the \emph{(competitive) facility location problem}~\citep{ahmed-atamturk,c-facility}:
\begin{equation}\label{eq:facility-location}
\max\bigg\{\sum_{j\in M}f_j\bigg(\sum_{i\in S}p_{i,j}\bigg):\ c(S)\leq B,\ S\subseteq N\bigg\}
\end{equation}
where $M$ is the set of customers and $N$ is the set of locations for opening up facilities. It can be easily observed that any nonnegative linear combination of functions that satisfy the extended diminishing returns property also has the property, which means that the objective in~\eqref{eq:facility-location} also satisfies the property. The \emph{saturated coverage function}~\citep{data-selection1,data-selection2}, the \emph{feature-based submodular function}~\citep{data-selection1,data-selection2}, and the \emph{Na\"ive Bayes submodular function}~\citep{data-selection3} used for data subset selection are all of the form~\eqref{eq:facility-location}.

\paragraph{(2) Top-$r$:} $f(x)= \sum_{i=1}^r x_{(i)}$ where $x_{(i)}$ is the $i\textsuperscript{th}$ largest value in $\{x_1,\ldots,x_n\}$. In~\eqref{eq:knapsack-problem}, instead of maximizing the total production, one might be interested in finding a set whose \emph{representative} items have high profits. When $r$ items can be appointed as representatives, based on their profits, we can optimize $f(x)= \sum_{i=1}^r x_{(i)}$ where $x=(p_1z_1,\ldots,p_n z_n)$ over $z\in\{0,1\}^n$. \cite{kleinberg15} used the top-$r$ objective for the team formation problem. In particular, when $r=1$, the function is often referred to as the \emph{best-shot} function. Another related application is the \emph{data subset selection problem} in machine learning. The data subset selection problem using the \emph{nearest neighbor submodular function} in~\citep{data-selection3} can be formulated as follows:
\begin{equation}\label{eq:max-comb}
\max\bigg\{\sum_{j\in V}\max_{i\in S}\{w_{i,j}\}:\ c(S)\leq B,\ S\subseteq \Omega\bigg\}
\end{equation}
where $V$ is the set of entire data set and $w_{i,j}$ is the \emph{similarity} between $i$ and $j$. As the objective in~\eqref{eq:max-comb} is a nonnegative combination of the best-shot functions, it satisfies the extended diminishing returns property. The exemplar clustering formulation in~\citep{examplar} uses a function which is a special case of~\eqref{eq:max-comb}.

\paragraph{(3) Constant elasticity of substitution (CES) utility function: } $f(\vx)=(\sum_{i=1}^n x_i^r)^{1/r}$, for $r > 0$. This is a classical production function used in the economic theory. It is used to model the effect of capital and labor on the production as well as a utility function in consumer theory where $x_i$ corresponds to consumption of good $i$ and $f(x)$ corresponds to the aggregate consumption. The CES function is used by~\cite{CES} for the team formation problem where $r$ captures the degree of substitutability of the task performed by the players. The items are perfect substitutes when $r$ approaches infinity and perfect complements when $r$ approaches zero. We may think of the CES function as $r$ approaches infinity to be a soft version of the maximum function. Any CES function with $r\geq 1$ satisfies the extended diminishing returns property.

\paragraph{(4) Success probability:} $f(x)=1-\prod_{i=1}^n(1-p(x_i))$, where $p:\mathbb{R}_+\to[0, 1]$ is an increasing function and $p(0) = 0$. An application of the success probability is a variant of the \emph{maximum weighted coverage problem}~\citep{coverage}. Consider $f(x)=\sum_{u\in U}w(u)\left(1 - \prod_{u\in B_i}(1-p(x_i))\right)$, where $U$ is the ground set, $B_1,\ldots, B_m$ are $m$ subsets of $U$, and $p:\mathbb{R}_+\to[0, 1]$ is an increasing function that satisfies $p(0) = 0$. Here, each $B_i$ has value $x_i$ and the probability of $B_i$ covering its elements is given by $p(x_i)$. When the weight of each element $u$ is given by $w(u)\geq0$, the expected weight of the elements covered by $\{B_i:i\in S\}$ is precisely $u(S):=\E[f(X_S)]$.

Although various important classes of submodular functions, as listed above, satisfy the extended diminishing returns property, we remark that not all submodular functions enjoy this property. We will show in the next section a test score algorithm that guarantees a constant-factor approximation for any submodular function with the extended diminishing returns property. In contrast,~\cite{kleinberg15} proved that no test score algorithm with a constant-factor approximation guarantee exists for a variant of set coverage functions over a graph and the matroid rank function, implying in turn that these functions do not satisfy the extended diminishing returns property.

\subsection{Test score based algorithms} We consider the class of test score algorithms for finding an approximate solution to the budgeted stochastic utility maximization problem in (\ref{eq:problem-def}). This class of algorithms uses as input scalar test scores $a_1, \ldots, a_n$ representing values of items, the cost values of items $c_1, \ldots, c_n$, and the budget parameter $B$. Given this input, the algorithm is required to output a set of items $S$ in $\mathcal{F}$ in a polynomial computation time. A key step in deriving approximation guarantees for test score algorithms is to define test scores $a_1, \ldots, a_n$ appropriately so that they well capture the contribution of an item to a set value according to given valuation function $f$, as well as account for underlying budget constraint defined by $\mathcal{F}$. Our aim is to show that there exist test score algorithms that can achieve a constant-factor approximation for the budgeted stochastic utility maximization problem in (\ref{eq:problem-def}). Aiming for a constant-factor approximation is justified by known inapproximability results for the class of monotone submodular functions \citep{Kleinberg18}, and prior work that has shown existence of a test score algorithm that guarantees a constant-factor approximation for maximizing a utility function of the form (\ref{eq:utility}), subject to a cardinality constraint, for valuation functions $f$ that satisfy the extended diminishing returns property \citep{sekar2019test-score}.	
\section{A test score %
algorithm and approximation guarantees}\label{sec:algorithm}

In Section~\ref{sec:score}, we define our test score measure, which extends the definition of replication test score to the setting of heterogeneous item costs. Then we discuss a greedy algorithm based on the modified version of replication test score in Section~\ref{sec:algo}. In fact, the algorithm chooses items based on \emph{estimates} of their replication test scores. We provide a condition on the required estimation accuracy to guarantee a good approximation. In Sections~\ref{sec:apx0}--\ref{sec:curvature}, we analyze the approximation quality of our test score-based greedy algorithm. Theorem~\ref{thm:constant1} provides a worst-case approximation ratio for all valuation functions satisfying the extended diminishing returns property and all possible %
values of item costs. Theorem~\ref{thm:constant2} provides a more refined analysis and describes how the approximation ratio depends on the variations in the item costs. Lastly, Theorems~\ref{thm:parametrized-guarantee1} and~\ref{thm:parametrized-guarantee2} give approximation guarantees that depends on the \emph{curvature} of the underlying valuation function.

\subsection{Replication test scores for general budget constraints}\label{sec:score}

In order to define replication test scores, we need to extend some of our definitions. Let $\tilde{\Omega}$ be an extended ground set of items, which for each item $i\in \Omega$ contains an infinite collection of replicas denoted as $i^{(1)}, i^{(2)}, \ldots$. Let $f$ be a set function, for any $d\geq 1$, mapping each set of values $\{x_1, \ldots, x_d\}$ to $f(\{x_1, \ldots, x_d\})\in \mathbb{R}_+$, where $x_i\in \mathbb{R}_+$ for each $i=1,\ldots, d$. With these new definitions, $u$ is a set function with the domain being the power set $2^{\tilde{\Omega}}$.

For a given valuation function $f$ and item distributions $P_1, \ldots, P_n$, we define the \emph{replication test score} of each item $i\in \Omega$ as
\begin{equation}
\label{eq:replication-test-score}
r_i:=u(\{i^{(1)},i^{(2)},\ldots,i^{(k_i)}\})=\E[f(\{X_i^{(1)},X_i^{(2)}\ldots,X_i^{(k_i)}\})]
\end{equation}
where 
$k_i:=\lfloor{B}/{c_i}\rfloor$
and
$X_i^{(1)},\ldots,X_i^{(k_i)}$ are $k_i$ independent and identically distributed random variables with distribution $P_i$.  We may interpret the replication test score of an item as the expected performance of a virtual set of as many replicas of this item as possible while satisfying the budget constraint. Note that the definition of replication test score in (\ref{eq:replication-test-score}) accounts for both the distribution of item value and its cost. By the definition of replication test score in~\eqref{eq:replication-test-score}, when the cost of an item is low, many copies of the item can be packed and the score can be potentially large, even if the value of a single copy is small. In particular, when $f$ is simply the summation of its arguments, the replication test score of an item $i\in \Omega$ is nothing but $\lfloor{B}/{c_i}\rfloor\E[X_i]$. Up to the scaling factor $B$, this value is close to $\E[X_i]/c_i$, the quantity often referred to as the \emph{density} of item $i$ for the knapsack problem. With this in mind, the reader would notice similarities between our test score based greedy algorithm and the greedy algorithm for the knapsack problem.

In the original definition by~\cite{sekar2019test-score}, precisely $k$ copies are taken for each item, where $k$ is the cardinality constraint. Therefore,~\eqref{eq:replication-test-score} is consistent with the definition in \cite{sekar2019test-score}. One might think of a different extension; instead of taking a varied number of copies for an item, one can take a fixed number, $B$, of copies inside the expectation and dividing it by the item cost. However, a part of our analysis fails with that definition.

\subsection{A greedy test score algorithm}\label{sec:algo}

Having access to the replication test scores of individual items, our goal is to find a ``good'' combination of items. By a good combination, we mean a set of items $S\subseteq \Omega$ whose utility $u(S)$ is at least a constant factor of the optimal utility. Specifically, we want to find a set $S\subseteq \Omega$ such that $u(S)\geq \alpha \cdot u(OPT)$ where $\alpha$ is a fixed constant in $[0,1]$ and $OPT$ is a set of items that is an optimal solution to~\eqref{eq:problem-def}. The restriction here is, our decision of which set to choose is made based solely on the replication test scores of items, costs of items, and the available budget. 

We consider a natural greedy algorithm based on the replication test scores and show that it returns a good combination of items under some conditions. In fact, we allow for estimation noise and assume that only \emph{estimates} of item replication test scores are given. Let $\hat r_i$ be the estimate of the replication test score of item $i\in \Omega$. We construct a set $S^*$ by greedily selecting items in decreasing order of the estimated replication test scores until the remaining budget cannot afford any additional item. Without loss of generality, we assume that
\begin{enumerate}
	\item $\hat r_1\geq \hat r_2\geq\cdots\geq \hat r_n$,
	\item $k$ is the smallest index in $\Omega$ satisfying $c_1+c_2+\cdots+c_{k+1}>B$, and
	\item $r_1,\ldots,r_{k+1} \geq (1-\epsilon)  \max\{r_i: i\in \Omega\setminus \{1,\ldots, k\}\}$.
\end{enumerate}
Under these assumptions, we have $\left\{1,\ldots,k\right\}\subseteq S^*$. Upon obtaining $S^*$, we construct another set $S^{**}$ in the following way. We start by adding the first item rejected by the greedy algorithm for $S^*$. Then, with the remaining items, we do the greedy addition until the remaining budget cannot afford to take any additional item. By construction, we know that $\left\{k+1\right\}\subseteq S^{**}$. Finally, we report the better of $S^*$ and $S^{**}$. The pseudo-code of the algorithm is given in Algorithm~\ref{algo:simple-greedy}.

\begin{algorithm}[t]
	\caption{Greedy test score algorithm}\label{algo:simple-greedy}
	\begin{algorithmic}
		\Require $\hat r_1\geq \hat r_2\geq\cdots\geq \hat r_n$ and $k\in\Omega$ is the index satisfying {$c(\{1,\ldots,k\})\leq B < c(\{1,\ldots, k+1\})$}.
		\State Start with $S^*=\{1,\ldots, k\}$, and for $i=k+2,\ldots,n$, add $i$ to $S^*$ whenever possible
		\State Start with $S^{**}=\{k+1\}$, and for $i=1,\ldots,k,k+2,\ldots,n$, add $i$ to $S^{**}$ whenever possible
		\State Return the better of the two sets.
	\end{algorithmic}
\end{algorithm}

Algorithm~\ref{algo:simple-greedy} is different from the greedy algorithm used in the previous works on test score algorithms, including \cite{kleinberg15} and \cite{sekar2019test-score}, 
designed for the special case of cardinality constraints. Therein, the algorithm simply selects a set of items of given cardinality that have highest test scores. A na\"ive extension of this algorithm to allow for more general budget constraints is to greedily select a set of items that have highest test score values subject to the total cost of these selected items being within given budget. However, there exist instances for which the outcome of this na\"ive greedy algorithm has an arbitrarily poor value. For instance, imagine two items such that item 1 has $c_1=B$ and $\E[X_1]=M$ for some large integer $M$ while item 2 has $c_2=B/(M+1)$ and $\E[X_2]=1$. Hence, we cannot afford to select both items. Since $k_1=1$ and $k_2=M+1$, for the modular valuation function, we have $r_1=\E[X_1^{(1)}]=M$ and $r_2=\E[X_2^{(1)}+\cdots+X_2^{(M+1)}]=M+1$. Thus, the na\"ive greedy algorithm would pick item 2 instead of item 1. However, the value of item 1 is $M$ times that of item 2, and $M$ can be arbitrarily large.

We mitigate this by constructing another set of items, which contains the first item rejected by the na\"ive greedy algorithm due to the budget restriction, and selecting the better of the two sets. The idea of maintaining two sets has been considered for the 0,1 knapsack problem~\citep{knapsack-01} and (deterministic) submodular function maximization subject to a knapsack constraint~\citep{kdd-knapsack}. A novelty of our approach is in using this idea along with selecting items using replication test scores. 

Another novelty of our algorithm is in using estimated values of test scores, unlike the previous work which used test scores defined as expected values with respect to distributions of item values. This allows us to accommodate practical scenarios when prior distributions of item values are unknown, but test scores are computed as a statistic from observed historical data. Our setting is more general in accommodating the case when test scores, defined as expected values with respect to prior distributions, are known as a special case with no estimation noise. For our algorithm to have a guarantee on performance, the test score estimates need to satisfy certain error bounds. What we want is to (approximately) find the best $k+1$ items and to detect the worst item among the best $k+1$ items. This can be done by estimating each individual replication test score within a small estimation error and ranking the items accordingly. In Section~\ref{sec:sampling}, we provide bounds on sufficient sample sizes. Specifically, we will prove that the number of samples required to achieve $\epsilon$-multiplicative accuracy grows as a function of $1/\epsilon$ and $k_i$ for item $i\in \Omega$. %

In the remainder of this section, we provide performance guarantees for Algorithm~\ref{algo:simple-greedy} under different assumptions about group valuation function and values of item costs. 

\subsection{Approximation guarantees}\label{sec:apx0}

We show that Algorithm~\ref{algo:simple-greedy} guarantees a constant-factor approximation for the budgeted stochastic utility maximization problem, for any utility function corresponding to a valuation function that satisfies conditions (C1)-(C4) stated in Section~\ref{sec:problem}. The approximation ratio we give depends on the noise factor $\epsilon$ in a smooth way.

\begin{theorem}\label{thm:constant1}
	Assume that $f$ satisfies conditions (C1)-(C4), 
	and let $u$ be the corresponding utility function defined as in~\eqref{eq:utility}. Then, the output $A$ of Algorithm~\ref{algo:simple-greedy} is such that 
	$$u(A)\geq \frac{(1-\epsilon)p}{(1-\epsilon)p+2q}u(OPT)$$
	where $p=1-1/e$ and $q=5.307$.
\end{theorem}

The proof of Theorem~\ref{thm:constant1} uses the framework for deriving approximation ratios for utility maximization problems, introduced in \cite{sekar2019test-score} for cardinality constraints, extended to the more general case of knapsack constraints. This framework is based on approximating the utility function with lower and upper bound ``sketch" functions. Specifically, the proof uses the lower bound and upper bound sketch functions $\underline{v}(S) = \min\{r_i: i\in S\}$ and $\overline{v}(S) = \max\{r_i: i\in S\}$, respectively. For approximating the utility function with the two sketch functions, an important quantity in our analysis is 
\begin{equation}
d(S) = \sum_{i\in S} \frac{1}{k_i}
\label{equ:relcost}
\end{equation}
which we refer to as the \emph{relative cost} of set $S$. Recalling the definition $k_i = \lfloor B / c_i \rfloor$, $d(S)$ can be roughly interpreted as the ratio of the total cost of items in set $S$ and the budget value. Then we show that
$$\left(1-e^{-d(S)}\right)\cdot\underline{v}(S) \leq u(S)\leq \left(1+d(S)+2\sqrt{d(S)}\right)\cdot\overline{v}(S)$$ for all $S\subseteq \Omega$.

The proof follows by showing that for the approximation guarantee of Theorem~\ref{thm:constant1} it suffices that the values of functions $1-e^{-d(S)}$ and $1+d(S)+2\sqrt{d(S)}$ are lower and upper bounded by constants asserted in the theorem. In fact, we prove that $d(S)\leq 1.7$ for any set $S$ which $c(S)\leq B$, from which we argue that $p\leq 1-e^{-d(S)}$ and $1+d(S)+2\sqrt{d(S)}\leq q$ hold for some strictly positive constants $p$ and $q$. %

As mentioned before, the approximation ratio in Theorem~\ref{thm:constant1} depends smoothly on the value of the estimation error parameter $\epsilon$. More specifically, it is a decreasing concave function of $\epsilon$, achieving value zero for the extreme point $\epsilon = 1$. Since the ratio $p/q$ is small, the approximation ratio approximately decreases linearly with parameter $\epsilon$.

For the special case when the items have identical costs and there is no estimation error, the bound of Theorem~\ref{thm:constant1} reads as $u(A)\geq u(OPT)/17.8$. \cite{kleinberg15}'s algorithm based on replication test scores for the case of top-$r$ function achieves approximation ratio of $\simeq1/30$, and our algorithm for broader classes of functions under heterogeneous costs achieves a better approximation ratio. \cite{sekar2019test-score}'s algorithm  achieves a bound of ratio 1/7.3, but it only works under unit costs and no estimation error. 

In the following section, we will refine our analysis to capture situations where the item costs are close to being uniform (we refer to as \emph{regular}), thereby providing a tighter approximation guarantee. The result in Theorem~\ref{thm:constant1} is derived by using a worst-case bound for the relative cost $d(S)$ which is achieved by a specific case when the item costs are not regular. This extreme case cannot occur in the case of equal item costs. In the regime where the item costs are regular, the factor $q=5.307$ in the utility bound of Theorem~\ref{thm:constant1} can be reduced, and in particular, we can recover the approximation guarantee of \cite{sekar2019test-score} which works for the setting of identical costs.

\subsection{A tighter approximation guarantee for bounded variation of item costs}\label{sec:apx1}

We derive a tighter approximation guarantee for Algorithm~\ref{algo:simple-greedy} than the approximation guarantee of Theorem~\ref{thm:constant1} for problem instances for which item costs satisfy a regularity condition. We use two different notions of regularity of item costs. Given a \emph{cost regularity parameter} $\beta\in[0,1)$, we say that item costs $c_1,\ldots,c_n$ are
\begin{description}
	\item[(C1)] \emph{$\beta$-small} if $c_1,\ldots, c_n\leq \beta B$,
	\item[(C2)] \emph{$\beta$-regular} if $k_i\geq (1-\beta)k_j$ for all $i,j\in \Omega$.
\end{description}

Under the $\beta$-small condition, each item cost is at most $\beta$ factor of the total budget. For small values of $\beta$ parameter, each item cost is a small fraction of the total budget. The case when item costs are small in comparison to available budget was studied in the context of online knapsack problems~\citep{Naor1,Naor2}, knapsack secretary problems~\citep{knapsack-secretary}, keyword building problems \citep{keyword-bidding}, and online packing problems \citep{resource-allocation}.

On the other hand, the $\beta$-regularity condition ensures that parameters $k_i$ are within constant factors of each other. Indeed, this condition is equivalent to $(1-\beta) \leq k_i/k_j \leq 1/(1-\beta)$ for all $i,j\in \Omega$. Recalling the definition $k_i = \lfloor B/c_i\rfloor$, it is easy to observe that the $\beta$-regularity implies $(1-\beta)c_i \leq c_j B/(B-c_j)$ for all $i,j\in \Omega$. If each item cost is a small fraction of the total budget, then approximately the ratio of the costs of any two items are in $[1-\beta, 1/(1-\beta)]$. The smaller the value of parameter $\beta$, the more balanced the item costs are. The $0$-regular case corresponds to the case of equal item costs, which accommodates the special case of cardinality constraints.

In the following theorem, we provide a tighter approximation guarantee for Algorithm~\ref{algo:simple-greedy} than in Theorem~\ref{thm:constant1}, under the assumption that item costs satisfy either $\beta$-small or $\beta$-regular condition.

\begin{theorem}\label{thm:constant2}
	Assume that $f$ satisfies conditions (C1)-(C4),
and let $u$ be the corresponding utility function defined as in~\eqref{eq:utility}. If the item costs are $\beta$-small or $\beta$-regular for some $\beta\in[0,1/2]$, then the output $A$ of Algorithm~\ref{algo:simple-greedy} is such that
	$$u(A)\geq \frac{(1-\epsilon)p}{(1-\epsilon)p+q}u(OPT)$$
	where $p=1-1/e$ and $q=\left(4+{3\beta}/{(1-\beta)}\right)\left(1+{\beta}/{(1-2\beta)}\right)$.
\end{theorem}

The difference with Theorem~\ref{thm:constant1} is in removing factor $2$ in the constant $q$ and having $q$ parametrized with the cost regularity parameter $\beta$. The approximation ratio is a continuous decreasing function in $\beta$. The degradation of the approximation ratio with the value of parameter $\beta$ is according to a concave function of $\beta$, achieving value zero for the extreme case $\beta = 1/2$.

For the case of identical item costs, i.e. the $0$-regular case, it follows from Theorem~\ref{thm:constant2} that $u(A)\geq (1-\epsilon)\left(1-1/e\right)/\left((1-\epsilon)\left(1-1/e\right)+4\right)u(OPT)$ where $A$ is the output solution of Algorithm~\ref{algo:simple-greedy}. If, in addition, there is no estimation error, this bound becomes $u(A)\geq \left(1-1/e\right)/\left(5-1/e\right)u(OPT)$, which recovers the result of~\cite{sekar2019test-score}.

The proof of Theorem~\ref{thm:constant2} is established by bounding the relative cost in (\ref{equ:relcost}) using the cost regularity parameter $\beta$. We show that if $c_1,\ldots, c_n$ are $\beta$-small or $\beta$-regular, then we have $1-\beta/(1-\beta)\leq d(S)\leq 1+\beta/(1-\beta)$.
Using these bounds alongside with approximation ratio guarantees derived by using sketch function arguments, yields the statement of the theorem. 

\subsection{Parametrizing by the curvature}\label{sec:curvature}

Algorithm~\ref{algo:simple-greedy} has similarities with the greedy approximation algorithm for the 0,1 knapsack problem. In fact, the only difference is that the greedy algorithm for the 0,1 knapsack problem selects items based on their \emph{density} values, where the density of an item is defined as the ratio of its reward and its cost, whereas Algorithm~\ref{algo:simple-greedy} is based on using replication test scores. We mentioned in Section~\ref{sec:score} that when the valuation function $f$ is the linear total production function, the replication test score of an item is (roughly) proportional to its density. It is folklore that the greedy algorithm for the 0,1 knapsack problem achieves $1/2$-factor approximation; see~\citep{knapsack-01}. Although we already have Theorems~\ref{thm:constant1} and~\ref{thm:constant2} providing constant-factor approximation guarantees for Algorithm~\ref{algo:simple-greedy}, there is still a gap between their guaranteed factors and $1/2$. Since Algorithm~\ref{algo:simple-greedy} works similarly as the greedy approximation algorithm for the 0,1 knapsack problem, it is natural to ask whether Algorithm~\ref{algo:simple-greedy} achieves a better performance guarantee for the (stochastic) 0,1 knapsack problem.

Motivated by this, we consider the notion of \emph{curvature} that parametrizes how close a given function is to being linear or modular. A monotone function $f:\mathcal{X}\to \mathbb{R}_+$, with $\mathcal{X}\subseteq \mathbb{R}_+^d$ is said to have \emph{curvature} $\alpha$ if for all $i\in \{1,\ldots,d\}$, $z > 0$, and $x,y\in \mathcal{X}$ such that $x\leq y$ and $x_i=y_i = 0$, it holds
\begin{equation}\label{curvature}
(1-\alpha)(f(x+ze_i)-f(x))\leq f(y+ze_i) - f(y).
\end{equation}
Note that if $f$ is in addition submodular, we have
\begin{equation}\label{curvature'}
(1-\alpha)(f(x+ze_i) - f(x))\leq f(y+ze_i)-f(y)\leq f(x+ze_i) - f(x)
\end{equation}
for any combination of a scalar $z$, and vectors $x,y\in \mathcal{X}$ such that $x\leq y$ and $x_i = y_i =0$. The notion of curvature was introduced by \cite{ConfortiCornuejols} initially for set functions. Our definition of curvature extends to valuation functions and is consistent with the original notion defined for set functions; if $f$ has curvature $\alpha$, then $\alpha$ is precisely the curvature of the corresponding utility function $u$.

Let $\alpha$ be the curvature of a valuation function $f$. If $f$ is a monotone submodular function, then $\alpha\in[0,1]$. Furthermore, if $f$ is modular, then $\alpha=0$.

We have the following theorem providing an approximation guarantee of Algorithm~\ref{algo:simple-greedy} for valuation functions of given curvature value. 

\begin{theorem}\label{thm:parametrized-guarantee1}
	Assume that $f$ satisfies conditions (C1)-(C4),
and let $u$ be the corresponding utility function defined as in~\eqref{eq:utility}. Then, the output $A$ of Algorithm~\ref{algo:simple-greedy} is such that
	\begin{equation}\label{parametrized-guarantee1}
	u(A)\geq (1-\epsilon)(1-\alpha)\frac{1-e^{-\alpha d([k+1])}}{\alpha d([k+1])}\cdot\frac{5}{17}u(OPT)
	\end{equation}
    where $k$ is defined as in the pseudo-code of Algorithm~\ref{algo:simple-greedy}, and $d([k+1])$ denotes the relative cost of set $[k+1]=\{1,\ldots, k+1\}$ defined as in~\eqref{equ:relcost}.
\end{theorem}

To prove Theorem~\ref{thm:parametrized-guarantee1}, we define the \emph{average score} of $S\subseteq \Omega$ defined by
$$v(S):=\frac{\sum_{i\in S}{r_i}/{k_i}}{\sum_{i\in S}{1}/{k_i}}.$$ 
Recall that we used the maximum score $\overline{v}(S)$ and the minimum score $\underline{v}(S)$ to approximate the utility function. Instead, we use the average score function to bound the utility function with bounded curvature. Note that $\underline{v}(S)\leq v(S)\leq \overline{v}(S)$. We show that when the curvature is $\alpha$, 
$$\frac{1-e^{-\alpha d(S)}}{\alpha}\cdot {v}(S) \leq u(S)\leq \frac{d(S)}{1-\alpha}\cdot {v}(S)$$ for all $S\subseteq \Omega$. Since $\alpha\leq 1$, the lower bound is always at least $1-e^{-d(S)}$.

When $f$ is modular, i.e., $\alpha = 0$, from the proof of Theorem~\ref{thm:parametrized-guarantee1}, it follows that
$$
u(A)\geq (1-\epsilon)\frac{5}{17}u(OPT)
$$
where $A$ is the output of Algorithm~\ref{algo:simple-greedy}. This confirms our intuition that Algorithm~\ref{algo:simple-greedy} works better for the stochastic 0,1 knapsack problem than the problem with an arbitrary submodular function. Moreover, the factor of $5/17$ in~\eqref{parametrized-guarantee1} can be improved when item costs are close to being uniform. To achieve this, we repeat the idea of parametrizing variation in item costs introduced in Section~\ref{sec:apx1}, which yields the following theorem.

\begin{theorem}\label{thm:parametrized-guarantee2}
	Assume that $f$ satisfies conditions (C1)-(C4),
	and let $u$ be the corresponding utility function defined as in~\eqref{eq:utility}. If the item costs are $\beta$-small or $\beta$-regular for some $\beta\in[0,1]$, then
	\begin{equation}\label{parametrized-guarantee2}
	u(A)\geq (1-\epsilon)(1-\alpha)(1-\beta)\frac{1-e^{-\alpha(1-2{\beta})/{(1-\beta)}}}{\alpha}u(OPT)
	\end{equation}
	where $A$ denotes the output of Algorithm~\ref{algo:simple-greedy}.
\end{theorem}

When $f$ is modular and $\alpha=0$,~\eqref{parametrized-guarantee2} implies that 
$$
u(A)\geq (1-\epsilon)(1-2\beta)u(OPT)
$$
where $A$ is the output of Algorithm~\ref{algo:simple-greedy}. 
This bound is useful when the item costs are close to being uniform so that $\beta$ is small.

\section{Streaming utility maximization with random item values}\label{sec:streaming}

In this section we consider the streaming computation setting where items arrive sequentially and only a small size memory is available for keeping information about items. The fact that items are chosen based on individual test scores in our framework is compelling for the streaming case. It is favorable especially in stochastic environments when estimating the value of each combination of items is expensive. Thanks to its simplicity, Algorithm~\ref{algo:simple-greedy} can be easily adapted to the streaming setting. 

Our streaming greedy algorithm evaluates the test score of each arriving item and keeps only the top $\ell$ best test score items so that $c_{i_1}+\dots +c_{i_\ell} > B$ and $c_{i_1}+\dots +c_{i_{\ell-1}} \le B$ where $i_1 ,\dots {i_\ell}$ denote the best items such that $\hat{r}_{i_1} \ge \dots  \ge \hat{r}_{i_\ell}$. This can be done by managing a buffer storing the items with the highest replication test scores seen so far, and whenever we see a new item whose score is higher than an item stored in the buffer, we add the new item and remove some items with low scores to satisfy the above budget condition if necessary. Algorithm~\ref{algo:streaming-greedy} provides a detailed pseudo-code of the streaming greedy algorithm. This is a single pass algorithm that requires at most $\max_{1\le i \le n} 2B/c_i$ memory space and has the same guarantee as Algorithm~\ref{algo:simple-greedy}.  

\begin{algorithm}
	\caption{Streaming greedy algorithm}\label{algo:streaming-greedy}
	\begin{algorithmic}
		\Require Items coming in the order of $i_1,i_2,\ldots,i_n$.
		\State Let $R$ be a buffer, and set $R\leftarrow \emptyset$.
		\State \textbf{for} $j=1$ to $n$ \textbf{do:} UPDATE($R,i_j$) \Comment{Procedure UPDATE is given below}
		\State Let $i$ be the element with the lowest $\hat r_i$ value in $R$.
		\State Return the better of $R\setminus\{i\}$ and $\{i\}$.
		\newline
		\Procedure{Update}{$R,i$}\Comment{Update the buffer with a new element $i$}
		\State \textbf{if} $c(R)\leq B$, \textbf{then:} $R\leftarrow R\cup \{i\}$.
		\State \textbf{else:} 
		\State \quad Let $i_1,\ldots, i_\ell$ be the current items of $R$ that are ordered so that $\hat r_{i_1}\geq\cdots\geq\hat r_{i_\ell}$.
		\State \quad \textbf{if} $\hat r_i$ is strictly higher than any of $\hat r_{i_1},\ldots, \hat r_{i_\ell}$, \textbf{then:}
		\State \quad\quad $R\leftarrow R\cup \{i\}$.
		\State \quad\quad Let $j$ be the smallest index such that $c_i + c_{i_1}+\cdots+c_{i_j}>B$.
		\State \quad\quad Remove $i_{j+1},\ldots, i_{\ell}$ from $R$.
		\EndProcedure
	\end{algorithmic}
\end{algorithm}

\begin{theorem}\label{thm:streaming}
The streaming greedy algorithm, Algorithm~\ref{algo:streaming-greedy}, provides the same approximation guarantees as Algorithm~\ref{algo:simple-greedy}. Moreover, it runs in $O(n)$ time and requires a single-pass and memory of size $B^* := \max_{1\le i \le n} 2B/c_i$.
\end{theorem}

The state-of-the-art algorithms for (deterministic) streaming submodular maximization require to evaluate values of combinations of items. The %
appealing aspect of our greedy algorithm is that checking only individual item scores works even for the streaming setting, while still providing a constant-factor approximation. Moreover, our algorithm is efficient in terms of time and space complexity, as the best algorithms for the knapsack constraint~\citep{streaming-knapsack1,streaming-knapsack2,streaming-knapsack3} require additional $\mbox{polylog}(B^* )$ factors in the time and space complexity. %

\section{Sufficient sample sizes for estimating replication test scores}\label{sec:sampling}

In this section we derive a sufficient sample size for estimating a replication test score within a prescribed statistical estimation accuracy guarantee. We consider two different statistical estimation accuracy criteria including (1) estimation of individual replication test scores within a given relative error tolerance and a given probability of error, and (2) estimation of replication test scores such that a set of top replication test score items can be identified within a given error tolerance and a given probability of error. The latter is of particular interest for our test score algorithms to provide approximation guarantees shown in Sections~\ref{sec:algorithm} and \ref{sec:streaming}. The sample size bounds depend on different factors including the individual cost of an item, budget parameter, properties of the valuation function, and gaps between the sorted values of replication test scores. The results in this section are established by applying some carefully chosen concentration bounds that allow to capture the effect of the valuation function properties on the sufficient sample size.

\subsection{Estimating individual replication test scores}\label{sec:num_samples}

We consider the estimator $\hat{r}_i$ of $r_i$, the replication test score of item $i$, defined by
\begin{equation}
\hat{r}_i = \frac{1}{m_i}\sum_{j=1}^{m_i} f(\{X^{((j-1)k_i+1)}_i,\ldots, X^{(jk_i)}_i\})
\label{equ:gmk}
\end{equation}
where $m_i\geq 1$ and $X^{(1)}_i, \ldots X^{(m_i k_i)}_i$ are independent samples from distribution $P_i$. Note that the estimator $\hat{r}_i$ uses as input $T_i = m_i k_i$ samples, which we refer to as the \emph{sample size}. We refer to $m_i$ as the \emph{number of sample batches}. Note that $\hat r_i$ is an unbiased estimator of $r_i$.

We consider the statistical estimation accuracy criteria that requires estimation of a replication test score within a relative error tolerance and a given probability of error. Specifically, for any given \emph{relative error tolerance} parameter $\epsilon \in (0,1]$ and \emph{probability of error} parameter $\delta \in (0,1)$, we say that $\hat{r}_i$ is $(\epsilon,\delta)$-accurate if $|\hat{r}_i-r_i|\leq \epsilon r_i$ with probability at least $1-\delta$. 

Our goal is to find a sufficient number of samples $T_i$ for $\hat{r}_i$ to be $(\epsilon,\delta)$-accurate. This requires to study the probability of deviation of $\hat{r}_i$ from its expected value $r_i$. We next derive two sufficient sample sizes by using different probability of deviation inequalities. 

A sufficient sample size can be derived for valuation functions that have bounded values on their domains. Under this assumption, the summation terms in \eqref{equ:gmk} are independent random variables with bounded values and expected value $r_i$. Hence, we can apply Hoeffding's inequality to obtain a sufficient sample size as given in the following proposition. 

\begin{proposition} Assume that $f$ is bounded on the domain $\mathcal{X}^{k_i}$ and let $\|f\|_{i,\infty}:=\sup_{x\in\mathcal{X}_i^{k_i}}f(x)$
is bounded, where $\mathcal{X}_i$ denotes support of distribution $P_i$. Then, $\hat{r}_i$ is $(\epsilon,\delta)$-accurate for any sample size $T_i$ such that 
\begin{equation}
T_i \geq \frac{1}{2}\frac{k_i||f||_{i,\infty}^2}{\epsilon^2 r_i^2}\log\left(\frac{1}{\delta}\right).\label{equ:mk-hoeffding}
\end{equation}
\label{pro:sample-hoeffding}
\end{proposition}
Note that the sufficient sample size in Proposition~\ref{pro:sample-hoeffding} grows linearly with parameter $k_i$ for some valuation functions $f$. In particular, this is the case for the total production, top-$r$, CES, and success probability valuation functions discussed in Section~\ref{sec:problem}. This dependence on $k_i$ results from using Hoeffding's inequality to the sum of random variables in \eqref{equ:gmk}, which does not exploit how each of these random variables depends on sampled item values. We can account for this by applying another probability of deviation inequality, namely McDiarmid's inequality, which captures the marginal effect of each individual item value. %

\begin{proposition}\label{pro:sample-macdiarmid}
 Let $f_1(x) : = f(x,0,\ldots,0)$, for $x\in \mathcal{X}_i$, and $\|f_1\|_{i,\infty}=\sup\{f_1(x):x\in\mathcal{X}_i\}$. Then,
\begin{equation}
T_i \geq \frac{1}{2}\frac{k_i^2 \|f_1\|_{i,\infty}^2}{\epsilon^2 r_i^2} \log\left(\frac{1}{\delta}\right).
\label{equ:mk-mcdiarmid}
\end{equation}
\label{pro:sample-mcdiarmid}
\end{proposition}

Note that the sufficient sample size in \eqref{equ:mk-mcdiarmid} corresponds to that in \eqref{equ:mk-hoeffding} except for replacing $\|f\|_{i,\infty}^2$ with $k_i\|f_1\|_{i,\infty}^2$. When applying Hoeffding's inequality, we consider $f(\{X^{((j-1)k_i+1)}_i,\ldots, X^{(j k_i)}_i\})$ as a random variable with only properties used that this is a bounded random variable and that it has  mean $r_i$. Essentially, the $k_i$ random variables $X^{((j-1)k_i+1)}_i,\ldots, X^{(j k_i)}_i$ are treated as a whole and their variations are captured by a single term $\|f\|_{i,\infty}^2$, where $\|f\|_{i,\infty}=\sup_{x\in\mathcal{X}_i^{k_i}}f(x)$ measures the maximum possible change of $f$ caused by changes in the $k_i$ random variables. As there are $m_i$ copies, we end up with $m_i\|f\|_{i,\infty}^2$. In contrast, McDiarmid's inequality takes the $m_ik_i$ random variables in $\hat{r}_i$, defined in \eqref{equ:gmk}, individually. Assuming the submodularity of $f$, the marginal squared variation by a single variable is bounded by $\|f_1\|_{i,\infty}^2$, so a total variation can be measured by $m_i k_i\|f_1\|_{i,\infty}^2$ as there are $m_i k_i$ variables.
	
We show below that the sufficient sample size in Proposition~\ref{pro:sample-mcdiarmid} %
improves the dependence on $k_i$ for some valuation functions. %
This is a matter of whether or not $k_i\|f_1\|_{i,\infty}^2$ is smaller than $\|f\|_{i,\infty}^2$.

If the curvature of $f$ is $\alpha$, then $r_i \geq %
k_i(1-\alpha)\E[f_1(X_i)]$.
By Proposition~\ref{pro:sample-mcdiarmid}, if $T_i$ satisfies
\begin{equation}
T_i \geq \frac{1}{2}\frac{\|f_1\|_{i,\infty}^2}{(1-\alpha)^2\epsilon^2 \E[f_1(X_i)]^2}\log\left(\frac{1}{\delta}\right),
\label{equ:curvature}
\end{equation}
then $\hat{r}_i$ is $(\epsilon,\delta)$-accurate. Note that if the curvature $\alpha$ is constant, then the bound in~\eqref{equ:curvature} does not dependent on $k_i$. Therefore, if $f$ is additive, the sufficient sample size is independent of $k_i$.

If $f$ is the total production function $f(x) = g(x_1 + \cdots + x_{k_i})$ for $g$ an increasing concave function, then by the strong law of large numbers,
$$
\frac{k_i^2 \|f_1\|_{i,\infty}^2}{r_i^2}
\sim \frac{k_i^2 g(\bar{\mathcal{X}}_i)^2}{g(k_i\E[X_i])^2} \hbox{ for large } k_i\quad\text{where $\bar{\mathcal{X}}_i:= \sup \mathcal{X}_i$.}
$$
Note that if $g$ is such that $g(k_ix)\geq k_i^{1-\gamma/2}g(x)$ for all $x\geq 0$, for some $\gamma \in [0,1)$, then the dependence of the bound~\eqref{equ:mk-mcdiarmid} on $k_i$ is sublinear. For instance, for $g(x)=x^a$ for $a \in (0,1)$, the dependence on $k_i$ is proportional to $k_i^{2(1-a)}$ asymptotically for large $k_i$, which is sublinear in $k_i$ for any $a \in (1/2,1]$.

For the case of CES function $f(x) = (\sum_{i=1}^{k_i} x_i^r)^{1/r}$ for $r\geq 1$, by the strong law of large numbers, we have
	$$
\frac{k_i^2\|f_1\|_{i,\infty}^2}{r_i^2} \sim 
	\frac{\bar{\mathcal{X}}_i^2}{\E[X_i^r]^{2/r}}k_i^{2(1-1/r)} \hbox{ for large } k_i.
	$$
Hence, the scaling with $k_i$ is sublinear if and only if $1\leq r<2$, for asymptotically large $k_i$.

\subsection{Top set identification}\label{sec:error_prob}

For the approximation guarantees of our test score algorithms, it is required that after ordering and relabeling the items in $\Omega$ so that $\hat r_1\geq\hat r_2\geq\cdots\geq \hat r_n$, the following \emph{$\epsilon$-top set accuracy} condition holds: $r_1,\ldots, r_{k+1}\geq (1-\epsilon)\max\{r_i:i\in \Omega \setminus \{1,\ldots, k\}\}$, where $k$ is the item satisfying $c_1+\cdots +c_k\leq B <c_1+\cdots +c_{k+1}$. We say that the order given by the estimated replication test scores is \emph{$(\epsilon,\delta)$-top set accurate} if the $\epsilon$-top set accuracy condition holds with probability at least $1-\delta$. 

Let $\sigma(1),\sigma(2),\ldots,\sigma(n)$ be the ``true" ordering of items $1,\ldots,n$, i.e., $r_{\sigma(1)}\geq r_{\sigma(2)}\geq\cdots\geq r_{\sigma(n)}$,
and let $k^*$ be the index satisfying $c_{\sigma(1)}+\cdots +c_{\sigma(k^*)}\leq B <c_{\sigma(1)}+\cdots +c_{\sigma(k^*)}+c_{\sigma(k^*+1)}$.
 The following proposition shows that $(\epsilon,\delta)$-top set accuracy can be guaranteed by noisy estimates of individual replication test scores.

\begin{proposition}\label{error-prob1} Let $T_i$ denote the number of samples for testing item $i$. For any $\epsilon,\delta \in(0,1)$, the $(\epsilon,\delta)$-test set accuracy holds provided that
\begin{equation}
T_i \ge 2 \frac{ \min\{k_i \| f\|_{i,\infty}^2,k_i^2 \| f_1\|_{i,\infty}^2\}}{\epsilon^2 r_{\sigma(k^*+1)}^2} \log\left(\frac{2n}{\delta}\right) \hbox{ for all } i\in \Omega.
\label{equ:error-prob1}
\end{equation}
\end{proposition}
Proposition~\ref{error-prob1} implies that we can guarantee that the ranking obtained from the estimated replication test scores is $(\epsilon,\delta)$-top set accurate by ensuring that individual replication test scores are estimated within a certain statistical estimation error. %

If $\epsilon = 0$, then the ordering of items induced by the estimated replication test scores is required to be accurate in the sense that $r_1,\ldots, r_k\geq \max\{r_i:i\in \Omega\setminus \{1,\ldots,k\}\}$, i.e. the $0$-top set accuracy condition holds. In this case, we can remove the factor of $1-\epsilon$ in our approximation guarantees. That said, we are interested in finding a sufficient sample size under which the ordering induced by estimated replication test scores is accurate with probability at least $1-\delta$. 

If the gaps between the replication test scores of distinct items are large, it is intuitive that items are easy to be correctly ordered based on the estimated values of replication test scores. It turns out that the probability of error for finding an accurate ordering becomes smaller as the gaps between $r_{\sigma(k^*)}$ and $r_{\sigma(k^*+1)}$, and $r_{\sigma(k^*+1)}$ and $r_{\sigma(k^*+2)}$, are larger, which is demonstrated by the following proposition.

\begin{proposition}\label{error-prob2}
Assume that $r_{\sigma(k^*)}-r_{\sigma(k^*+1)}\geq 2\Delta$ and $r_{\sigma(k^*+1)}-r_{\sigma(k^*+2)}\geq 2\Delta$ for some $\Delta>0$. Then, for any $\delta \in (0,1)$, the ordering according to estimated replication test scores is accurate with probability at least $1-\delta$, under the condition 
$$
T_i \ge \frac{1}{2}\frac{ \min\{k_i \| f\|_{i,\infty}^2, k_i^2\| f_1\|_{i,\infty}^2\}}{\Delta^2}\log\left(\frac{2n}{\delta}\right) \hbox{ for all } i \in \Omega.
$$
\end{proposition}
The sufficient sample size in Proposition~\ref{error-prob2} depends on the minimum gap between the replication test score value of the item with the $k^*$-th largest replication test score value and its adjacent items in the ordering of items with respect to the replication test score values. This is intuitive as the goal is to correctly identify the set of $k^*$ items with largest replication test scores values.

\section{Numerical experiments}\label{sec:num}

In this section, we present our experimental results to assess the numerical performance of our test-score-based greedy ($\texttt{TSG}$) algorithm (Algorithm~\ref{algo:simple-greedy}) for the budgeted stochastic utility maximization problem. We evaluated the efficiency of our algorithm under various assumptions on item value distributions and the costs of items, different group valuation functions, and varying the size of test score estimation sample. We have performed carefully designed experiments with two sets of test instances: (1) instances generated by synthetic simulated data for item values and costs, and (2) instances obtained by using data from the Academia.StackExchange Q\&A platform. In the StackExchange platform, users post tagged questions, to which other users or experts can submit answers. Hence, experts providing answers to questions can be naturally viewed as items, and the group value can be measured by the popularity or quality of answers submitted by the experts selected in a group. We report our results from the synthetic instances in Section~\ref{sec:synthetic} and results from the instances with the StackExchange data in Section~\ref{sec:stackexchange}. Overall, our numerical results show that $\texttt{TSG}$ performs at least as well as a benchmark algorithm for most instances. Our code is publicly available in \url{https://github.com/dabeenl/TSG-experiments}.

\paragraph{Benchmark algorithm} We chose as a benchmark the algorithm proposed by~\cite{kdd-knapsack}, namely, the \emph{Cost Effective Lazy Forward ($\texttt{CELF}$)} algorithm, against which we compare the performance of $\texttt{TSG}$. $\texttt{CELF}$, developed for submodular maximization subject to a knapsack constraint, works in a similar way to the greedy algorithm of~\cite{cardinality3} for the cardinality-constrained case and achieves $(1-1/e)/2$-factor approximation. There exist other algorithms that guarantee higher approximation ratios~\citep{Sviridenko,Yoshida,Ene}, but they are less practical as they involve enumerating a massive number of sets or computing over the multilinear extension. It is important to note that $\texttt{CELF}$ requires heavy value oracle calls, and we observed that $\texttt{TSG}$ runs much faster than $\texttt{CELF}$. However, we focus on the values of outputs produced by the algorithms.

\subsection{Simulation with synthetic data}\label{sec:synthetic}

\subsubsection{Generating instances}

Each instance consists of $n=100$ items and the budget $B$ is always set to 30. We examine four types of valuation functions, the max, the CES utility function with degree 2 (CES-2), the modular function $f(x)=x_1 +\cdots + x_n$, and the square root function $f(x)=\sqrt{x_1+\cdots + x_n}$ (see Section~\ref{sec:problem}). 

\paragraph{Item value distributions} For each item $i\in\Omega$, we sample its mean value $\mu_i=\E[X_i]$ from the uniform distribution over the unit interval $[0,1]$. We consider three types of distributions of item values. We take the Bernoulli, Exponential, and Pareto distributions, which represent bounded-support, light-tailed, and heavy-tailed distributions, respectively. In particular, we use the Type I Pareto distribution and tested various values $\{1.05, 1.5, 1.95, 3\}$ of the shape parameter.

\paragraph{Item costs} For each item $i\in\Omega$, we set $c_i = 1+ \lambda\mu_i$ where $\lambda$ is chosen from $\{0,B/10,2B/10,\ldots,9B/10\}$. Hence, $c_i$ grows linearly with the mean value $\mu_i=\E[X_i]$. When $\lambda=0$, the budget restriction reduces to the cardinality constraint.

\paragraph{Sample size}

For each item $i\in\Omega$, we generate $N\in\{50,100,150,200,250\}$ training samples of its random performance value $X_i$. We use these $N$ samples to estimate the replication test score $r_i$; we obtain $\lfloor N / k_i\rfloor$ disjoint batches of $k_i$ samples, each of which gives a realization of $r_i$, and $\hat r_i$ is obtained as the arithmetic mean of the $\lfloor N / k_i\rfloor$ values. When evaluating the value of a group $S\subseteq\Omega$ in the middle of running $\texttt{CELF}$, we obtain $N$ realizations of the value of $S$ so that no sample of an item is used for different realizations. The output of an algorithm is evaluated by a separate set of 50,000 test samples. 

\paragraph{Number of instances} A setting is defined by a choice of objective function, a choice of value distribution, %
a choice of coefficient $\lambda$ to define the costs of items from $\{0,B/10,2B/10,\ldots,9B/10\}$, and a choice of sample size $N$ from $\{50, 100, 150, 200, 250\}$. We generate 100 randomly generated instances for each problem setting.

\subsubsection{Results}

\begin{figure}[h!]
	\begin{center}
		\includegraphics[width=6in]{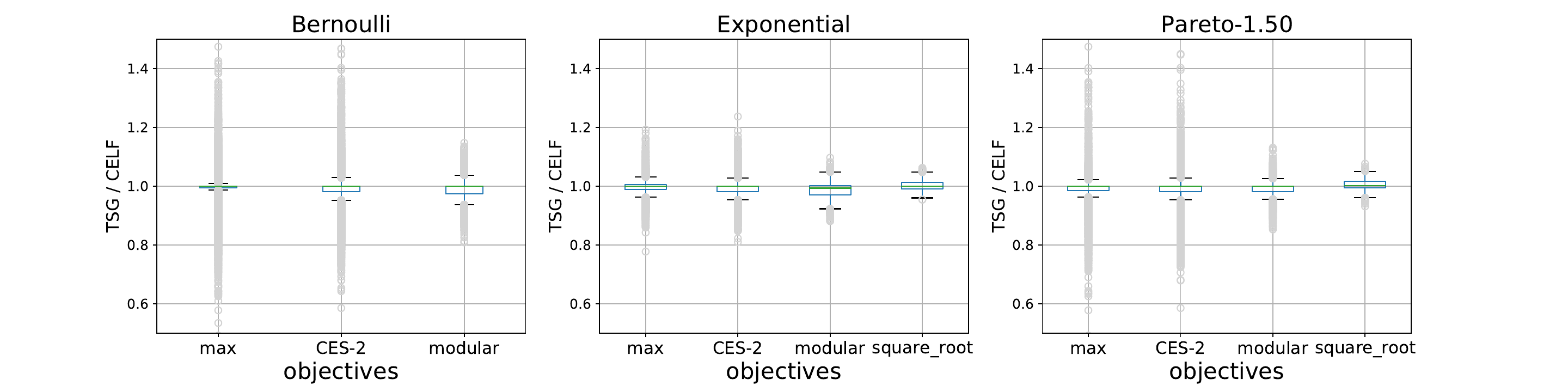}
		\caption{The ratio of the output value of~$\texttt{TSG}$ and that of $\texttt{CELF}$ for various objective functions and item value distributions.}\label{fig:synthetic-summary}
	\end{center}
\end{figure}

Before we see how certain factors affect the performance of $\texttt{TSG}$ against $\texttt{CELF}$, we aggregate the results from all instances and provide Figure~\ref{fig:synthetic-summary} showing the performance of $\texttt{TSG}$ for various objective functions and item value distributions.
In Figure~\ref{fig:synthetic-summary}, TSG/CELF denotes the ratio of the output value of~$\texttt{TSG}$ and that of $\texttt{CELF}$ for an instance. As the CES-2 function and the square root function have equal values for a binary vector, we omit the results for the square root function under the Bernoulli distribution. We can see that the ratio values are concentrated around 1, implying in turn that $\texttt{TSG}$ is as good as $\texttt{CELF}$ for most instances. %
Results under the Pareto distribution with parameters 1.05, 1.95, and 3 are similar and the corresponding plots are in Appendix~\ref{appendix:synthetic-summary}.

\begin{figure}[h!]
	\begin{center}
		\includegraphics[width=6in]{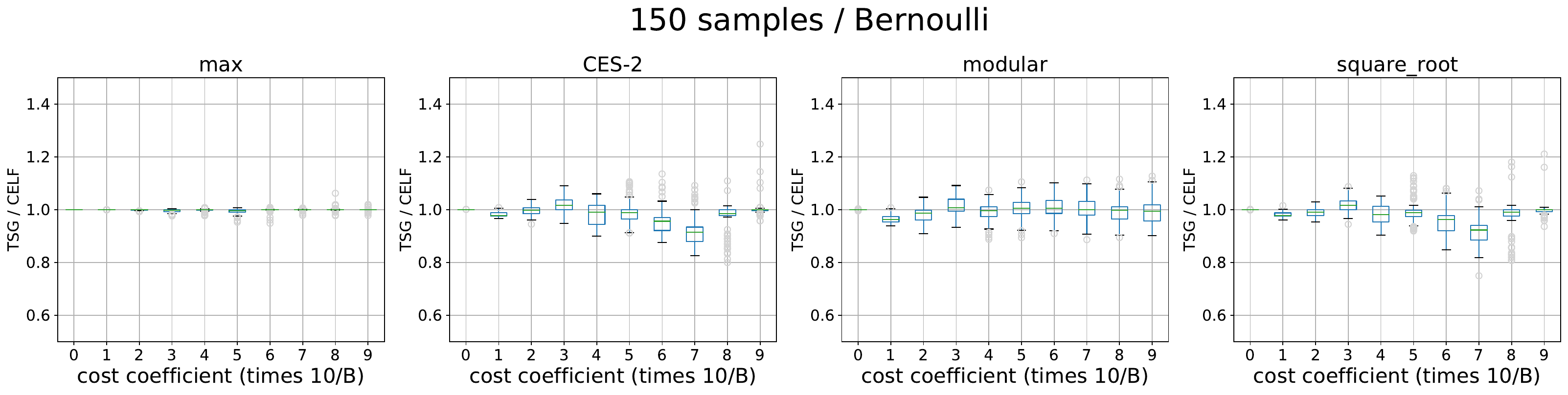}
		\includegraphics[width=6in]{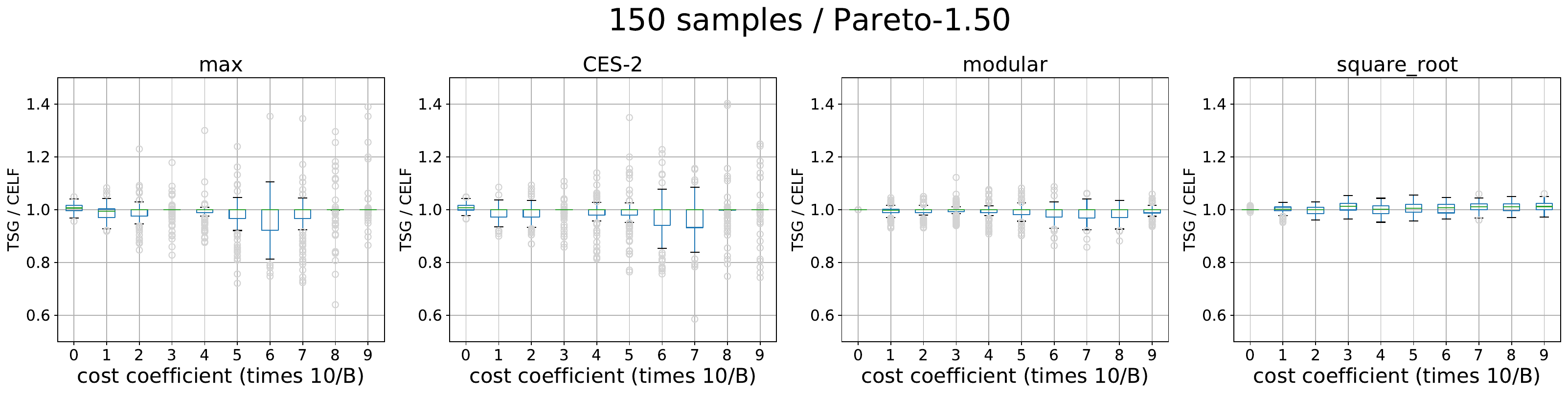}
		\caption{Results from $N=150$ and different values of the coefficient $\lambda\in \{0,B/10,2B/10,\ldots,9B/10\}$.}\label{fig:synthetic-cost}
	\end{center}
\end{figure}

\paragraph{Low to high costs} To examine how variations in item costs affect the performance of $\texttt{TSG}$, we provide Figure~\ref{fig:synthetic-cost} plotting ratio values under different values of the cost coefficient $\lambda$.
Overall, $\texttt{TSG}$ exhibits good performance compared to $\texttt{CELF}$, regardless of the size of $\lambda$. Results for the other distributions are similar and included in Appendix~\ref{appendix:synthetic-cost}.

\begin{figure}[h!]
	\begin{center}
		\includegraphics[width=6in]{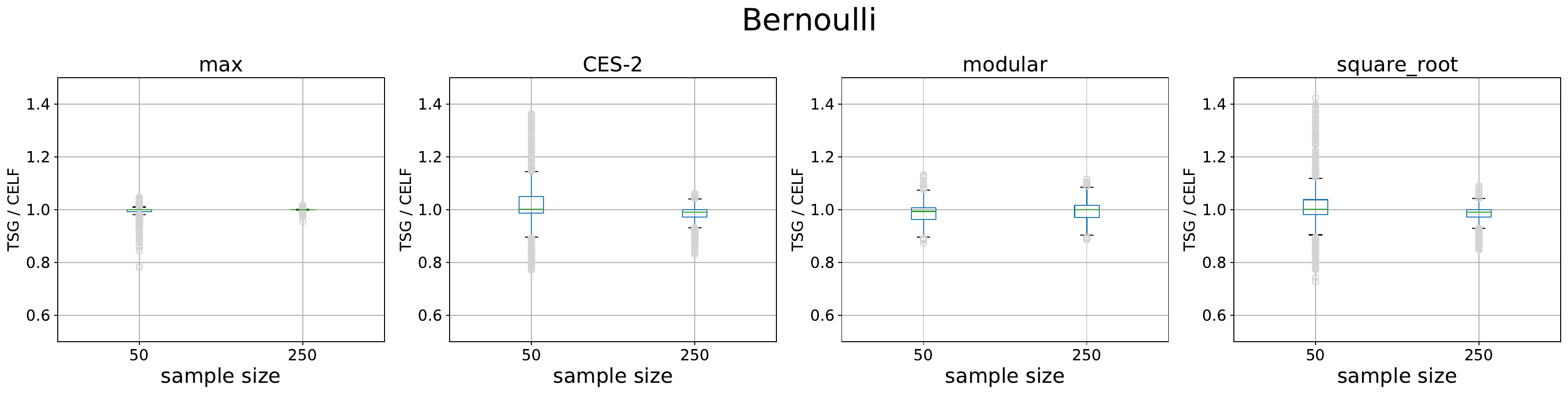}
		\includegraphics[width=6in]{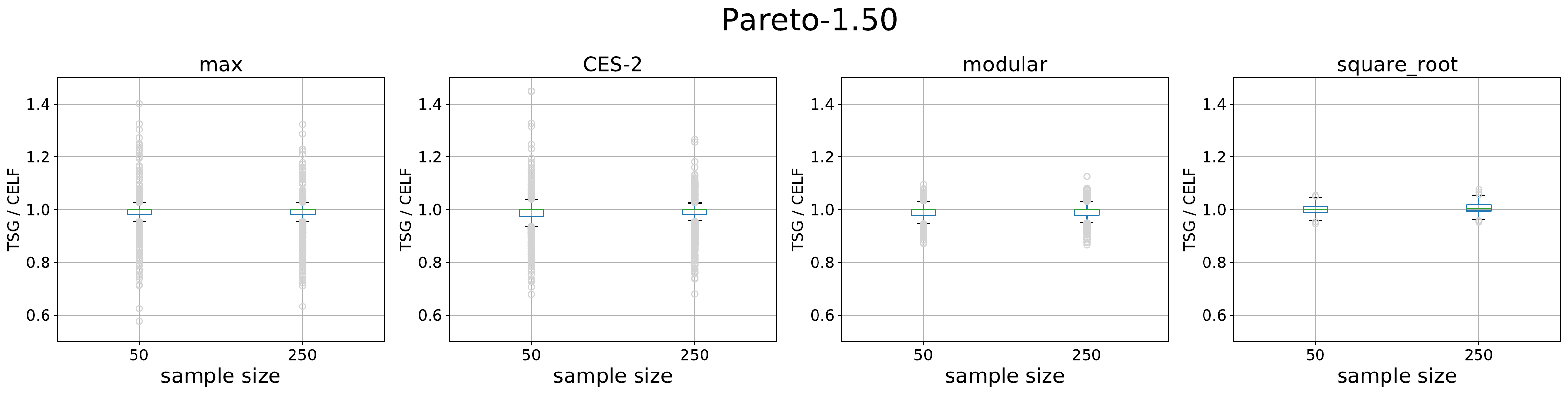}
		\caption{Results from testing different changing the sample size $N\in\{50,100,150,200,250\}$.}\label{fig:synthetic-sample}
	\end{center}
\end{figure}

\paragraph{Sample size and estimation noise} Figure~\ref{fig:synthetic-sample} plots the results of instances using 50 training samples and 250 training samples. It shows that $\texttt{TSG}$ performs well in both cases, although we see a slight improvement in value concentrations as the sample size increases. This indicates that $\texttt{TSG}$ is robust to estimation noise caused by lack of samples.
We observed similar concentration results for $N\in\{100,150,200\}$ and for the other distributions, and the corresponding plots are provided in Appendix~\ref{appendix:synthetic-sample}.

\paragraph{Costs independent from item values}

Note that item $i$'s cost $c_i = 1+ \lambda\mu_i$ is correlated its mean value $\mu_i$. To examine the setting where item costs and values are not correlated, we sample a new value $\mu_i^\prime$ from the uniform distribution $[0,1]$ and set $c_i=1+ \lambda\mu_i^\prime$ for $i\in\Omega$. We observed that the ratio values under this setting have a higher level of concentration around 1 than the instances with correlated costs, and we added plots showing this in Appendix~\ref{appendix:cor-ind}.

\subsection{Experiments with the StackExchange dataset}\label{sec:stackexchange}

\subsubsection{Dataset} We used the dataset from the Academia.StackExchange platform, retrieved on September 7, 2020, containing 33,082 questions and 78,059 answers. Each answer receives ``up-votes" and ``down-votes" where an up-vote indicates being liked by a user and a down-vote means being disliked. We use the number of up-votes and down-votes to measure the quality of an answer. In our experiments, we filtered out users who have not submitted many answers. We took only the users who have provided at least 130 answers, which amounts to a total of 89 users.

\subsubsection{Defining the value of a user} Our model of stochastic utility maximization assumes that each user $i\in \Omega$ has random performance value $X_i$. We assume that $X_i$ is realized when user $i$ submits an answer to a question and receives votes from other users. When the user submits answer $a$ to question $q$, the answer receives $u(a,q)$ up-votes and $d(a,q)$ down-votes. Then we regard
\begin{equation}\label{SE:measure}
s(a,q):=  \frac{u(a,q) + \alpha}{u(a,q)+ d(a,q) +\alpha+\beta }
\end{equation}
as a realization of value. By definition, $s(a,q)\in[0,1]$ and it increases as answer $a$ keeps getting more up-votes while receiving relatively few down-votes. Our justification for using $s(a,q)$ as a value measure is as follows.  We assume that the quality of answer $a$ for question $q$ is a latent variable $X_{a,q}$ with prior $\texttt{Beta}(x;\alpha,\beta)$, the Beta distribution with parameters $\alpha$ and $\beta$. Individual up-vote or down-vote inputs are independent Bernoulli variables with mean $X_{a,q}$, and we evaluate the posterior of $X_{a,q}$ having observed $u(a,q)$ and $d(a,q)$. This posterior is $\texttt{Beta}(x; u(a,q) + \alpha, d(a,q) + \beta)$. Then we define $s(a,q)$ to be the expected value of $X_{a,q}$ with respect to the posterior distribution, and $s(a,q)=\mathbb{E}\left[X_{a,q}\ | \ u(a,q),d(a,q)\right]$ is given by~\eqref{SE:measure}.

\begin{figure}[h!]
	\begin{center}
		\includegraphics[width=6in]{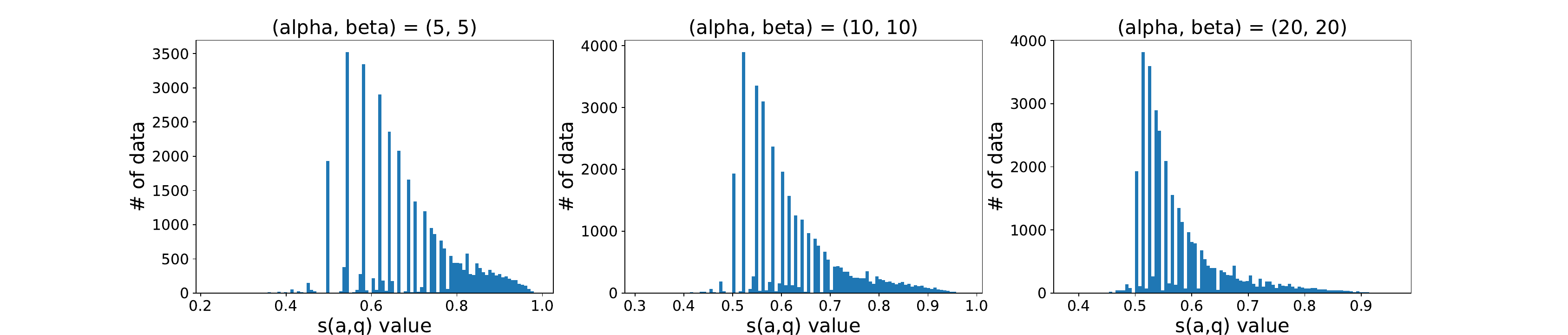}
		\includegraphics[width=6in]{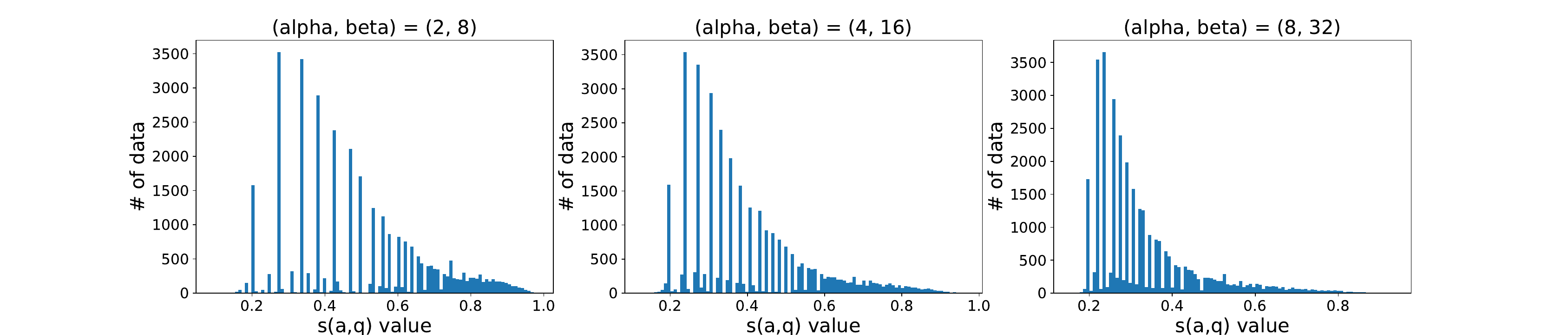}
		\caption{Distributions of $s(a,q)$ values for different choices of $(\alpha,\beta)$. Most values are around $\alpha/(\alpha,\beta)$ and the frequency drops as the value increases.}\label{fig:value-distribution}
	\end{center}
\end{figure}

In~\eqref{SE:measure}, $\alpha$ and $\beta$ can be interpreted as the numbers of ``virtual" up-votes and down-votes for an answer, respectively. We can also interpret $\alpha + \beta$ to roughly correspond to the actual number of votes for an answer needed for them to have an effect on the value of this answer. $s(a,q)$ starts with the initial value of $\alpha/(\alpha+\beta)$. Hence, the initial value is determined by the relative values of $\alpha$ and $\beta$. When $\alpha=\beta$, this value is $1/2$, indicating that an answer can be good or bad with equal chance. When $\alpha <\beta$, the value is less than $1/2$, which means that a user's initial judgement over an answer is generally bad and an answer should receive more up-votes to achieve a high value in comparison to the $\alpha=\beta$ case. Moreover, for $s(a,q)$ to get significantly greater than $\alpha/(\alpha+\beta)$, answer $a$ should receive a large number of up-votes compared to the magnitude of $\alpha$ and $\alpha+\beta$. We test the 6 choices of $(\alpha,\beta)$ shown in Table~\ref{tab:ab}.
\begin{table}[h!]
	\centering
	\caption{Settings of $(\alpha,\beta)$ considered in our experiments.}
	\begin{tabular}{c | c c c c c c} 
		$(\alpha,\beta)$ & $(5,5)$ & $(2,8)$ & $(10,10)$ & $(4,16)$ & $(20,20)$ & $(8,32)$\\ [0.5ex] 
		\hline
		Interpretation & $\begin{array}{c}\text{balanced} \\\text{small}\end{array}$ & $\begin{array}{c}\text{conservative} \\\text{small}\end{array}$ & $\begin{array}{c}\text{balanced} \\\text{medium}\end{array}$ & $\begin{array}{c}\text{conservative} \\\text{medium}\end{array}$ & $\begin{array}{c}\text{balanced} \\\text{large}\end{array}$ & $\begin{array}{c}\text{conservative} \\\text{large}\end{array}$
	\end{tabular}
	\label{tab:ab}
\end{table}
The pairs with $\alpha/\beta=1/2$ are called ``balanced" while the ones with $\alpha/\beta=1/4$ are called ``conservative". Depending on the size of $\alpha+\beta$, we call $(\alpha)$ small, medium, or large. We computed the $s(a,q)$ value of every question submitted by the 89 users. Figure~\ref{fig:value-distribution} shows how the values are distributed for different choices of $\alpha$ and $\beta$.

\subsubsection{Item costs}

We may interpret item cost parameters as payments made to users. This mimics the situation in paid labor online platforms where monetary incentives are used to elicit user contributions. Intuitively, the higher the skill of a user, the larger the payment for this user. We define the item cost parameters to be increasing, linear functions of user values with a budget truncation. Specifically, for  each user $i\in\Omega$, we set the cost $c_i$ of selecting the user to $c_i=\min\{1+\lambda\hat\mu_i, B\}$ where $\lambda$ is chosen from $\{0,B/10,2B/10,\ldots,9B/10\}$. Therefore, $c_i$ is a (truncated) linear function of the estimated mean value $\hat\mu_i$. Again, when $\lambda=0$, the budget restriction reduces to the cardinality constraint.

\subsubsection{Setting up instances}

Again, we test the four valuation functions, the max, the CES-2, the modular, and the square root functions. As mentioned, we take the $n=89$ users who have provided at least $N=130$ answers. For each user $i$, we compute the values, defined by~\eqref{SE:measure}, of all answers that he/she submitted, which give rise to samples of the user's performance value $X_i$. Then we take the sample mean $\hat \mu_i$ of $X_i$ to estimate the mean value $\mu_i=\E[X_i]$. The budget $B$ is set to $0.3\sum_{i\in\Omega}\hat \mu_i$, that is, 30\% of the total sum of users' sample means.

We generate instances by sampling randomly $N=130$ answers of each user, 100 of which are for computing the replication test scores for $\texttt{TSG}$ and running $\texttt{CELF}$ while the other 30 samples are separated for the test sample set. As the synthetic data setting, we use the test sample set to evaluate the outputs of $\texttt{TSG}$ and $\texttt{CELF}$.

\subsubsection{Results}

\begin{figure}[h!]
	\begin{center}
		\includegraphics[width=6in]{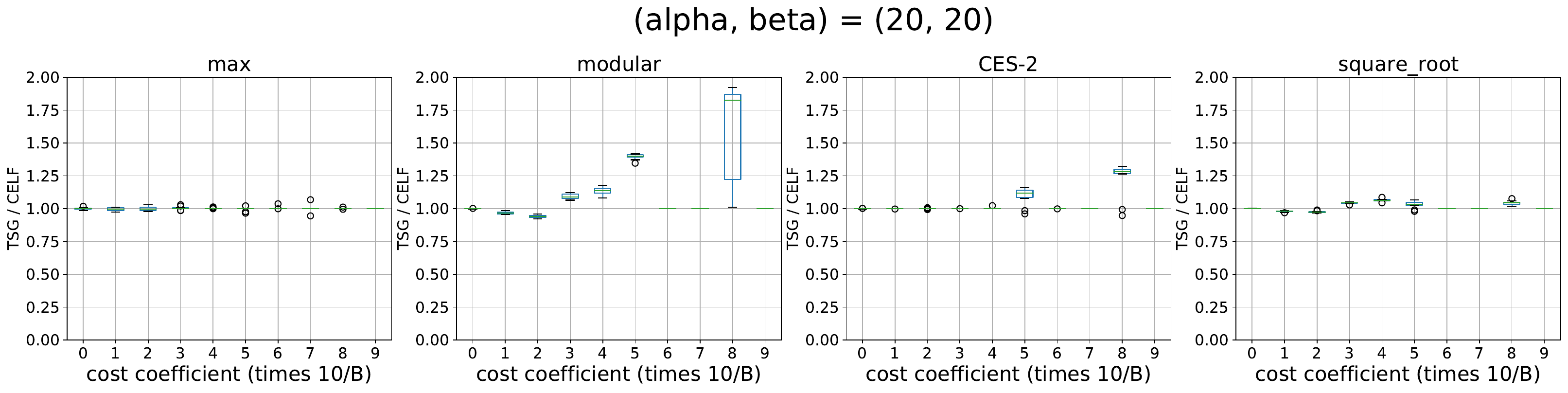}
		\includegraphics[width=6in]{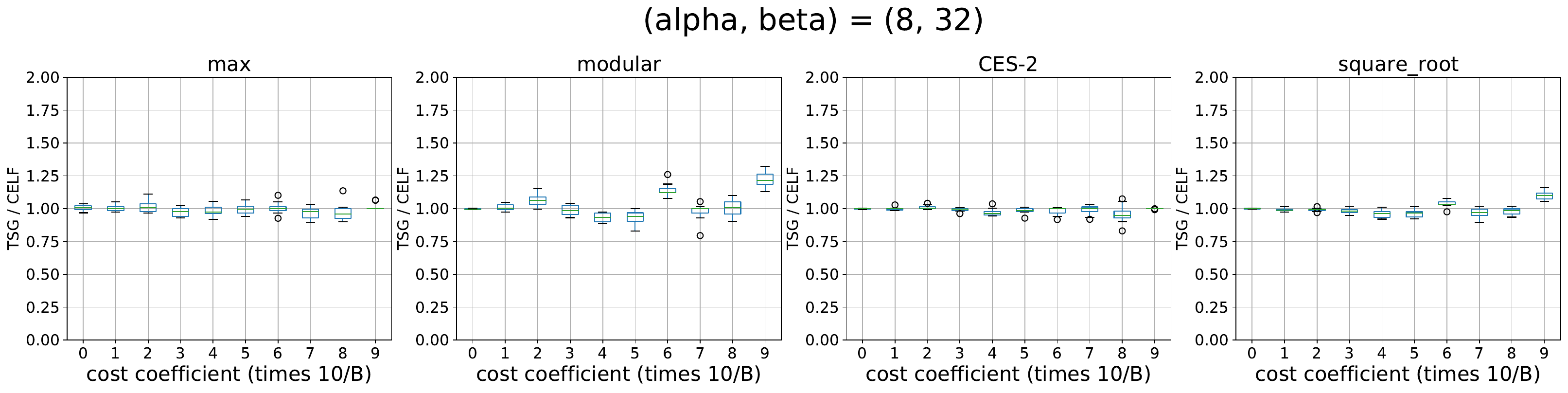}
		\caption{Comparing the output value of $\texttt{TSG}$ and that of $\texttt{CELF}$ for $(\alpha,\beta)=(20,20),(8,32)$ and for different values of $\lambda$ (the cost coefficient).}\label{fig:SE-ratios}
	\end{center}
\end{figure}

Figure~\ref{fig:SE-ratios} shows the results for different values of the cost coefficient $\lambda$. 
Overall, $\texttt{TSG}$ performs well compared to $\texttt{CELF}$, and in fact, $\texttt{TSG}$ often shows strictly better performance than $\texttt{CELF}$. Figure~\ref{fig:SE-ratios} contains the results for $(\alpha,\beta)=(20,20),(8,32)$, but the results for the other values of $(\alpha,\beta)$ are similar and included in Appendix~\ref{appendix:se-ratios}.

\begin{figure}[h!]
	\begin{center}
		\includegraphics[width=6in]{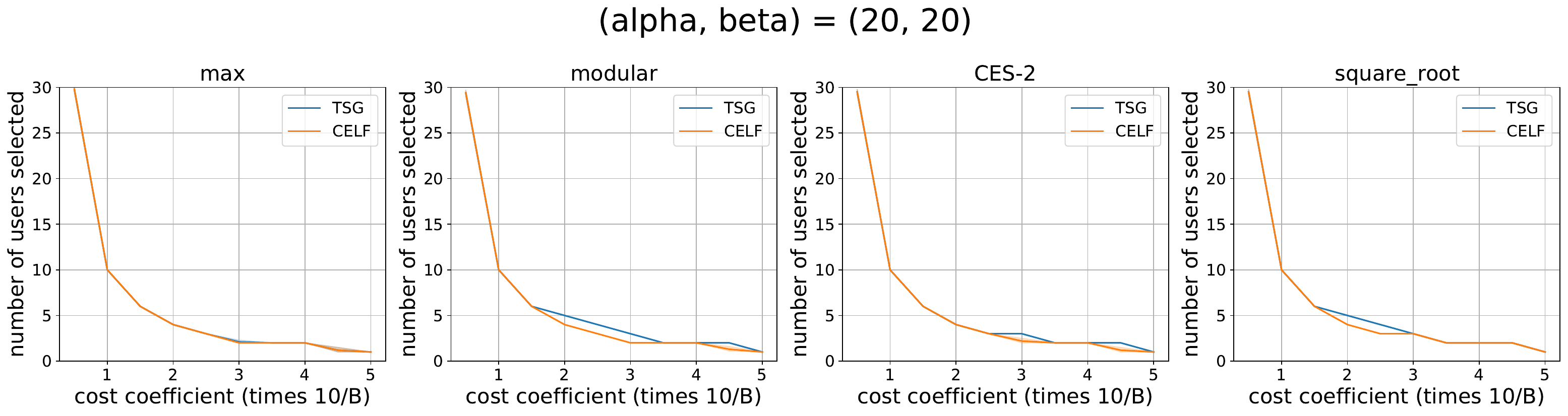}
		\includegraphics[width=6in]{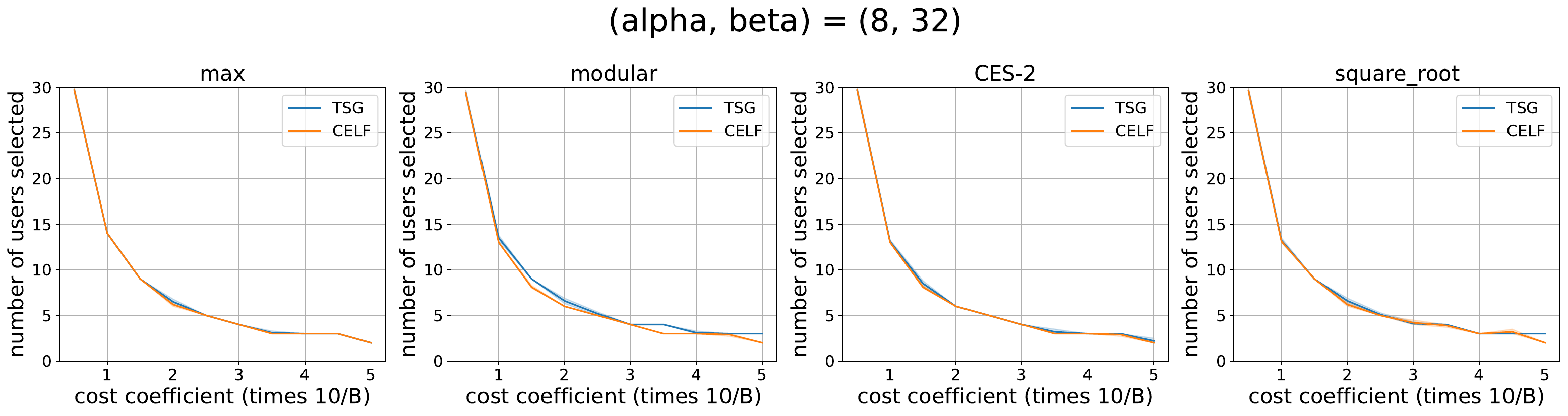}
		\caption{Comparing the number of users selected by $\texttt{TSG}$ and that by $\texttt{CELF}$. Showing the results for $(\alpha,\beta)=(20,20), (8,32)$ and for different values of $\lambda$ (the cost coefficient).}\label{fig:SE-num-users}
	\end{center}
\end{figure}

Note that when the ratio given by $\texttt{TSG}$ and $\texttt{CELF}$ is strictly 1, the value is often concentrated around a specific number. We have observed that the number of users selected by $\texttt{TSG}$ is often higher than by the number of users selected by $\texttt{CELF}$ (Figure~\ref{fig:SE-num-users}). This can be an explanation for why $\texttt{TSG}$ achieves a strictly better performance than $\texttt{CELF}$ in some cases. In particular, as the modular function takes the summation of the values of items selected, the more users get selected, the higher the value of the modular function becomes, especially when item values are not too different.
Plots for the other values of $(\alpha,\beta)$ can be checked in Appendix~\ref{appendix:se-num-users}.

\section{Conclusion and future work}

We studied a budgeted stochastic utility function maximization problem, where items have independent random values and arbitrary cost values. We devised new replication test scores that account for an item's value distribution and cost, and its contribution to a group value. We showed a greedy algorithm, whose input only uses estimated values of replication test scores and cost values of items, and the budget value, which guarantees a constant-factor approximation for a broad class of utility functions. We also provided a refined analysis to capture the regime of regular item costs and the case of small curvature. Lastly, we considered the streaming stochastic utility maximization subject to a knapsack constraint, for which, a slight modification of the greedy algorithm achieves the same approximation guarantee. We also derived sufficient sample sizes for estimating the replication test scores, which provide guidelines for practical algorithms, as well as insights into how many observations are needed for to achieve a certain approximation guarantee, for a given group valuation function and budget value. Our numerical results demonstrate the efficiency of our test score algorithm in a wide range of scenarios. 

For future work, it may be interesting to study test score algorithms for other classes of utility functions and constraints. For example, it may be of interest to consider profit maximization problems where the objective function is defined as the difference of the utility and the cost. Another direction is to consider nonlinear cost functions, unlike our current setting where adding an item incurs a fixed cost. As studied by~\cite{nonlinear-cost}, we may assume a non-modular but monotone set function for defining the cost of each group of items. The notion of replication test scores naturally extends to non-modular cost functions, so it is an interesting question whether an algorithm based on replication test scores can provide a constant-factor approximation.

\ACKNOWLEDGMENT{This research was supported, in part, by the Institute for Basic Science (IBS-R029-C1, IBS-R029-Y2).}

\bibliographystyle{plainnat} 
\bibliography{mybibfile}

\newpage
\begin{APPENDICES}

\section{Approximation guarantees}\label{sec:sketch-proof}

We first present several lemmas in Section~\ref{sec:sketching}, then present the proof of Theorem~\ref{thm:constant1} in Section~\ref{sec:constant1}, and then show proofs of the lemmas in Section~\ref{sec:proofs}.

\subsection{Sketching the utility by test scores}\label{sec:sketching}

By construction, items in $S^*$ and $S^{**}$ have high replication test scores. Our intuition is that a set consisting of items with high replication test scores has a high utility. To substantiate this idea, we come up with a way to measure or approximate the utility of a given set with the replication test scores of the items in the set. For every given $S\subseteq \Omega$, we define $\underline{v}(S)$ and $\bar v(S)$ as
\begin{equation}\label{minmax-scores}
\underline v(S):=\min\{r_i:\ i\in S\}\quad\text{and}\quad \bar v (S):=\max\{r_i: \ i\in S\}.
\end{equation}
We refer to $\underline{v}(S)$ and $\bar v(S)$ as the \emph{minimum score} and \emph{maximum score} of $S$, respectively. We will need one more notion of score a set, that is, the \emph{average score} of $S\subseteq \Omega$ defined by
\begin{equation}\label{average-score}
v(S):=\frac{\sum_{i\in S}{r_i}/{k_i}}{\sum_{i\in S}{1}/{k_i}}.
\end{equation}
We will use the three score functions defined in~\eqref{minmax-scores} and~\eqref{average-score} to approximate the utility of a given set. By definition, the following relations between the score functions hold:
\begin{equation}\label{minmax-score}
\underline{v}(S)\leq v(S)\leq \bar v(S),\hbox{ for all } S\subseteq \Omega.
\end{equation}
 
An important mathematical tool used in this section is the notion of \emph{sketch}. We say that a pair of two nonnegative set functions $v_1,v_2:2^{\Omega}\to\mathbb{R}_+$ is a \emph{sketch of $u$ with factors $p,q:\mathbb{R}_+\to\mathbb{R}_+$}  if
\begin{equation}\label{sketch}
p(d(S))v_1(S)\leq u(S)\leq q(d(S))v_2(S).
\end{equation}
Recall that $d(S)$ is defined as $\sum_{i\in S}1/k_i$, which we call the relative cost of $S$. For simplicity, we use notations $p(S):=p(d(S))$ and $q(S):=q(d(S))$ hereinafter.
Hence, a sketch of $u$ can be viewed as an approximation of $u$, and how well it approximates $u$ depends on its factors. We will use the score functions $\underline{v}$, $\bar v$, and $v$ to construct sketches of $u$. To be specific, we consider two sketches of $u$, $(\underline{v},\bar{v})$ and $(v,v)$. Once we figure out right factors for each of the two sketches, we will use them to provide approximation guarantees for Algorithm~\ref{algo:simple-greedy}. 

We first show approximation guarantees that can be obtained from the sketch given by $(\underline v,\bar v)$. Recall that items $1,\ldots,k+1$ are the ones with the highest $k+1$ estimated replication test scores and that the cost of $S^*=\{1,\ldots,k\}$ is under budget $B$ but that of $[k+1]=S^*\cup\{k+1\}$ exceeds $B$. Algorithm~\ref{algo:simple-greedy} selects either $S^*$ or $S^{**}$ where $S^{**}$ is constructed by first adding item $k+1$ and then greedily selecting other items.
\begin{lemma}\label{lemma:minmax-score}
	Let $u:2^{\Omega}\to\mathbb{R}_+$ be a set function, and let $\underline v,\bar v$ be the minimum and maximum score functions defined as in~\eqref{minmax-scores}. If $(\underline v,\bar v)$ is a sketch of $u$ with factors $p,q:2^{\Omega}\to\mathbb{R}_+$, then
	\begin{equation}\label{minmaxbound1}
	u(S^*)\geq \frac{(1-\epsilon)p([k])}{(1-\epsilon)p([k])+q(\OPT\setminus S^*)} u(\OPT).
	\end{equation}
	Moreover,
	\begin{equation}\label{minmaxbound2}
	\max\left\{u(S^*),\ u(S^{**})\right\}\geq \frac{(1-\epsilon)p([k+1])}{(1-\epsilon)p([k+1])+2q(\OPT\setminus S^*)} u(\OPT).
	\end{equation}
\end{lemma}

Here, if $q(\OPT\setminus S^*)/p([k])$ or $q(\OPT\setminus S^*)/p([k+1])$ is a constant,~\eqref{minmaxbound1} or~\eqref{minmaxbound2} provides a constant approximation guarantee on Algorithm~\ref{algo:simple-greedy}. We will argue that $(\underline{v},\bar v)$ is a ``good" sketch of $u$ by showing that $q(\OPT\setminus S^*)/p([k])$ or $q(\OPT\setminus S^*)/p([k+1])$ is bounded.

Next, we consider the sketch given by $(v,v)$ and study approximation guarantees derived from it. Recall that $k_i=\lfloor{B}/{c_i}\rfloor$ counts the maximum number of replicas of item $i$ we can select under budget $B$, and note that the value of ${1}/{k_i}$ is close to the quantity ${c_i}/{B}$, which measures the proportion of budget committed to item $i$. %

\begin{lemma}\label{lemma:average-score}
	Let $u:2^{\Omega}\to\mathbb{R}_+$ be a set function, and let $v:2^{\Omega}\to\mathbb{R}_+$ be the average score function defined as in~\eqref{average-score}. If $(v,v)$ is a sketch of $u$ with factors $p,q:2^{\Omega}\to\mathbb{R}_+$, then
	\begin{equation}\label{avgbound1}
	u(S^*)\geq (1-\epsilon)\frac{p(S^*)}{q(\OPT)}\cdot\frac{d(\OPT)}{\max\{d(S^*),d(\OPT)\}}\cdot u(\OPT).
	\end{equation}
	Moreover,
	\begin{equation}\label{avgbound2}
	\max\left\{u(S^*),\ u(S^{**})\right\}\geq\frac{(1-\epsilon)}{2}\cdot\frac{p([k+1])}{q(\OPT)}\cdot\frac{d(\OPT)}{\max\{d([k+1]),d(\OPT)\}}\cdot u(\OPT).
	\end{equation}
\end{lemma}

Here, to understand how strong the guarantees~\eqref{avgbound1} and~\eqref{avgbound2} are, we need to analyze the relative cost values $d(S^*)$, $d([k+1])$, and $d(\OPT)$. In fact, we will see that even the factors for $(\underline{v},\bar v)$ when applying Lemma~\ref{lemma:minmax-score} depend on the relative cost function. Hence, in any case, it is necessary to understand how small or large the relative cost of a given set can be, which we study in the next subsection.

For the utility functions satisfying the extended diminishing returns property, we will show that the minimum and maximum score functions give rise to a good sketch, to demonstrate which we use Lemma~\ref{lemma:minmax-score}. This will yield a constant-factor approximation guarantee of Algorithm~\ref{algo:simple-greedy} for the budget stochastic utility maximization problem. 

The following lemma shows that the utility can be lower bounded using the minimum score function.

\begin{lemma}\label{lemma:knapsack-lb}
	Let $f$ be a monotone submodular function satisfying the \emph{extended} diminishing returns property. If $u$ is the corresponding utility function defined as in~\eqref{eq:utility}, then
$$
u(S)\geq \left(1-e^{-d(S)}\right)\underline v(S), \hbox{ for all } S\subseteq \Omega.
$$
\end{lemma}

Lemma~\ref{lemma:knapsack-lb} follows from a more general lemma, Lemma~\ref{lemma:knapsack-lb'} given in Appendix~\ref{appendix:curvature}.
Next we upper bound the utility by the maximum score function. 
\begin{lemma}\label{lemma:ub}
	Let $f$ be a valuation function that is monotone, continuous and satisfies the extended diminishing returns property. If $u$ is the corresponding utility function defined as in~\eqref{eq:utility}, then 
	$$	u(S)\leq \left(1+d(S)+2\sqrt{d(S)}\right)\bar v(S) \hbox{ for all } S\subseteq \Omega.$$
\end{lemma}

By Lemmas~\ref{lemma:knapsack-lb} and~\ref{lemma:ub}, $(\underline v,\bar v)$ is a sketch of $u$ with factors $p$ and $q$ where
\begin{equation}\label{minmax:factors}
p(S)= 1-e^{-d(S)} \quad\text{and}\quad q(S)=1+d(S)+2\sqrt{d(S)}, \hbox{ for } S\subseteq \Omega.
\end{equation}

Recall that the guarantees~\eqref{minmaxbound1} and~\eqref{minmaxbound2} given in Lemma~\ref{lemma:minmax-score} do not explicitly have the relative cost terms, but the factors given in~\eqref{minmax:factors} themselves are dependent on the relative cost $d(S)$. Hence, to apply Lemma~\ref{lemma:minmax-score}, we need to measure the relative cost of a given set. We next show that, although the relative cost $d(S)$ is a variable quantity with respect to set $S$, it can be uniformly bounded by a constant.

\begin{lemma}\label{lemma:k-sum}
	For any $S\subseteq \Omega$ satisfying $c(S)\leq B$, it holds that
	$d(S)\leq  1.7$.
\end{lemma}

We can now prove Theorem~\ref{thm:constant1} by using Lemmas \ref{lemma:minmax-score} and \ref{lemma:k-sum} as shown next. 

\subsubsection{Proof of Theorem~\ref{thm:constant1}}\label{sec:constant1}

Since $d([k+1])\geq c([K+1])/B\geq 1$ and $1-e^{-x}$ is an increasing function of $x$, it follows that $p([k+1])\geq 1-1/e$. Since $\OPT\setminus S^*$ is a feasible subset, Lemma~\ref{lemma:k-sum} implies that $q(\OPT\setminus S^*)\leq 5.307$. Therefore, the result follows from~\eqref{minmaxbound2} in Lemma~\ref{lemma:minmax-score}.\Halmos
\endproof

\subsection{Proofs of lemmas}\label{sec:proofs}

\subsubsection{Proof of Lemma~\ref{lemma:minmax-score}}

Notice that
	\begin{equation}\label{sketch1}
	u(\OPT)\leq  u(S^*)+ u(\OPT\setminus S^*)\leq u(S^*)+ q(\OPT\setminus S^*)\bar v(\OPT\setminus S^*).
	\end{equation}
	Since $\OPT\setminus S^*$ contains none of $1,\ldots, k$, $\bar v(\OPT\setminus S^*)\leq\max \left\{r_i: i\in\Omega\setminus[k]\right\}$. By our assumption that $r_1,\ldots,r_{k+1} \geq (1-\epsilon)  \max\{r_i: i\in\Omega\setminus [k]\}$,
	\begin{equation}\label{sketch2}
	\underline v([k+1])\geq (1-\epsilon)\bar v(\OPT\setminus S^*).
	\end{equation}
	Since $u([k+1])\geq p([k+1])\underline v([k+1])$ and $u(S^*)+u(S^{**})\geq u([k+1])$, we derive from~\eqref{sketch1} and~\eqref{sketch2} that
	\begin{align}
	u(\OPT)&\leq u(S^*)+\frac{q(\OPT\setminus S^*)}{(1-\epsilon)p([k+1])}(u(S^*)+u(S^{**}))\notag\\
	&=\frac{(1-\epsilon)p([k+1])+q(\OPT\setminus S^*)}{(1-\epsilon)p([k+1])}u(S^*)+\frac{q(\OPT\setminus S^*)}{(1-\epsilon)p([k+1])}u(S^{**})\label{sketch3}.
	\end{align}
	Then \eqref{sketch3} implies that either $u(S^*)$ or $u(S^{**})$ must be greater than or equal to the right-hand side of~\eqref{minmaxbound2}.
	
	Next we prove~\eqref{minmaxbound1}. As $[k]$ is a subset of $[k+1]$, $\underline v([k])\geq \underline v([k+1])$ clearly holds, so by~\eqref{sketch2}, $\underline v([k])\geq (1-\epsilon)\bar v(\OPT\setminus S^*)$. Then $\bar v(\OPT\setminus S^*)\leq u([k])/((1-\epsilon)p([k]))$ because $p([k])\underline v([k])\leq u([k])$. Since $S^*$ contains $[k]$, $u([k])\leq u(S^*)$, which implies that $\bar v(\OPT\setminus S^*)\leq u(S^*)/((1-\epsilon)p([k]))$. Plugging this into~\eqref{sketch1}, we obtain
	\begin{equation}
	u(\OPT)\leq \frac{(1-\epsilon)p([k])+q(\OPT\setminus S^*)}{(1-\epsilon)p([k])}u(S^*)\notag,
	\end{equation}
	which is equivalent to~\eqref{minmaxbound1}.\Halmos
	\endproof

\subsubsection{Proof of Lemma~\ref{lemma:average-score}}

	Let $S$ be such that either $S=S^*$ or $S=[k+1]$. We shall show that 
	\begin{equation}\label{eq:comparison}
	u(S)\geq (1-\epsilon)\frac{p(S)}{q(\OPT)}\cdot\frac{d(\OPT)}{\max\{d(S),d(\OPT)\}}\cdot u(\OPT).
	\end{equation}
	Then the result follows as $u(S^*)+u(S^{**})\geq u([k+1])$ and thus $\max\{u(S^*),\ u(S^{**})\}\geq u([k+1])/2$.
	
	Let $r_{\min}=\min\{r_i:i\in S\}$. Then $r_{\min}\geq (1-\epsilon)r_j$ for any $j\in \OPT\setminus S$. We first consider the case $d(S)\geq d(\OPT)$. Note that
	\begin{align}
	\sum_{i\in S}\frac{r_i}{k_i}&=\sum_{i\in S\cap \OPT}\frac{r_i}{k_i} + \sum_{i\in S\setminus \OPT}\frac{r_i}{k_i}\notag\\
	&\geq\sum_{j\in S\cap \OPT}\frac{r_j}{k_j} + d(S\setminus \OPT)r_{\min}\notag\\
	&\geq\sum_{j\in S\cap \OPT}\frac{r_j}{k_j} + d(\OPT\setminus S)r_{\min}\notag\\
	&\geq\sum_{j\in S\cap \OPT}\frac{r_j}{k_j} + (1-\epsilon)\sum_{j\in \OPT\setminus S}\frac{r_j}{k_j}\notag\\
	&\geq (1-\epsilon)\sum_{j\in \OPT}\frac{r_j}{k_j}\label{eq:comparison1}
	\end{align}
	where the second inequality follows from the assumption that $d(S)-d(\OPT)\geq0$ and the third inequality is due to $r_{\min}\geq (1-\epsilon)r_j$ for $j\in \OPT\setminus S$. Next, we consider the case $d(S)\leq d(\OPT)$. Notice that
	\begin{align}
	d(\OPT)\sum_{i\in S}\frac{r_i}{k_i}&=d(\OPT)\sum_{i\in S\cap \OPT}\frac{r_i}{k_i}+d(\OPT)\sum_{i\in S\setminus \OPT}\frac{r_i}{k_i}\notag\\
	&= d(S)\sum_{i\in S\cap \OPT}\frac{r_i}{k_i}+\left(d(\OPT)-d(S)\right)\sum_{i\in S\cap \OPT}\frac{r_i}{k_i}+d(\OPT)\sum_{i\in S\setminus \OPT}\frac{r_i}{k_i}\notag\\
	&\geq d(S)\sum_{i\in S\cap \OPT}\frac{r_i}{k_i}+\left(d(\OPT)-d(S)\right)d(S\cap \OPT)r_{\min}+d(\OPT)d(S\setminus \OPT)r_{\min}\notag\\
	&=d(S)\sum_{i\in S\cap \OPT}\frac{r_i}{k_i} + d(S)d(\OPT\setminus S)r_{\min}\notag\\
	&\geq d(S)\sum_{i\in S\cap \OPT}\frac{r_i}{k_i} + (1-\epsilon)d(S)\sum_{j\in \OPT\setminus S}\frac{r_j}{k_j}\notag\\
	&\geq(1-\epsilon)d(S)\sum_{j\in \OPT}\frac{r_j}{k_j}\label{eq:comparison2}
	\end{align}
	where the first inequality is due to our assumption that $d(\OPT)-d(S)\geq0$ and the third inequality is due to $r_{\min}\geq (1-\epsilon)r_j$ for $j\in \OPT\setminus S$. Therefore,~\eqref{eq:comparison1} and~\eqref{eq:comparison2} imply that
	\begin{equation}\label{eq:comparison3}
	\frac{1}{d(S)}\sum_{i\in S}\frac{r_i}{k_i}\geq (1-\epsilon)\frac{1}{\max\left\{d(S),\ d(\OPT)\right\}}\sum_{j\in \OPT}\frac{r_j}{k_j}.
	\end{equation}
	It follows from~\eqref{eq:comparison3} that
	\begin{equation}\label{eq:comparison4}
	v(S)\geq (1-\epsilon)\frac{d(\OPT)}{\max\left\{d(S),\ d(\OPT)\right\}}v(\OPT).
	\end{equation}
	Since $(v,v)$ is a sketch of $u$ with factors $p$ and $q$, we have that $u(S)\geq p(S)v(S)$ and $u(\OPT)\leq q(\OPT)v(\OPT)$. Combining this and~\eqref{eq:comparison4} shows that~\eqref{eq:comparison} holds, as required.
\Halmos
\endproof

\subsubsection{Proof of Lemma~\ref{lemma:ub}}\label{sec:ub}

	We shall show that for any $\lambda>1$ and $S\subseteq\Omega$ satisfying $c(S)\leq B$, 
	$$u(S)\leq \left(\lambda+d(S)\left(1-\frac{1}{\lambda}\right)^{-1}\right)\bar v(S).$$
	In particular, when $\lambda=1+\sqrt{d(S)}$, the term $\lambda+d(S)\left(1-{1}/{\lambda}\right)^{-1}$ is minimized. 
	
	To this end, take a number $\lambda>1$ and a subset $S\subseteq\Omega$ satisfying $c(S)\leq B$. 
	Recall that $m$ is the number of arguments that $f$ takes and $n$ is the number of items. In Section~\ref{sec:score}, we have provided an extension of the valuation function defined in Section~\ref{sec:problem} that takes an arbitrary subset of $2^{\tilde{\mathbb{R}}}$ where $\tilde{\mathbb{R}}$ is an infinite collection of replicas of each element $i\in \mathbb{R}$. Then for any integer $d\geq 1$ and $x=(x_1,\ldots,x_d)\in\mathbb{R}^d$, we may define $f(x)$ as $f(\left\{x_1,\ldots, x_d\right\})$. Let $m=\lceil B\rceil$.
	Take a vector $\vz\in\mathbb{R}_+^{m}$ such that $f(\vz)=\lambda \bar v(S)$. If such a vector does not exist, it follows that $\lambda \bar v(S)$ is greater than $f_{m}^{\max}:=\max_{y\in\mathbb{R}_+^{m}}f(y)$ because $f$ is continuous, in which case, we take $z$ satisfying $f(z)=f_{m}^{\max}$. Therefore, we have
	\begin{equation}\label{eq:z-choice}
	f(z)=\min\{\lambda\bar v(S),f_{m}^{\max}\}.
	\end{equation}
    For ease of notation, for any vector $y=(y_1,\ldots,y_d)$ and a number $x$, we use notation $f(\{y,x\})$ to denote $f(\{y_1,\ldots,y_d,x\})$.
	By the extended diminishing returns property of $f$, we may assume that for all $y\in \mathbb{R}_+^j$ with $j\leq B$ such that $f(\vy)\leq f(\vz)$ and for all $x\geq 0$, we have $f(\{y,x\})-f(y)\geq f(\{z,x\})-{f(z)}$.
	
	Now take a set of fictitious items $T=\{q_1,\ldots,q_{m}\}$ such that $X_{q_j}=z_j$ with probability 1. By our choice of $T$, we have \begin{equation}\label{eq:T-choice}
	u(T)=\E[f(\{X_{q_1},\ldots,X_{q_m}\})]=f(z).
	\end{equation} 
	Notice that
	\begin{equation}\label{eq:ub-intermediate}
	u(S)\leq u(T\cup S)\leq u(T)+\sum_{i\in S}\left(u(T\cup\{i\})-u(T)\right)\leq \lambda\bar v(S) +\sum_{i\in S}\left(u(T\cup\{i\})-u(T)\right)
	\end{equation}
	where the first inequality is by the monotonicity of $u$, the second inequality is due to the submodularity of $u$, and the last inequality comes from our choice of $\vz$ and $T$ given in~\eqref{eq:z-choice} and~\eqref{eq:T-choice}. %
	
	Now we take an $i\in S$ and consider the term $u(T\cup\{i\})-u(T)$. Let $X_i^{(1)},\ldots,X_i^{(k_i)}$ be independent copies of $X_i\sim P_i$. Let $Y_i=(X_i^{(1)},\ldots,X_i^{(k_i-1)})$ and $W_i=X_i^{(k_i)}$. Then
	\begin{align}
	u(T\cup \{i\})-u(T)&=\mathbb{E}[f(\{\vz,W_i\})-f(z)]\notag\\
	&=\E[f(\{\vz,W_i\})-f(z)|\ f(Y_i)\leq f(z)]\label{eq:T-i-1}\\
	&\leq \E[f(\{Y_i,W_i\})-f(Y_i)|\ f(Y_i)\leq f(z)]\label{eq:T-i-2}\\
	&\leq \frac{u(\{i^{(1)},\ldots,i^{(k_i)}\})-u(\{i^{(1)},\ldots,i^{(k_i-1)}\})}{\bbP[f(Y_i)\leq f(z)]}\label{eq:T-i-3}
	\end{align}
	where~\eqref{eq:T-i-1} holds because $Y_i$ and $X_i$ are independent and~\eqref{eq:T-i-2} is due to the diminishing returns property of {$f$}. We can find an upper bound on the numerator in~\eqref{eq:T-i-3} as follows:
	\begin{align}
	u(\{i^{(1)},\ldots,i^{(k_i)}\})-u(\{i^{(1)},\ldots,i^{(k_i-1)}\})&\leq \frac{1}{k_i}\sum_{j=0}^{k_i-1}u(\{i^{(1)},\ldots,i^{(j)},i^{(k_i)}\})-u(\{i^{(1)},\ldots,i^{(j)}\})\label{eq:i-sum-diff}\\
	&=\frac{1}{k_i}\sum_{j=0}^{k_i-1}u(\{i^{(1)},\ldots,i^{(j)},i^{(j+1)}\})-u(\{i^{(1)},\ldots,i^{(j)}\})\notag\\
	&=\frac{1}{k_i}u(\{i^{(1)},\ldots,i^{(k_i)}\})\notag\\
	&=\frac{1}{k_i}r_i\notag\\
	&\leq \frac{1}{k_i}\bar v(S) \label{eq:numerator-ub}
	\end{align} 
	where~\eqref{eq:i-sum-diff} is due to the submodularity of $u$ and~\eqref{eq:numerator-ub} holds because $i\in S$ and $\bar v(S)=\max\{r_i:i\in S\}$. Thus, to upper bound the term $u(T\cup \{i\})-u(T)$, we need to find a lower bound on the probability term in the denominator in~\eqref{eq:T-i-3}. If $f(\vz)=f_m^{\max}$, then $f(Y_i)\leq f(\vz)$ always holds, and therefore, $\bbP[f(Y_i)\leq f(\vz)]=1$. If $f(\vz)=\lambda\bar v(S)$,
	\[
	\bbP[f(Y_i)\leq f(\vz)]\geq 1-\bbP[f(Y_i)\geq f(\vz)]\geq 1-\frac{\E[f(Y_i)]}{\lambda\bar v(S)}\geq 1-\frac{\E[f(\{Y_i,W_i\})]}{\lambda\bar v(S)}=1-\frac{r_i}{\lambda\bar v(S)}\geq 1-\frac{1}{\lambda}
	\]
	where the second inequality is implied by Markov's inequality and the last inequality is satisfied because $\bar v(S)=\max\{r_i:i\in S\}$. 
	Finally, we have argued so far that for all $i\in S$,
	\[
	u(T\cup \{i\})-u(T)\leq \frac{1}{k_i}\left(1-\frac{1}{\lambda}\right)^{-1}\bar v(S).
	\]
	Then, by~\eqref{eq:ub-intermediate}, we obtain
	\[
	u(S)\leq \lambda\bar v(S)+ d(S)\left(1-\frac{1}{\lambda}\right)^{-1}\bar v(S),
	\]
	as required.\Halmos

\subsubsection{Proof of Lemma~\ref{lemma:k-sum}}

	Since $k_i+1>{B}/{c_i}$, we have that
	$$
	\sum_{i\in S}\frac{1}{k_i+1}<\sum_{i\in S}\frac{c_i}{B}=\frac{c(S)}{B}\leq 1.
	$$
	If $k_i\geq 2$ for all $i\in S$, then $3k_i/2\geq k_i+1$ for $i\in S$, and thus, 
	$$
	d(S)\leq 1.5\sum_{i\in S}\frac{1}{k_i+1}\leq 1.5<1.7.
	$$
	Thus we may assume that $k_i=1$ for some $i\in S$. Without loss of generality, for ease of notation, we assume that $k_1=1$ and $1\in S$. Since $\lfloor{B}/{c_1}\rfloor=1$, it follows that $c_1 > B/2$, which means that $c_i < B/2$ and $k_i\geq 2$ for all $i\in S\setminus\{1\}$. If $k_i\geq 3$ for all $i\in S\setminus\{1\}$, then $4k_i/3 \geq k_i+1$ for $i\in S\setminus \{1\}$, which means that
	$$
	d(S)\leq 1 +4/3\cdot\sum_{i\in S\setminus\{1\}}\frac{1}{k_i+1}\leq 1+ 4/3\cdot1/2=5/3<1.7.
	$$
	Hence, we may assume that $k_i=2$ for some $i\in S$. Again, without loss of generality, we assume that $k_2=2$ and $2\in S$. Since $\lfloor{B}/{c_2}\rfloor=2$, we have $c_2 >B/3$, so $c_1+c_2 > {5B}/6$ and $c_i < B/6$ for $i\in S\setminus\{1,2\}$. This in turn implies that $k_i \geq 6$ for $i\in S\setminus\{1,2\}$, and thus, $7k_i/6 \geq k_i+1$ for $i\in S\setminus\{1,2\}$. Then it follows that
	$$
	d(S)\leq 1  +1/2+ {7}/{6}\cdot\sum_{i\in S\setminus\{1,2\}}\frac{1}{k_i+1}\leq 1+1/2+ 7/6\cdot1/6=61/36<1.7.
	$$
	Therefore, $d(S)\leq 1.7$ holds, as required.\Halmos
	\endproof

\section{Tighter approximation guarantee for bounded various of item costs}

\subsection{A lemma on the relative cost bounds}

\begin{lemma}\label{lemma:cost-regularity}
	Let $S\subseteq\Omega$ be a maximal subset satisfying $c(S)\leq B$, i.e. $c(S)+c_j > B$ for any $j\in \Omega\setminus S$. For $\beta\in[0,1]$, if $c_1,\ldots,c_n$ are $\beta$-small or $\beta$-regular, then 
	$$1-\frac{\beta}{1-\beta}\leq d(S)\leq 1+\frac{\beta}{1-\beta}.$$
\end{lemma}

In particular, Lemma~\ref{lemma:cost-regularity} implies that if the item costs are $\beta$-small or $\beta$-regular, then the utility function $u$ and the maximum score $\bar{v}$ satisfy the following relation $u(S)\leq \left(4+{3\beta}/{(1-\beta)}\right)\bar v(S).$

\proof{{\bf Proof of Lemma~\ref{lemma:cost-regularity}}}

	First, suppose that $c_1,\ldots,c_n$ are $\beta$-small. Then $c(S)+c_j > B$ for all $j\in\Omega\setminus S$ implies that $c(S)>(1-\beta)B$, so ${c(S)}/{B}>1-\beta$. As $d(S)\geq {c(S)}/{B}$, it follows that $d(S)\geq1-\beta\geq 1-\beta/(1-\beta)$. For proving $d(S)\leq1+{\beta}/{(1-\beta)}$, note that $k_i+1>{B}/{c_i}\geq {1}/{\beta}$ for $i\in\Omega$. Thus, $k_i\geq {1}/{\beta}-1$ for $i\in\Omega$. Moreover, we obtain from $k_i+1>{B}/{c_i}$ that ${1}/{k_i}<{c_i}/{B}(1+{1}/{k_i})$. Note that
	\[
	d(S)<\sum_{i\in S}\frac{c_i}{B}\left(1+\frac{1}{k_i}\right)\leq \sum_{i\in S}\frac{c_i}{B}\left(1+\frac{1}{\frac{1}{\beta}-1}\right)=\left(1+\frac{\beta}{1-\beta}\right)\frac{c(S)}{B}\leq 1+\frac{\beta}{1-\beta},
	\]
	as required.
	
	Next, suppose that $c_1,\ldots,c_n$ are $\beta$-regular. Let $k_{\max}$ and $k_{\min}$ denote $\max\{k_i:i\in S\}$ and $\min\{k_i:i\in S\}$, respectively. If $k_{\max}=k_{\min}$, then as $k_i\leq {B}/{c_i}<k_i+1$ for $i\in\Omega$, we have $t\leq B/c_i<t+1$ for $i\in \Omega$ where $t=k_{\max}=k_{\min}$. This in turn implies that ${B}/({t+1})<c_i\leq{B}/{t}$ for $i\in\Omega$. Since $B-c(S)\geq B-{|S|B}/{t}$, our maximal choice of $S$ implies that $|S|\geq t$. Moreover, since $0\leq B-c(S)< B-{|S|B}/{(t+1)}$, it follows that $|S|=t$. In this case, $d(S)=1$. Therefore, we may assume that $k_{\max}\geq k_{\min}+1$. As $c_1,\ldots, c_n$ are $\beta$-regular, we get that $\beta \geq 1-{k_{\min}}/{k_{\max}}\geq {1}/{k_{\max}}$, so $k_{\max}\geq{1}/{\beta}$. Again, as $k_{\min}\geq (1-\beta)k_{\max}$, we obtain $k_{\min}\geq {1}/{\beta}-1$. We first argue that for $j\in\Omega\setminus S$,
	\[
	d(S)\geq \frac{c(S)}{B}>1-\frac{c_j}{B}\geq 1-\frac{1}{k_j}\geq1-\frac{1}{k_{\min}}\geq1-\frac{1}{\frac{1}{\beta}-1}=1-\frac{\beta}{1-\beta},
	\]
	which proves the lower bound. Next we show the upper bound. Since ${B}/{c_i}<k_i+1$, we have ${1}/{k_i}<{c_i}/{B}(1+{1}/{k_i})$. Then
	\[
	d(S)<\sum_{i\in S}\frac{c_i}{B}\left(1+\frac{1}{k_i}\right)\leq \sum_{i\in S}\frac{c_i}{B}\left(1+\frac{1}{k_{\min}}\right)\leq\left(1+\frac{\beta}{1-\beta}\right)\frac{c(S)}{B}\leq 1+\frac{\beta}{1-\beta},
	\]
	as required.\Halmos
\endproof

\subsection{Proof of Theorem~\ref{thm:constant2}}

If $d([k])\geq 1$, we have $p([k])\geq 1-{1}/{e}$. Since $\OPT\setminus S^*$ is a feasible set, $q(\OPT\setminus S^*)\leq \left(4+{3\beta}/{(1-\beta)}\right)$ and thus $q(\OPT\setminus S^*)\leq \left(4+{3\beta}/{(1-\beta)}\right)({(1-\beta)}/{(1-2\beta)})$. Then the result follows from~\eqref{minmaxbound1} in Lemma~\ref{lemma:minmax-score}. Thus, we may assume that $d([k])\leq 1$. Since $c_1,\ldots,c_n$ are $\beta$-small or $\beta$-regular, by Lemma~\ref{lemma:cost-regularity}, $d([k])\geq {(1-2\beta)}{(1-\beta)}$. Notice that
$p([k])={d([k])}/{\alpha d([k])}\cdot\left(1-e^{-\alpha d([k])}\right)$. It can be checked that $\left(1-e^{-\alpha x}\right)/\alpha x$ decreases as $x\in[0,1]$ increases. Since $d([k])\leq 1$, $\left(1-e^{-\alpha d([k])}\right)/\alpha d([k])\geq \left(1-e^{-\alpha}\right)/\alpha\geq 1-{1}/{e}$, so $p([k])\geq d([k])\left(1-{1}/{e}\right)$. Since $q(\OPT\setminus S^*)\leq \left(4+{3\beta}/{(1-\beta)}\right)$ and $d([k])\geq {(1-2\beta)}/{(1-\beta)}$,~\eqref{minmaxbound1} in Lemma~\ref{lemma:minmax-score} implies the result, as required.\Halmos
\endproof

\section{Parametrizing by the curvature}

\subsection{Sketching the utility by test scores using curvature}\label{appendix:curvature}

Provided that the curvature of $g$ is $\alpha$, we can take this into account and parametrize approximation guarantees of Algorithm~\ref{algo:simple-greedy} by $\alpha$. Recall that Lemma~\ref{lemma:knapsack-lb} provides a lower bound on the utility as a function of the minimum score. Next we show that we can lower bound the utility by a function of the average score and the curvature.

\begin{lemma}\label{lemma:knapsack-lb'}
	Let $f$ be a valuation function that is $\alpha$-monotone and $\gamma$-submodular for $\alpha,\gamma\in[0,1]$. If $u$ is the corresponding utility function defined as in~\eqref{eq:utility}, then 
	$$
	u(S)\geq \frac{1-e^{-\alpha d(S)}}{\alpha} v(S) \hbox{ for all } S\subseteq \Omega.
	$$
\end{lemma}
We defer the proof of this lemma to Section~\ref{sec:lb}. 

Note that Lemma~\ref{lemma:knapsack-lb} directly follows from Lemma~\ref{lemma:knapsack-lb'}. That is because, as the relative cost $d(S)$ is nonnegative, the factor ${\left(1-e^{-\alpha d(S)}\right)}/{\alpha}$ increases as $\alpha$ decreases. Hence, the minimum value of ${\left(1-e^{-\alpha d(S)}\right)}/{\alpha}$ is attained at $\alpha=1$, so ${\left(1-e^{-\alpha d(S)}\right)}/{\alpha}\geq 1-e^{-d(S)}$. Moreover, its maximum value is $d(S)$ attained at $\alpha=0$, and it can be easily observed that the maximum value is $d(S)$.

An upper bound on the utility using the average score can be obtained as follows:
\begin{lemma}\label{lemma:knapsack-ub1}
	Let $f$ be a valuation function with curvature $\alpha$ for some $\alpha\in[0,1]$. If $u$ is the corresponding utility function defined as in~\eqref{eq:utility}, then  
	$$
	u(S)\leq \frac{1}{1-\alpha}d(S) v(S) \hbox{ for all } S\subseteq \Omega.
	$$
\end{lemma}

We are now in a position to prove Theorems~\ref{thm:parametrized-guarantee1} and \ref{thm:parametrized-guarantee2} as shown next. 

\subsubsection{Proof of Theorem~\ref{thm:parametrized-guarantee1}} 

With the factors $p,q$ given as in~\eqref{average:factors}, it follows from~\eqref{avgbound2} of Lemma~\ref{lemma:average-score} that
\begin{equation}\label{parametrized-S**}
\max\{u(S^*),\ u(S^{**})\}\geq (1-\epsilon)(1-\alpha)\frac{1-e^{-\alpha d([k+1])}}{2\alpha \cdot\max\{d([k+1]),d(\OPT)\}}\cdot u(\OPT).
\end{equation}
If $d([k+1])\leq d(\OPT)$, then~\eqref{parametrized-S**} implies~\eqref{parametrized-guarantee1} as $1/2\geq5/17$. Thus, we may assume that $\max\{d([k+1]),d(\OPT)\}=d(\OPT)$. Then it is sufficient to argue that $d([k+1])/d(\OPT)\geq 10/17$, which is indeed true because $d([k+1])\geq \sum_{i\in [k+1]}{c_i}/{B}\geq1$ and $d(\OPT)\leq 17/10$ by Lemma~\ref{lemma:k-sum}.\Halmos
\endproof

\subsubsection{Proof of Theorem~\ref{thm:parametrized-guarantee2}}  \eqref{avgbound1} of Lemma~\ref{lemma:average-score} implies that
\begin{equation}\label{parametrized-S*}
u(S^*)\geq (1-\epsilon)(1-\alpha)\frac{1-e^{-\alpha d(S^*)}}{\alpha\cdot \max\{d(S^*),d(\OPT)\}}\cdot u(\OPT).
\end{equation}

Since $f$ is monotone, we may assume that adding any additional item to $\OPT$ would make it violate the budget restriction. In addition, our choice of $S^*$ given by Algorithm~\ref{algo:simple-greedy} directly implies that $c(S^*)+c_j>B$ for any $j\in \Omega\setminus S^*$. So, applying Lemma~\ref{lemma:cost-regularity} to $\OPT$ and $S^*$ gives us that for $S\in\{\OPT,S^*\}$, $1-{\beta}/{(1-\beta)}\leq d(S)\leq 1+{\beta}/{(1-\beta)}$. Moreover, as $1-e^{-\alpha x}$ is an increasing function in $x$, we have $1-e^{-\alpha d(S^*)}\geq 1-e^{-\alpha(1-2{\beta})/{(1-\beta)}}$. Therefore,~\eqref{parametrized-guarantee2} follows, as required.\Halmos
\endproof

\subsection{Proofs of lemmas}\label{appendix:curvature-proofs}

\subsubsection{Proof of Lemma~\ref{lemma:knapsack-lb'}}\label{sec:lb}

Suppose that $S=\{e_1,\ldots, e_\ell\}$ and the elements of $S$ are ordered so that $r_{e_1}\leq r_{e_2}\leq \cdots \leq r_{e_\ell}$. For $j\in [\ell]$, let $S_j\subseteq\Omega$ be defined as $S_j:=\{e_1,\ldots,e_j\}$ and $R_j\subseteq\Omega$ be defined as
\[
R_j:=\{e_j^{(1)},\ldots,e_j^{(k_{e_j})}\}.
\]
Thus, $u(R_j)$ is precisely the replication test score $r_{e_j}$. Since $R_j\cap S_{j-1}=\emptyset$,
\begin{equation}\label{R_j-1}
u(R_j) = u(R_j\cup S_{j-1}) - \sum_{t\in [j-1]}\left(u(R_j\cup S_{t})-u(R_j\cup S_{t-1})\right).
\end{equation}
Since {$f$} satisfies the diminishing returns property, it follows that
\begin{equation}\label{R_j-2}
u(R_j\cup S_{j-1})\leq u(S_{j-1}) + \sum_{e\in R_j} \left(u(\{e\}\cup S_{j-1})-u(S_{j-1})\right).
\end{equation}
Moreover, since the curvature of {$f$} is $\alpha$, we obtain the following from~\eqref{curvature'}:
\begin{equation}\label{R_j-3}
u(R_j\cup S_{t})-u(R_j\cup S_{t-1})\geq (1-\alpha)\left(u(S_{t})-u(S_{t-1})\right), \hbox{ for all } t\in[j-1].
\end{equation}
Combining~\eqref{R_j-1}--\eqref{R_j-3}, we obtain
\begin{equation}\label{R_j-4-knapsack}
u(R_j)\leq \alpha\sum_{t\in[j-1]}\left(u(S_{t})-u(S_{t-1})\right) +{k_{e_j}}\left(u(S_{j})-u(S_{j-1})\right).
\end{equation}
Denoting $u(S_{j})-u(S_{j-1})$ by $x_j$, we have $u(S)=\sum_{j\in[\ell]}x_j$ and we can rewrite~\eqref{R_j-4-knapsack} as
\begin{equation}\label{R_j-5-knapsack}
\alpha(x_1+\cdots+x_{j-1}) +{k_{e_j}}x_j\geq r_{e_j}.
\end{equation}
Now we multiply both sides of~\eqref{R_j-5-knapsack} by $\prod_{t=j+1}^\ell\left(1-{\alpha}/{k_{e_t}}\right)/{k_{e_j}}$. Then
\begin{equation}\label{R_j-6-knapsack}
\frac{\alpha}{k_{e_j}}\prod_{t=j+1}^\ell\left(1-\frac{\alpha}{k_{e_t}}\right)(x_1+\cdots+x_{j-1}) +\prod_{t=j+1}^\ell\left(1-\frac{\alpha}{k_{e_t}}\right)x_j\geq \frac{1}{k_{e_j}}\prod_{t=j+1}^\ell\left(1-\frac{\alpha}{k_{e_t}}\right)r_{e_j}.
\end{equation}
After summing up~\eqref{R_j-6-knapsack} over all $j\in[\ell]$, the coefficient of $x_j$ in the left-hand side is equal to
$$\prod_{t=j+1}^\ell\left(1-\frac{\alpha}{k_{e_t}}\right)+\sum_{t=j+1}^\ell \frac{\alpha}{k_{e_t}} \prod_{s=t+1}^\ell\left(1-\frac{\alpha}{k_{e_s}}\right).$$
Expanding $\prod_{t=j+1}^\ell\left(1-{\alpha}/{k_{e_t}}\right)$, we know that it equals $1-\sum_{t=j+1}^\ell{\alpha}/{k_{e_t}}\prod_{s=t+1}^\ell\left(1-{\alpha}/{k_{e_s}}\right)$, implying in turn that the coefficient of $x_j$ is precisely 1. Therefore, the coefficients of $x_1,\ldots, x_\ell$ are all the same as 1. Thus, summing up~\eqref{R_j-6-knapsack} over all $j\in[\ell]$, we obtain
\begin{equation}
\sum_{j\in[\ell]}x_j \geq \sum_{j\in[\ell]} \frac{1}{k_{e_j}}\prod_{t=j+1}^\ell\left(1-\frac{\alpha}{k_{e_t}}\right)r_{e_j}.
\end{equation}
Now we claim that
\begin{equation}\label{R_j-7-knapsack}
\sum_{j\in[\ell]} \frac{1}{k_{e_j}}\prod_{t=j+1}^\ell\left(1-\frac{\alpha}{k_{e_t}}\right)r_{e_j}\geq\frac{1}{\alpha}\left(1-\prod_{j\in [\ell]}\left(1-\frac{\alpha}{k_{e_j}}\right)\right)\frac{\sum_{j\in [\ell]}r_{e_j}/k_{e_j}}{\sum_{j\in [\ell]}1/k_{e_j}}.
\end{equation}
After moving the right-hand side term of~\eqref{R_j-7-knapsack} to the left-hand side, the coefficient of $r_{e_j}$ becomes $(M_j - L)/k_{e_j}$ where
\[
M_j:=\prod_{t=j+1}^\ell\left(1-\frac{\alpha}{k_{e_t}}\right),\ j\in[\ell]\quad\text{and}\quad L:=\frac{1}{\alpha}\left(1-\prod_{j\in [\ell]}\left(1-\frac{\alpha}{k_{e_j}}\right)\right)\frac{1}{\sum_{j\in [\ell]}1/k_{e_j}}.
\]
Hence,~\eqref{R_j-7-knapsack} can be rewritten as
\begin{equation}\label{R_j-7-knapsack-re}
\sum_{j\in[\ell]} \frac{1}{k_{e_j}}(M_j-L) r_{e_j}\geq 0.
\end{equation}
First, observe that
\begin{equation*}
	\sum_{j\in[\ell]} \frac{1}{k_{e_j}}M_j=\sum_{j\in[\ell]}\frac{1}{k_{e_j}}\prod_{t=j+1}^\ell\left(1-\frac{\alpha}{k_{e_t}}\right)=\frac{1}{\alpha}\left(1-\prod_{j\in [\ell]}\left(1-\frac{\alpha}{k_{e_j}}\right)\right)= L\sum_{j\in[\ell]} \frac{1}{k_{e_j}},
\end{equation*}
implying in turn that $\sum_{j\in[\ell]}(M_j-L)/{k_{e_j}}=0$. Since $\sum_{j\in[\ell]}(M_j-L)/{k_{e_j}}=0$ and $M_j$ increases as $j$ increases, there must exists $t\in[\ell]$ such that $M_t\geq L > M_{t-1}$. By our assumption that $r_{e_\ell}\geq\cdots\geq r_{e_1}$, it follows that
\[
\sum_{j=t}^{\ell} \frac{1}{k_{e_j}}(M_j-L) r_{e_j} \geq r_{e_t}\sum_{j=t}^\ell\frac{1}{k_{e_j}}(M_j-L) = -r_{e_t}\sum_{j\in[t-1]}\frac{1}{k_{e_j}}(M_j-L)\geq -\sum_{j\in[t-1]}\frac{1}{k_{e_j}}(M_j-L)r_{e_j},
\]
which shows that~\eqref{R_j-7-knapsack-re} holds, and therefore, so does~\eqref{R_j-7-knapsack}. Hence, we have proved that 
$$u(S)\geq \frac{1}{\alpha}\left(1-\prod_{i\in S}\left(1-\frac{\alpha}{k_i}\right)\right)\frac{\sum_{i\in S}r_i/k_i}{\sum_{i\in S}1/k_i}.$$ Notice that
$$\prod_{i\in S}\left(1-\frac{\alpha}{k_i}\right)\leq \exp\left(-\sum_{i\in S}\frac{\alpha}{k_i}\right),$$
implying in turn that $1-\prod_{i\in S}\left(1-{\alpha}/{k_i}\right)\geq 1-e^{-\alpha\sum_{i\in S}{1}/{k_i}}=1-e^{-\alpha d(S)}$. The claim follows as $\sum_{i\in S}{r_i}/{k_i}= \underline{v}(S)\sum_{i\in S}{1}/{k_i}$, as required.\Halmos

\subsubsection{Proof of Lemma~\ref{lemma:knapsack-ub1}}
	For each $i\in\Omega$, since the curvature of {$f$} is $\alpha$, $$r_i=\sum_{j=1}^{k_i}\left(u\{i^{(1)},\ldots, i^{(j)}\}-u\{i^{(1)},\ldots, i^{(j-1)}\}\right)\geq k_i(1-\alpha)u(\{i\}).$$
	Hence, it follows that $u(S)\leq \sum_{i\in S}u(\{i\})\leq {1}/{(1-\alpha)}\cdot\sum_{i\in S}r_i/k_i=({d(S)}/{(1-\alpha)})v(S)$.\Halmos
	\endproof
Therefore, by Lemmas~\ref{lemma:knapsack-lb'} and~\ref{lemma:knapsack-ub1}, $(v,v)$ is a sketch of $u$ with factors $p$ and $q$ where
\begin{equation}\label{average:factors}
p(S)=\frac{1-e^{-\alpha d(S)}}{\alpha} \text{ and } q(S)=\frac{d(S)}{1-\alpha},\hbox{ for } S\subseteq\Omega.
\end{equation}

\section{Streaming utility maximization}

\subsection{Proof of Theorem~\ref{thm:streaming}}

It is clear that the algorithm requires a single-pass, and the number of items stored in the buffer is always at most $1+B/\min_{i\in\Omega}c_i$. %
Hence, it remains to show that the algorithm has the same approximation guarantees as Algorithm~\ref{algo:simple-greedy}. Let us assume that $\hat r_1\geq \hat r_2\geq\cdots\geq \hat r_n$ and $k$ is the smallest index in $\Omega$ such that 
$$
c_1+c_2+\cdots+c_{k+1}>B
$$
and
$$
r_1,\ldots,r_{k+1} \geq (1-\epsilon)  \max\{r_i: i\in\Omega\setminus [k]\}.
$$
Then the output of Algorithm~\ref{algo:streaming-greedy} is the better of $R\setminus \{k+1\}$ and $\{k+1\}$ where $R\setminus\{k+1\}$ contains $[k]$. For ease of notation, let $R^*$ denote $R\setminus \{k+1\}$. Recall that, by Lemmas~\ref{lemma:knapsack-lb} and~\ref{lemma:ub}, $(\underline v,\bar v)$ is a sketch of $u$ with factors $p$ and $q$ where
\[
p(S)=1-e^{-d(S)}\quad\text{and}\quad q(S)=1+d(S)+2\sqrt{d(S)},\quad \hbox{ for } S\subseteq\Omega
\]
where $d(S)=\sum_{i\in S}\frac{1}{k_i}$ is the relative cost of $S$.

The rest of the proof is similar to that of Lemma~\ref{lemma:minmax-score}. Notice that
\begin{equation}\label{sketch1'}
u(\OPT)\leq  u(R^*)+ u(\OPT\setminus R^*)\leq u(R^*)+ q(\OPT\setminus R^*)\bar v(\OPT\setminus R^*).
\end{equation}
Since $\OPT\setminus R^*$ contains none of $1,\ldots, k$, we have $\bar v(\OPT\setminus R^*)\leq\max \left\{r_i: i\in\Omega\setminus[k]\right\}$. By our assumption that $r_1,\ldots,r_{k+1} \geq (1-\epsilon)  \max\{r_i: i\in\Omega\setminus [k]\}$,
\begin{equation}\label{sketch2'}
\underline v([k+1])\geq (1-\epsilon)\bar v(\OPT\setminus R^*).
\end{equation}
Since $u([k+1])\geq p([k+1])\underline v([k+1])$ and $u(R^*)+u(\{k+1\})\geq u([k+1])$, we derive from~\eqref{sketch1'} and~\eqref{sketch2'} that
\begin{align}
u(\OPT)&\leq u(R^*)+\frac{q(\OPT\setminus R^*)}{(1-\epsilon)p([k+1])}(u(R^*)+u(\{k+1\}))\notag\\
&=\frac{(1-\epsilon)p([k+1])+q(\OPT\setminus R^*)}{(1-\epsilon)p([k+1])}u(R^*)%
+\frac{q(\OPT\setminus R^*)}{(1-\epsilon)p([k+1])}u(\{k+1\})\label{sketch3'}.
\end{align}
Then it follows from \eqref{sketch3'} that
\begin{equation}\label{minmaxbound1'}
u(R^*)\geq \frac{(1-\epsilon)p(R^*)}{(1-\epsilon)p(R^*)+q(\OPT\setminus R^*)}\cdot u(\OPT).
\end{equation}
Moreover,
\begin{equation}\label{minmaxbound2'}
\max\left\{u(R^*),\ u(\{k+1\})\right\}\geq \frac{(1-\epsilon)p([k+1])}{(1-\epsilon)p([k+1])+2q(\OPT\setminus R^*)}\cdot u(\OPT).
\end{equation}
Together with the bounds on the relative cost $d(S)$ for $S$ satisfying $c(S)\leq B$,~\eqref{minmaxbound1'} and~\eqref{minmaxbound2'} provide the same approximation guaurantees as the ones given by Theorems~\ref{thm:constant1} and~\ref{thm:constant2}.\Halmos

\section{Sample size bounds%
}\label{appendix:sampling}

\subsection{Proof of Proposition~\ref{pro:sample-hoeffding}}

The proof is based on the Hoeffding's inequality stated in the following theorem.

\begin{theorem}[Hoeffding's inequality] Assume $X_1, \ldots, X_m$ are independent random variables whose distributions have supports in $[a_1,b_1], \ldots, [a_m,b_m]$, respectively. Then, for any $t > 0$,
	$$
	\Pr\left[\frac{1}{m}\sum_{j=1}^m X_j - \E\left[\frac{1}{m}\sum_{j=1}^m X_j\right]\right] \leq \exp\left(-\frac{2m^2 t}{\sum_{j=1}^m (b_j-a_j)^2}\right).
	$$
\end{theorem}
We can apply Hoeffding's inequality to our problem as $\hat{r}_i = \frac{1}{m_i}\sum_{j=1}^{m_i} Z_j$ where $Z_j : = f(X_i^{((j-1)k_i)},\ldots, X_i^{jk_i})$ are independent random variables with values in $[0,||f||_{i,\infty}]$.

\subsection{Scaling of the sufficient sample size in Proposition~\ref{pro:sample-hoeffding} with $k_i$ }

\paragraph{Total production function} For $f(x)=g(x_1 + \cdots + x_{k_i})$ where $g$ is an increasing concave function, we have $||f||_{i,\infty} = g(k_i\bar{\mathcal{X}_i})$. By Jensen's inequality, $$\E[g(X^{(1)}_i+\cdots +X^{(k_i)}_i)]\leq g(k_i\E[X_i]).$$ Hence, 
$$
\left(\frac{||f_1||_{i,\infty}}{r_i}\right)^2 
= \left(\frac{g(k_i\bar{\mathcal{X}}_i)}{\E[g(X^{(1)}_i+\cdots +X^{(k_i)}_i)]}\right)^2
\geq \left(\frac{g(k_i\bar{\mathcal{X}}_i)}{g(k_i\E[X_i])}\right)^2\geq 1
$$
where the second inequality is because $\bar{\mathcal{X}}_i\geq \E[X_i]$ and $g$ is non-decreasing. 

By the strong law of large numbers, $\E[g(X^{(1)}_i+\cdots +X^{(k_i)}_i)]\sim g(k_i\E[X_i])$ for large $k_i$ as $g$ is continuous. Hence,
$$
\left(\frac{||f||_{i,\infty}}{r_i}\right)^2 
\sim \left(\frac{g(k_i\bar{\mathcal{X}}_i)}{g(k_i\E[X_i])}\right)^2, \hbox{ for large } k_i.
$$
Note that
$$
\frac{g(k_i\bar{\mathcal{X}}_i)}{g(k_i\E[X_i])}=\frac{g(k_i\bar{\mathcal{X}}_i)}{k_i\bar{\mathcal{X}}_i}\cdot \frac{k_i\E[X^{(1)}_i]}{g(k_i\E[X_i])}\cdot \frac{\bar{\mathcal{X}}_i}{\E[X_i]}\leq \frac{\bar{\mathcal{X}}_i}{\E[X_i]}
$$
where the inequality follows from the observation that $g(x)/x$ is decreasing in $x$ and $\bar{\mathcal{X}}_i\geq \E[X_i]$. Therefore, the sufficient sample size scales linearly with $k_i$.

\paragraph{Best-shot production function} For $f(x) = \max\{x_1,\dots, x_{k_i}\}$, we have $||f||_{i,\infty} = \bar{\mathcal{X}}_i$. Note that 
$$
\E[X_i]\leq\E[\max\{X^{(1)}_i,\ldots ,X^{(k_i)}_i\}]\leq\bar{\mathcal{X}}_i.
$$
Then it follows that
$$
\left(\frac{\bar{\mathcal{X}}_i}{\E[X_i]}\right)^2\geq \left(\frac{||f||_{i,\infty}}{r_i}\right)^2 
= \left(\frac{\bar{\mathcal{X}}_i}{\E[\max\{X^{(1)}_i,\ldots ,X^{(k_i)}_i\}]}\right)^2\geq 1.
$$
In fact, $\E[\max\{X^{(1)}_i,\ldots ,X^{(k_i)}_i\}]\sim \bar{\mathcal{X}}_i$ for large $k_i$. This implies that the sufficient sample size scales linearly with $k_i$.

\paragraph{Top-$r$ function} For $f(x) = x_{(1)} + \cdots + x_{(r)}$ where $x_{(1)}, \ldots, x_{(r)}$ are the highest $r$ values in $\{x_1, \ldots,x_{k_i}\}$ for $1\leq r \leq k_i$, we have $||f||_{i,\infty} = r\bar{\mathcal{X}}_i$. Let $Y_{(1)},\ldots, Y_{(r)}$ be the highest $r$ values in $\left\{X^{(1)}_i,\ldots, X^{(k_i)}_i\right\}$. Then $r_i=\E[Y_{(1)}+\cdots +Y_{(r)}]$. Note that
$$
r\E[X_i]=\frac{r}{k_i}\E[X^{(1)}_i+\cdots+X^{(k)}_i]\leq\E[Y_{(1)}+\cdots +Y_{(r)}]\leq r\bar{\mathcal{X}}_i
$$
where the second inequality is asymptotically tight for large $k_i$.
Hence,
$$
\left(\frac{\bar{\mathcal{X}}_i}{\E[X_i]}\right)^2\geq\left(\frac{||f||_{i,\infty}}{r_i}\right)^2 
= \left(\frac{r\bar{\mathcal{X}}_i}{\E[X^{(1)}_i+\cdots +X^{(r)}_i]}\right)^2 
\geq 1,
$$
which implies that the sufficient sample size scales linearly with $k_i$.

\paragraph{CES function} For $f(x) = (\sum_{i=1}^{k_i} x_i^r)^{1/r}$ for $r\geq 1$, we have $||f||_{i,\infty} = k_i^{1/r}\bar{\mathcal{X}}_i$. By Jensen's inequality, 
$$
\E\left[\left(\sum_{j=1}^{k_i} (X^{(j)}_i)^r\right)^{1/r}\right]\leq \left(\E\left[\sum_{j=1}^{k_i} (X^{(j)}_i)^r\right]\right)^{1/r}={k_i}^{1/r}\E[(X_i)^r]^{1/r}.
$$
Hence, 
$$
\left(\frac{||f||_{i,\infty}}{r_i}\right)^2 
= \left(\frac{{k_i}^{1/r}\bar{\mathcal{X}}_i}{\E\left[\left(\sum_{j=1}^{k_i} (X^{(j)}_i)^r\right)^{1/r}\right]}\right)^2
\geq  \left(\frac{\bar{\mathcal{X}}_i}{\E[(X_i)^r]^{1/r}}\right)^2
$$
where the inequality is tight for asymptotically large $k_i$. Therefore, again, the sufficient sample size scales linearly with $k_i$.

\paragraph{Success probability function} For $f(x) = 1-\prod_{j\in[k_i]}(1-p(x_j))$ where $p:\mathbb{R}_+\to [0,1]$ is an increasing function with $p(0)=0$, we have $||f||_{i,\infty} =1$. Note that
$$
\prod_{j\in[k_i]}(1-p(x_j))\leq \exp\left(-\sum_{j\in[k_i]}p(x_j)\right).
$$
Since $X^{(1)}_i,\ldots, X^{(k_i)}_i$ are independent and identically distributed random variables, $r_i \geq 1- \E[\exp\left(-p(X_i)\right)]^{k_i}$. Since $k_i\geq 1$ and $\exp\left(-p(x)\right)\leq 1$ for all $x\geq 0$, it follows that $r_i \geq 1- \E[\exp\left(-p(X_i)\right)]$. Hence, we have
$$
\left(\frac{1}{1- \E[\exp\left(-p(X_i)\right)]}\right)^2\geq\left(\frac{||f||_{i,\infty}}{r_i}\right)^2 \geq 1.
$$
Therefore, the sufficient sample size scales linearly with $k_i$.

\subsection{Proof of Proposition~\ref{pro:sample-mcdiarmid}}

We use the McDiarmid's inequality stated in the following theorem.

\begin{theorem}[McDiarmid's inequality] Let $X_1, \dots, X_m$ be independent random variables with distributions whose supports are contained in range $\mathcal{X}$ and $h:\mathcal{X}^m \to \mathbb{R}$ be a function satisfying that for all $x_1,\dots, x_m \in \mathcal{X}$ and for all $z\in \mathcal{X}$
	$$
	|h(x_1,\dots,x_m) - h(x_1,\dots, x_{j-1},z,x_{j+1},\dots,x_m)| \le c_j \hbox{ for all } j\in [m].
	$$
	For any $t>0$,
	$$
	\mathbb{P}[ h(X_1,\dots, X_m) - \E [h(X_1,\dots, X_m)] \ge t] \leq \exp\left( -\frac{2t^2}{\sum_{j=1}^m c_j^2}\right).
	$$
	\label{thm:mcdiarmid}
\end{theorem}
We can apply MCDiarmid's inequality to our problem as $\hat{r}_i = h(X_i^{(1)}, \ldots, X_i^{(m_i k_i)})$ where 
$$
h(x_1, \ldots, x_{m_i k_i}) := \frac{1}{m_i}\sum_{i=1}^{m_i} f(\{x_{((j-1)k_i)},\ldots, x_{jk_i}\})
$$
and by sumodularity of $f$, for all $x_1,\dots, x_m \in \mathcal{X}_i$ and for all $z\in \mathcal{X}_i$, we have $|h(x_1, \ldots, x_m) - h(x_1, \ldots, x_{j-1}, z, x_{j+1}, \ldots, x_m)|\leq \max_{x\in \mathcal{X}_i} f(x, 0, \ldots, 0)$.

\subsection{Proof of Proposition~\ref{error-prob1}}

Let $B_\epsilon$ be the bad event that the ordering induced by estimated test scores is not $\epsilon$-top set accurate.

We first argue that the following holds:
$$
\mathbb{P}[B_\epsilon]\leq \sum_{i=1}^n \mathbb{P}\left[|\hat r_i - r_i | \le \frac{\epsilon}{2} r_{\sigma(k^*+1)}\right].
$$
Let $\pi(1),\pi(2),\ldots,\pi(n)$ be a permutation of items $1,\ldots,n$ such that $\hat r_{\pi(1)}\geq \hat r_{\pi(2)}\geq \cdots \geq \hat r_{\pi(n)}$, and let $k_\pi$ be the index satisfying $c_{\pi(1)}+\cdots +c_{\pi(k_\pi)}\leq B <c_{\pi(1)}+\cdots +c_{\pi(k_\pi)}+c_{\pi(k_{\pi}+1)}$. As $\hat r_1,\hat r_2,\ldots, \hat r_n$ are random variables, so are $\pi(1),\ldots, \pi(n)$ and $k_\pi$. With this notation, we know that the $\epsilon$-top set accuracy condition is violated if, and only if, there exists a pair of distinct items $(i,j)$ with $r_i<(1-\epsilon)r_j$ such that either
\begin{equation}\label{bad-pair}
i\in \{\pi(1),\ldots,\pi(k_\pi)\},\ j\in\{\pi(k_\pi+1),\ldots,\pi(n)\}\quad \text{or}\quad i=\pi(k_\pi+1),\ j\in\{\pi(k_\pi+2),\ldots,\pi(n)\}.
\end{equation}
If~\eqref{bad-pair} is satisfied for some pair of distinct items $(i,j)$ with $r_i<(1-\epsilon)r_j$, then it automatically implies that there is a pair of distinct items $(i,j)$ with $r_i<(1-\epsilon)r_j$ such that $\hat r_i\geq \hat r_j$. 

When $|\hat r_i - r_i | \le \frac{\epsilon}{2} r_{\sigma(k^*+1)}$ for all $i$, $\pi(1),\dots, \pi(n)$ satisfies that
\begin{equation}
r_{\pi(i)} + \epsilon r_{\sigma(k^*+1)} \ge r_{\pi(j)} \quad \mbox{for all} \quad i<j. \label{eq:rpicond1}
\end{equation}
Then, since $c_{\sigma(1)}+\cdots +c_{\sigma(k^*)}\leq B <c_{\sigma(1)}+\cdots +c_{\sigma(k^*)}+c_{\sigma(k^*+1)}$,
\begin{equation}
r_{\pi(i)} \ge (1-\varepsilon) r_{\sigma(k^* +1)}  \quad \mbox{for all}\quad i \le k_\pi +1. \label{eq:rpicond2}
\end{equation}
When there exists $i \leq k_\pi +1$ such that $r_{\pi(i)} < (1-\varepsilon) r_{\sigma(k^* +1)}$, $\sigma(j) \in \{\pi(1),\dots, \pi(i-1) \}$ for all $j \le k^*+1$. Since $B <c_{\sigma(1)}+\cdots +c_{\sigma(k^*)}+c_{\sigma(k^*+1)}$,  $c_{\pi(1)}+\cdots +c_{\pi(k_\pi)} > B$ which contradicts the definition of $\pi(k_\pi)$.
From \eqref{eq:rpicond1} and \eqref{eq:rpicond2}, when $|\hat r_i - r_i | \le \frac{\epsilon}{2} r_{\sigma(k^*+1)}$ for all $i$,
$$\frac{\min_{i\in \{\pi(1),\ldots,\pi(k_\pi+1)\}} r_i}{\max_{j\in \{\pi(k_\pi + 1),\ldots,\pi(n)\}} r_j} \ge 1-\epsilon. $$
Therefore, it follows that $\mathbb{P}[B_\epsilon]\leq \sum_{i=1}^n \mathbb{P}[|\hat r_i - r_i | \le \frac{\epsilon}{2} r_{\sigma(k^*+1)}].$

From \eqref{equ:mk-hoeffding}, with $T_i \ge 2 k_i \left(\frac{  \| f\|_{i,\infty}}{\epsilon r_{\sigma(k^*+1)}}\right)^2 \log\left(\frac{2n}{\delta}\right)$ or from \eqref{equ:mk-mcdiarmid}, with $T_i \ge 2\left(\frac{  k_i \| f_1\|_{i,\infty}}{\epsilon r_{\sigma(k^*+1)}}\right)^2 \log\left(\frac{2n}{\delta}\right)$, 
$$ 
\mathbb{P}\left[|\hat r_i - r_i | \geq \frac{\epsilon}{2} r_{\sigma(k^*+1)}\right] \le \frac{\delta}{n} .
$$
From the union bound, we can conclude Proposition~\ref{error-prob1}.
\endproof

\subsection{Proof of Proposition~\ref{error-prob2}}

Recall that $r_{\sigma(1)}\geq r_{\sigma(2)}\geq\cdots\geq r_{\sigma(n)}$ and $k^*$ is the index satisfying $c_{\sigma(1)}+\cdots +c_{\sigma(k^*)}\leq B <c_{\sigma(1)}+\cdots +c_{\sigma(k^*)}+c_{\sigma(k^*+1)}$. Suppose that
\begin{equation}\label{true-ordering'}
\hat r_{\sigma(1)},\ldots, \hat r_{\sigma(k^*)} > \hat r_{\sigma(k^*+1)}>\hat r_{\sigma(k^*+2)},\ldots, \hat r_{\sigma(n)}.
\end{equation}
Since $r_{\sigma(1)},\ldots, r_{\sigma(k^*+1)} \geq\max\{r_{\sigma(i)}: i\in\Omega\setminus[k^*]\}$, if~\eqref{true-ordering'} holds, then the ranking by the estimated replication test scores is accurate. The contrapositive of this is that, if the ranking is not accurate, then there exists a pair of distinct items $(i,j)$ with $i>j$ such that one of the following two holds:
\begin{itemize}
	\item $j\leq k^*$, $i\geq k^*+1$, and $\hat r_{\sigma(j)}\leq \hat r_{\sigma(i)}$, 
	\item $j=k^*+1$, $i\geq k^*+2$, and $\hat r_{\sigma(j)}\leq \hat r_{\sigma(i)}$.
\end{itemize}
When $|\hat r_{\sigma(i)} - r_{\sigma(i)} | < \Delta$ for all $i$, there is no such pair. 

First, consider $(i,j)$ with $j\leq k^*$, $i\geq k^*+1$. Then, when $|\hat r_{\sigma(i)} - r_{\sigma(i)} | < \Delta$,
\begin{align} \label{pair:2}
\begin{aligned}
\hat r_{\sigma(j)} - \hat r_{\sigma(i)} &= \hat r_{\sigma(j)}-r_{\sigma(j)} - \hat r_{\sigma(i)} +r_{\sigma(i)}  +r_{\sigma(j)} - r_{\sigma(i)}\\
&\geq  \hat r_{\sigma(j)}-r_{\sigma(j)} - \hat r_{\sigma(i)} +r_{\sigma(i)} + 2\Delta\\
 &= \Delta -(r_{\sigma(j)}-\hat r_{\sigma(j)}) +\Delta - (\hat r_{\sigma(i)}- r_{\sigma(i)}) \\
&>0 
\end{aligned}
\end{align}
where~\eqref{pair:2} holds because $r_{\sigma(j)}-r_{\sigma(i)}\geq r_{\sigma(k^*)}-r_{\sigma(k^*+1)}\geq 2\Delta$. 

Next, consider $i\geq k^*+2$. Then, when $|\hat r_{\sigma(i)} - r_{\sigma(i)} | < \Delta$,
\begin{align}\label{pair:3}
\begin{aligned}
\hat r_{\sigma(k^*+1)}- \hat r_{\sigma(i)}&=\hat r_{\sigma(k^*+1)}-r_{\sigma(k^*+1)} - \hat r_{\sigma(i)} +r_{\sigma(i)}+r_{\sigma(k^*+1)} - r_{\sigma(i)}\\
&\geq\hat r_{\sigma(k^*+1)}-r_{\sigma(k^*+1)} - \hat r_{\sigma(i)} +r_{\sigma(i)}+2\Delta \\
&= \Delta - (r_{\sigma(k^*+1)}-\hat r_{\sigma(k^*+1)}) +  \Delta -(\hat r_{\sigma(i)}- r_{\sigma(i)}) \\
&>0
\end{aligned}
\end{align}
where~\eqref{pair:3} holds because $r_{\sigma(k^*+1)}-r_{\sigma(i)}\geq r_{\sigma(k^*+1)}-r_{\sigma(k^*+2)}\geq 2\Delta$. 

Therefore,
\begin{align}\label{eq:condition}
\begin{aligned}
\mathbb{P}[B_\epsilon]&\leq  \mathbb{P}\left[\left\{\min_{j\leq k^*}\hat r_{\sigma(j)}\leq \max_{i\geq k^*+1}\hat r_{\sigma(i)}\right\}\cup \left\{\hat r_{\sigma(k^*+1)}\leq\max_{i\geq k^*+2} \hat r_{\sigma(i)}\right\}\right] \\
&\leq\mathbb{P}\left[\max_{i\in \Omega}|\hat r_{\sigma(i)} - r_{\sigma(i)} | \geq \Delta \right] \\
&\leq\sum_{i\in \Omega} \mathbb{P}\left[|\hat r_{\sigma(i)} - r_{\sigma(i)} | \geq \Delta \right].
\end{aligned}
\end{align}

From \eqref{equ:mk-hoeffding}, with $T_i \ge \frac{k_i}{2} \left(\frac{  \| f\|_{i,\infty}}{\Delta}\right)^2 \log\left(\frac{2n}{\delta}\right)$ or from \eqref{equ:mk-mcdiarmid}, with $T_i \ge \frac{k_i^2}{2}\left(\frac{ \| f_1\|_{i,\infty}}{\Delta}\right)^2 \log\left(\frac{2n}{\delta}\right)$, 
\begin{equation} 
\mathbb{P}[|\hat r_i - r_i | \geq \Delta] \le \frac{\delta}{n} .\label{eq:concent-delta} \end{equation}

From \eqref{eq:condition} and \eqref{eq:concent-delta}, we conclude Proposition~\ref{error-prob2}.
\endproof

\section{Additional numerical results
}\label{appendix:num}

\subsection{Supplementary plots for Figure~\ref{fig:synthetic-summary}}\label{appendix:synthetic-summary}

Figure~\ref{fig:synthetic-summary-supplement} shows the results from all instances under the Pareto distribution with parameters 1.05, 1.95, and 3.00. 
\begin{figure}[h!]
	\begin{center}
		\includegraphics[width=6in]{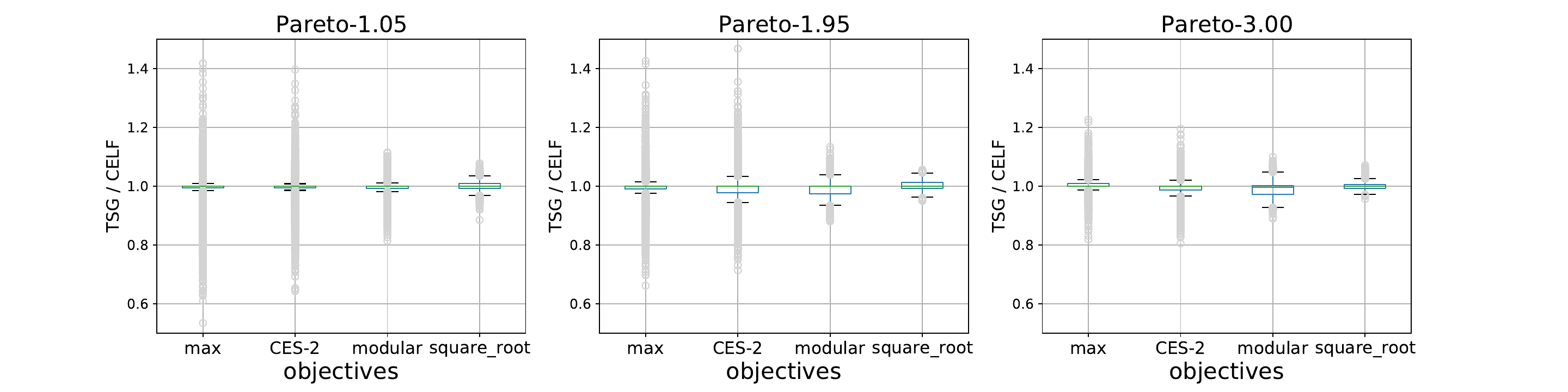}
		\caption{The ratio of the output value of~$\texttt{TSG}$ and that of $\texttt{CELF}$ for various objective functions and item value distributions.}\label{fig:synthetic-summary-supplement}
	\end{center}
\end{figure}
We can observe that the ratio values for most instances are concentrated around 1.

\subsection{Supplementary plots for Figure~\ref{fig:synthetic-cost}}\label{appendix:synthetic-cost}

Figure~\ref{fig:synthetic-cost2} shows the results for the instances under the Exponential distribution and the Pareto distribution with parameter 1.05 for different values of the cost coefficient $\lambda$.
\begin{figure}[h!]
	\begin{center}
		\includegraphics[width=6in]{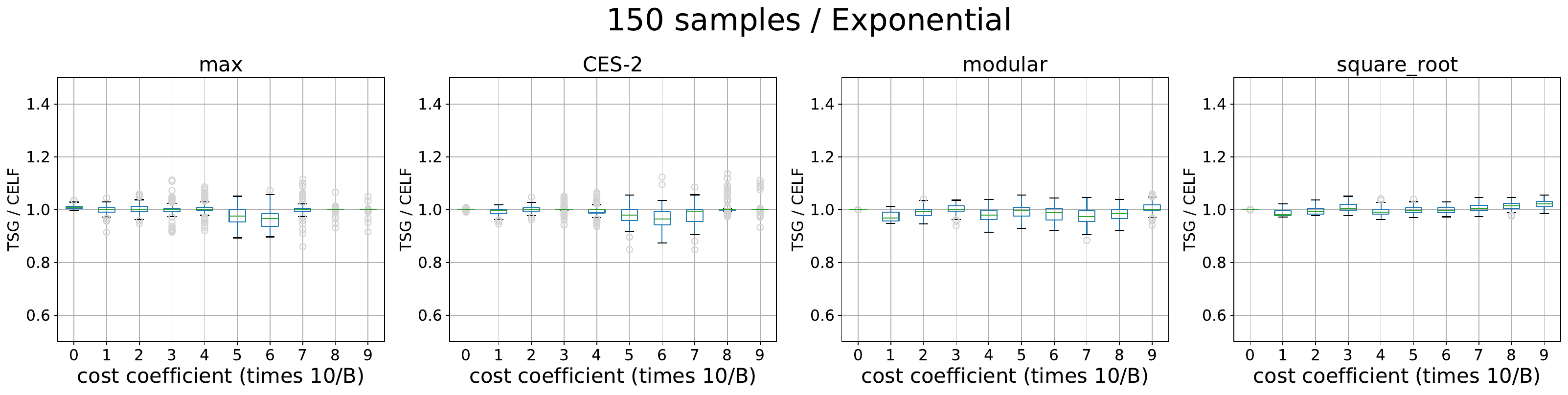}
		\includegraphics[width=6in]{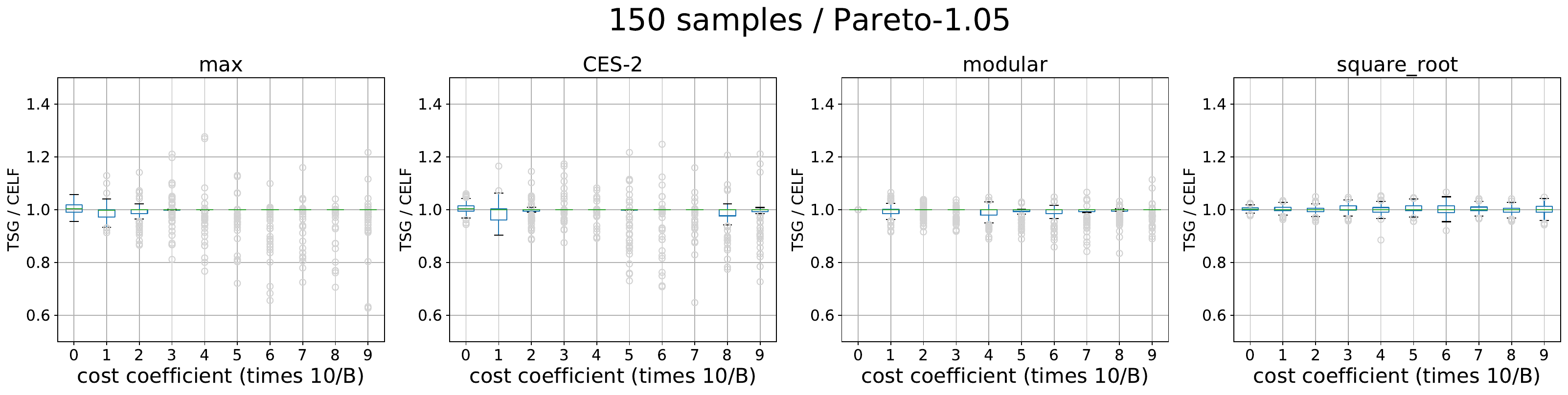}
		\caption{Results from $N=150$ and different values of the coefficient $\lambda\in \{0,B/10,2B/10,\ldots,9B/10\}$.}\label{fig:synthetic-cost2}
	\end{center}
\end{figure}
Figure~\ref{fig:synthetic-cost3} shows the results from the instances under the Pareto distribution with parameter 1.95 and 3 for various different values of the cost coefficient $\lambda$.
\begin{figure}[h!]
	\begin{center}
		\includegraphics[width=6in]{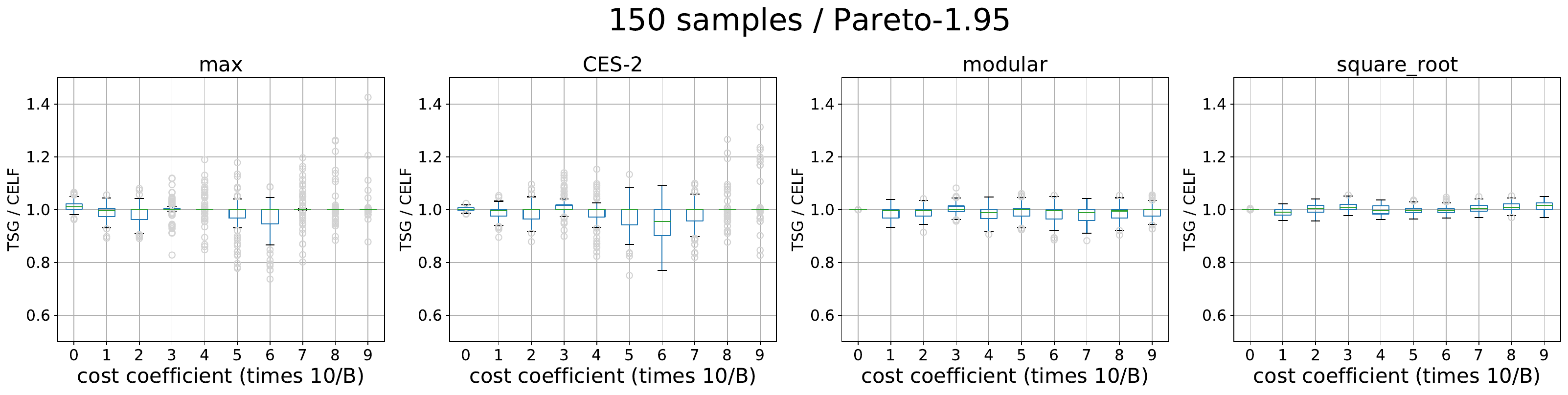}
		\includegraphics[width=6in]{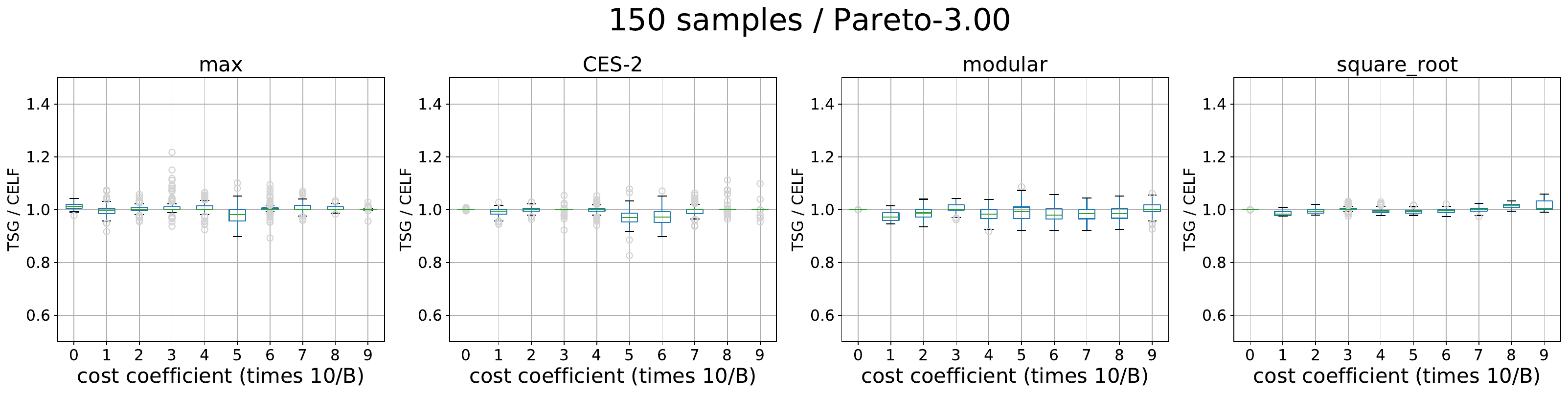}
		\caption{Results from $N=150$ and different values of the coefficient $\lambda\in \{0,B/10,2B/10,\ldots,9B/10\}$.}\label{fig:synthetic-cost3}
	\end{center}
\end{figure}

\subsection{Supplementary plots for Figure~\ref{fig:synthetic-sample}}\label{appendix:synthetic-sample}

Figure~\ref{fig:synthetic-sample2} shows the results for instances using different numbers of training samples under the Bernoulli distribution, the Exponential distribution, and the Pareto distribution with parameter 1.05.
\begin{figure}[h!]
	\begin{center}
		\includegraphics[width=6in]{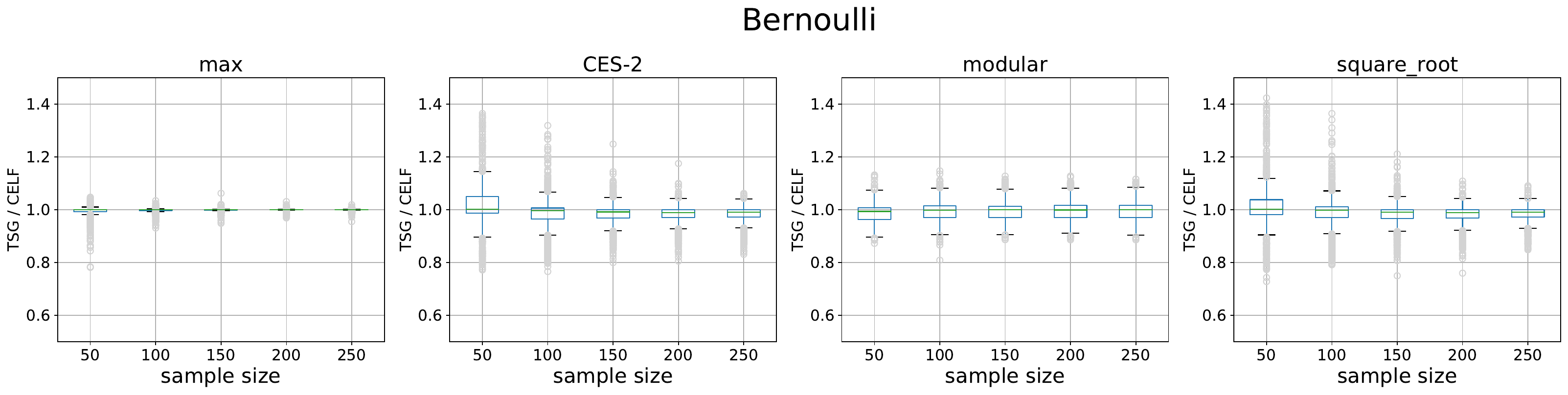}
		\includegraphics[width=6in]{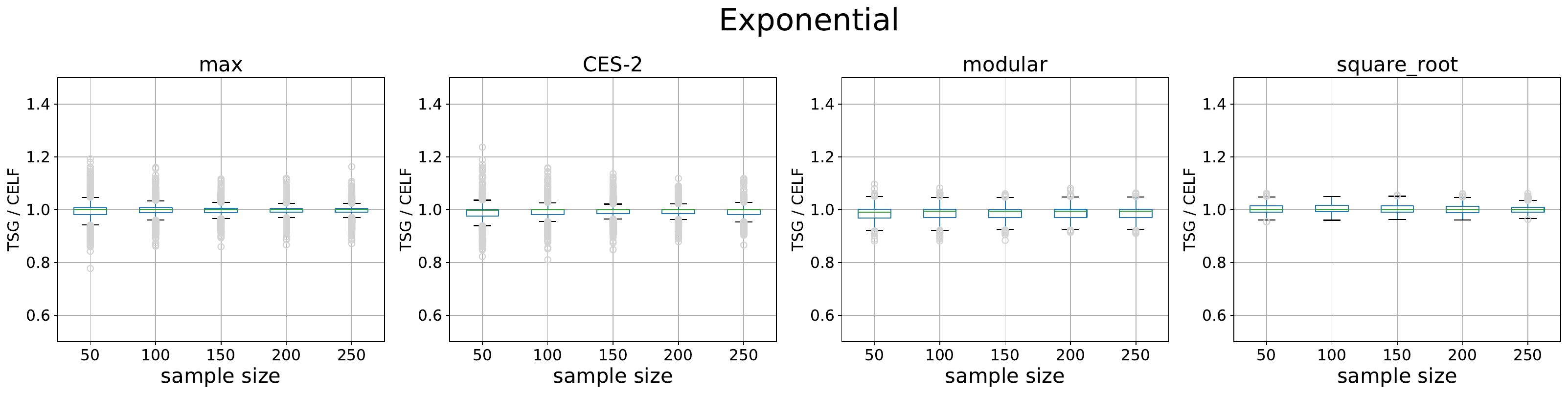}
		\includegraphics[width=6in]{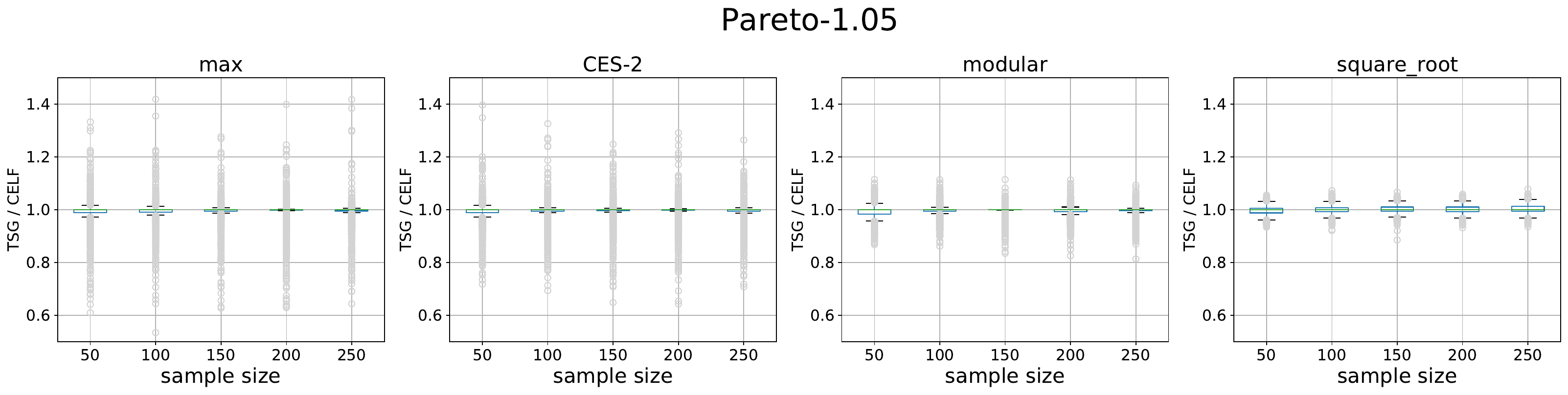}
		\caption{Results from testing different changing the sample size $N\in\{50,100,150,200,250\}$.}\label{fig:synthetic-sample2}
	\end{center}
\end{figure}
Figure~\ref{fig:synthetic-sample3} shows the results from instances using different numbers of training samples under the Pareto distribution with parameters 1.5, 1.95, and 3.
\begin{figure}[h!]
	\begin{center}
		\includegraphics[width=6in]{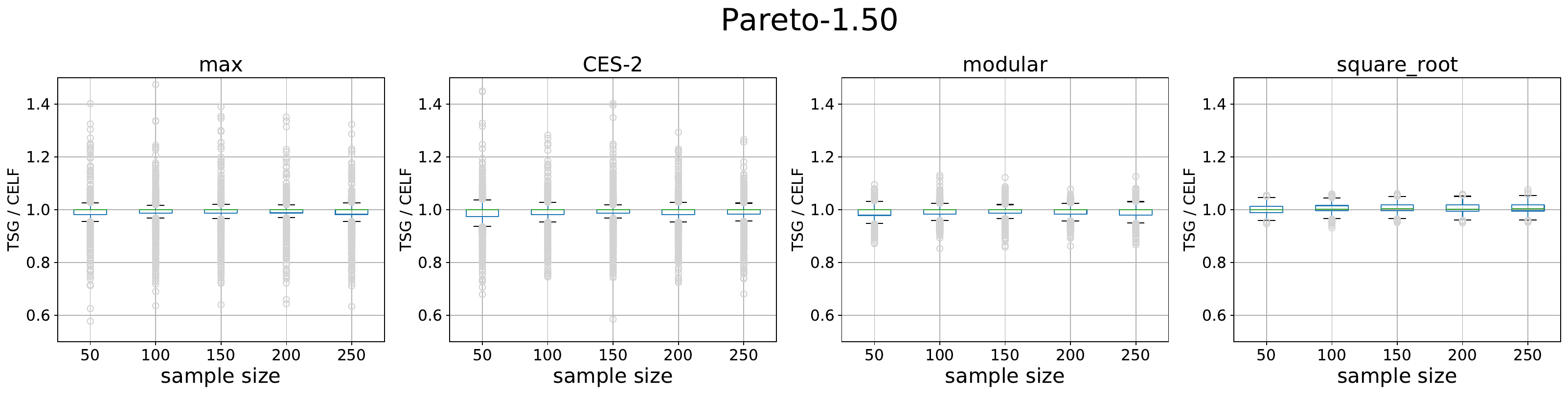}
		\includegraphics[width=6in]{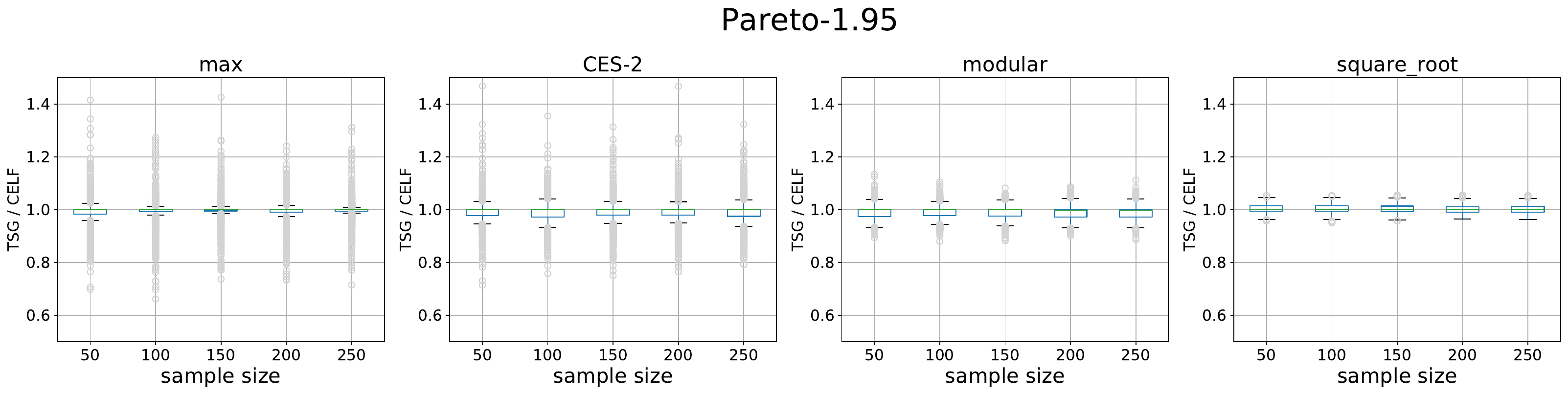}
		\includegraphics[width=6in]{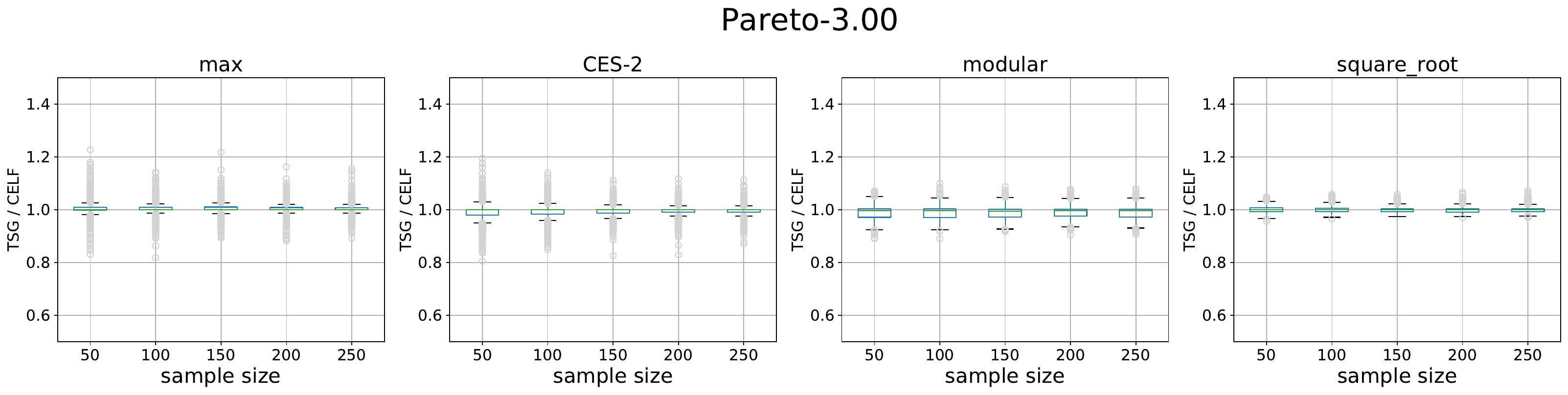}
		\caption{Results from testing different changing the sample size $N\in\{50,100,150,200,250\}$.}\label{fig:synthetic-sample3}
	\end{center}
\end{figure}

\subsection{Correlated costs versus independent costs}\label{appendix:cor-ind}

To show that the setting of independent costs exhibits a higher level of concentrations than the setting of correlated costs, we provide Figures~\ref{fig:cor-ind1} and~\ref{fig:cor-ind2}. These figures compare the distribution of the ratio values under correlated costs and the distribution of the ratio values under independent costs. In particular, Figures~\ref{fig:cor-ind1} and~\ref{fig:cor-ind2} show the corresponding cumulative distribution functions. Figure~\ref{fig:cor-ind1} shows the results under the Bernoulli distribution, the Exponential distribution, and the Pareto distribution with parameter 1.05.
\begin{figure}[h!]
	\begin{center}
		\includegraphics[width=6in]{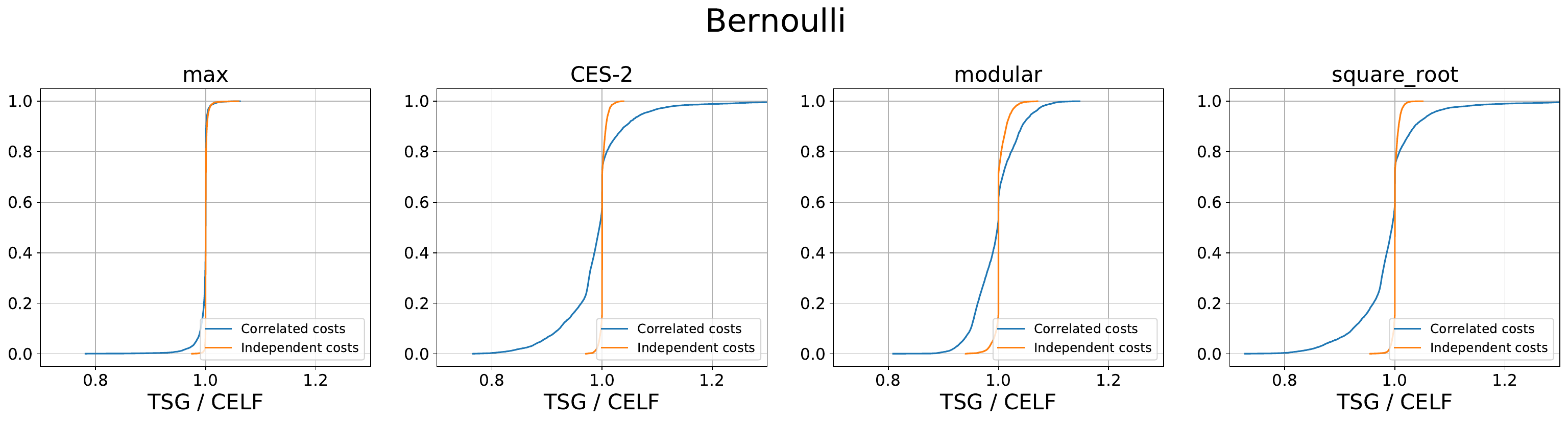}
		\includegraphics[width=6in]{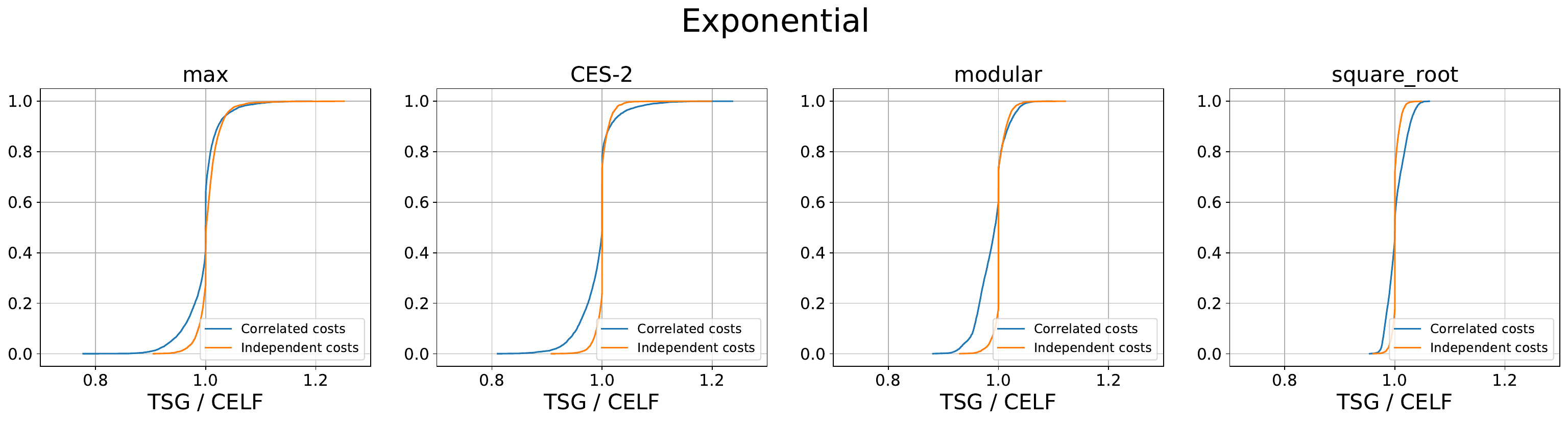}
		\includegraphics[width=6in]{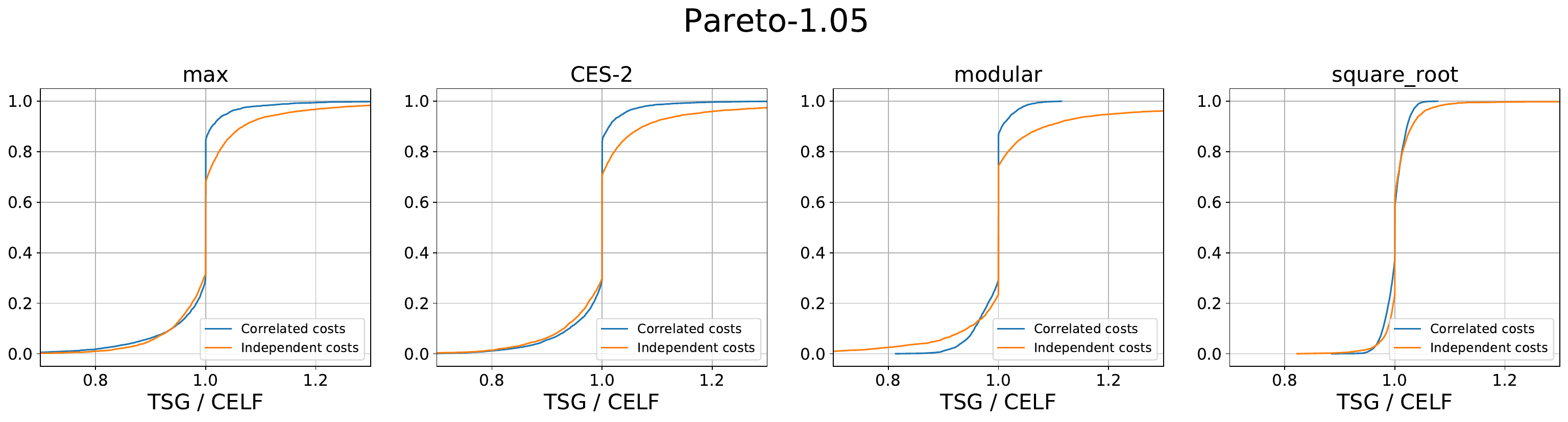}
		\caption{Comparing the output value of $\texttt{TSG}$ and that of $\texttt{CELF}$ under correlated costs and independent costs.}\label{fig:cor-ind1}
	\end{center}
\end{figure}
Figure~\ref{fig:cor-ind1} shows the results under the Pareto distribution with parameters 1.5, 1.95, and 3.
\begin{figure}[h!]
	\begin{center}
		\includegraphics[width=6in]{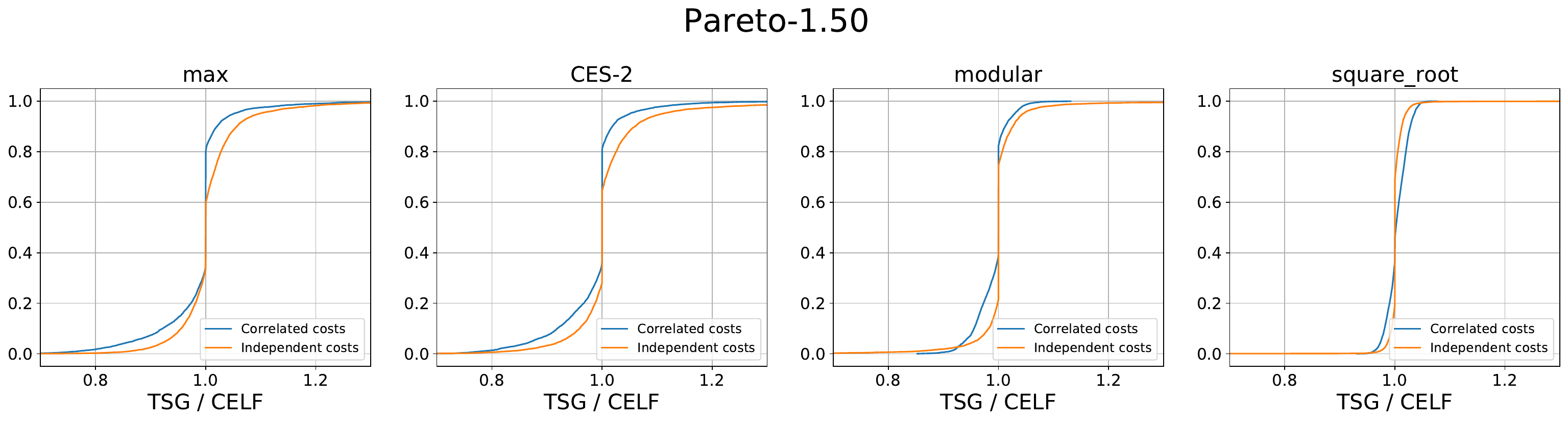}
		\includegraphics[width=6in]{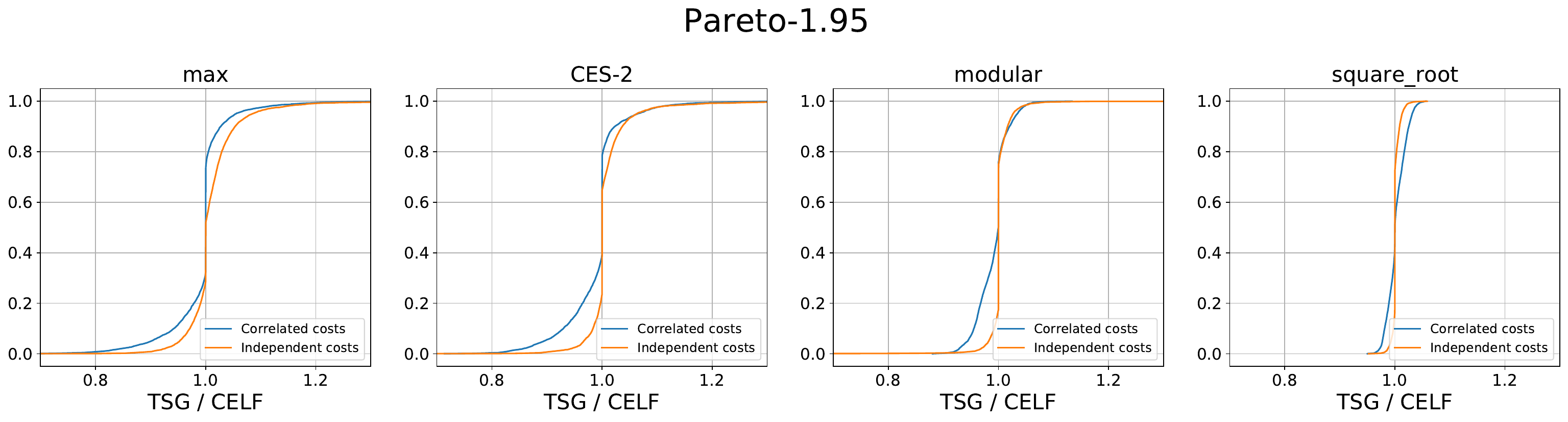}
		\includegraphics[width=6in]{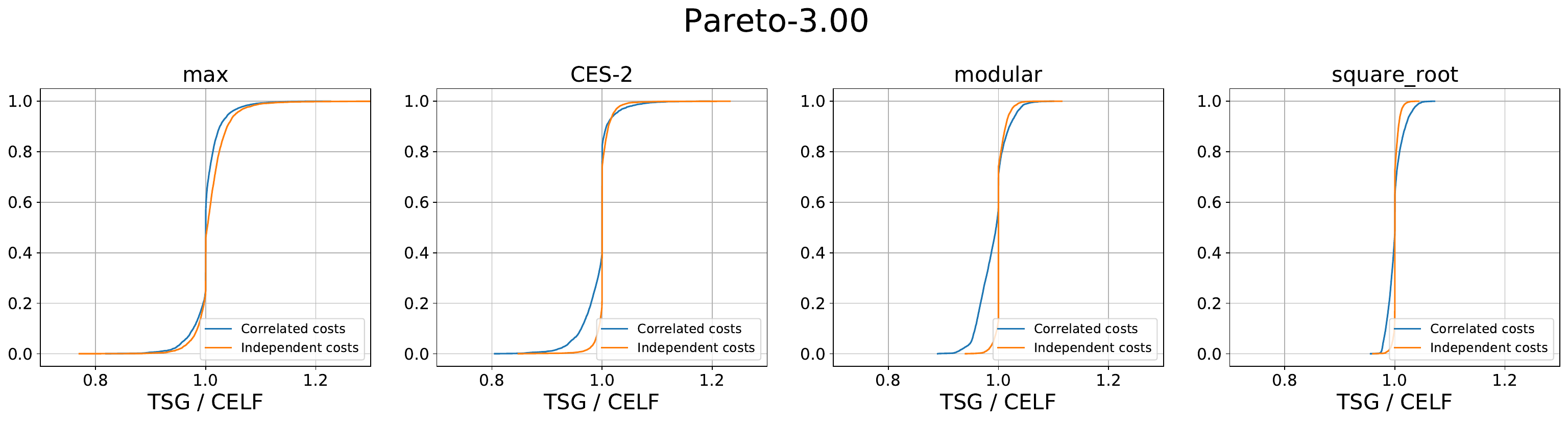}
		\caption{Comparing the output value of $\texttt{TSG}$ and that of $\texttt{CELF}$ under correlated costs and independent costs.}\label{fig:cor-ind2}
	\end{center}
\end{figure}
We can observe from Figures~\ref{fig:cor-ind1} and~\ref{fig:cor-ind2} that the setting of independent costs exhibits a higher level of concentrations than the setting of correlated costs.

\subsection{Supplementary plots for Figure~\ref{fig:SE-ratios}}\label{appendix:se-ratios}

Figure~\ref{fig:SE-ratios2} shows the results for different values of the cost coefficient $\lambda$ for $(\alpha,\beta)=(5,5),(2,8)$.
\begin{figure}[h!]
	\begin{center}
		\includegraphics[width=6in]{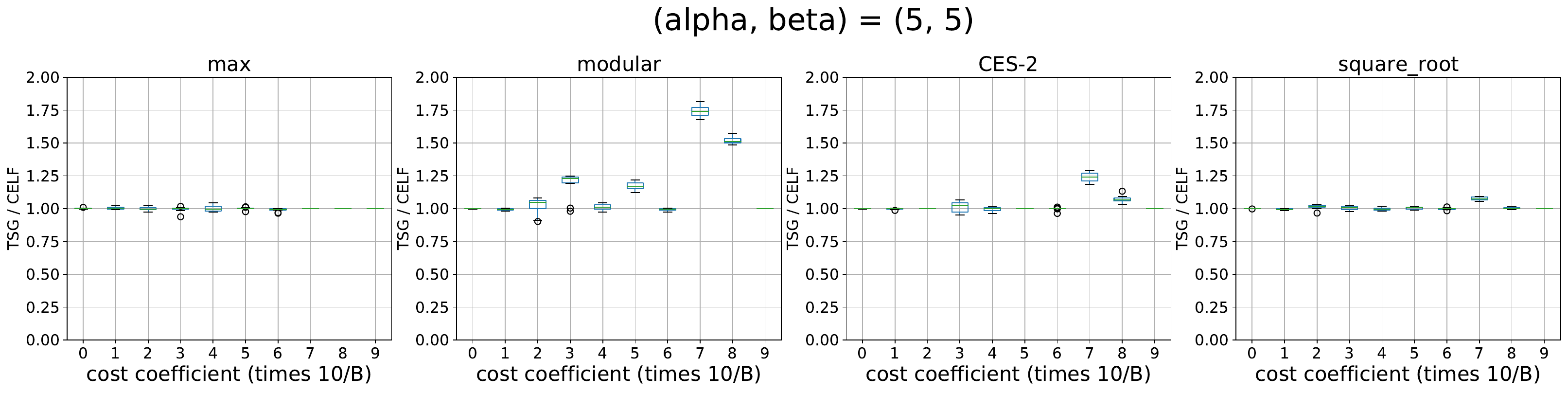}
		\includegraphics[width=6in]{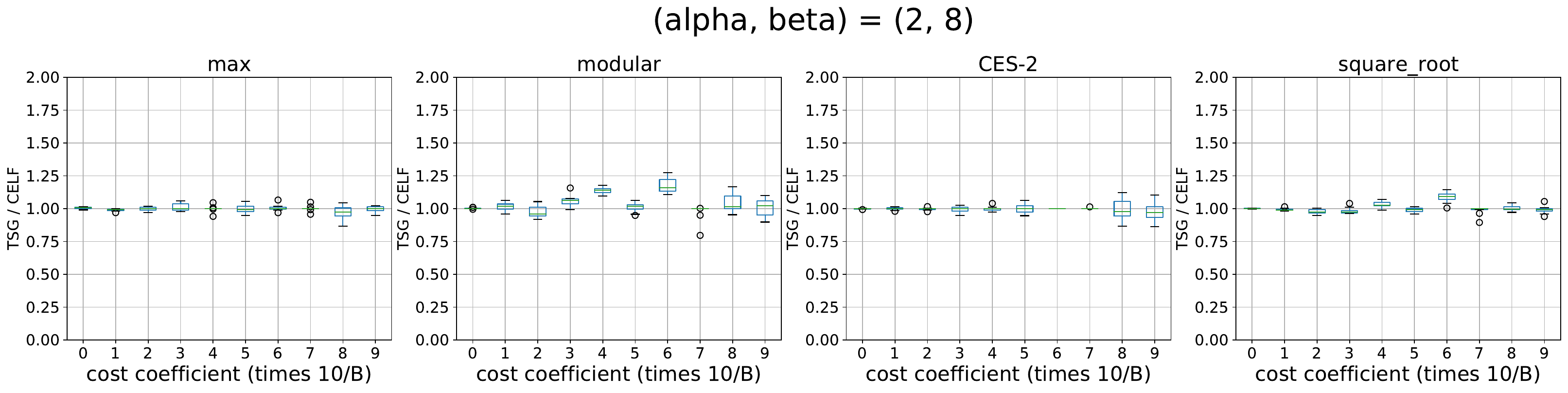}
		\caption{Comparing the output value of $\texttt{TSG}$ and that of $\texttt{CELF}$ for $(\alpha,\beta)=(5,5),(2,8)$ and for different values of $\lambda$ (the cost coefficient).}\label{fig:SE-ratios2}
	\end{center}
\end{figure}
Figure~\ref{fig:SE-ratios3} shows the results for different values of the cost coefficient $\lambda$ for $(\alpha,\beta)=(10,10),(4,16)$.
\begin{figure}[h!]
	\begin{center}
		\includegraphics[width=6in]{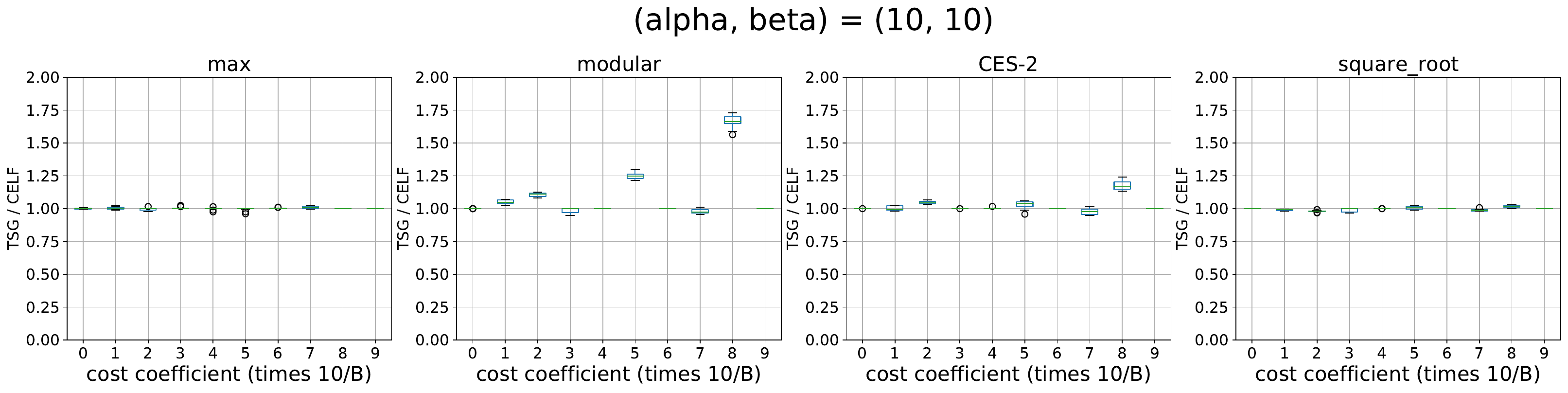}
		\includegraphics[width=6in]{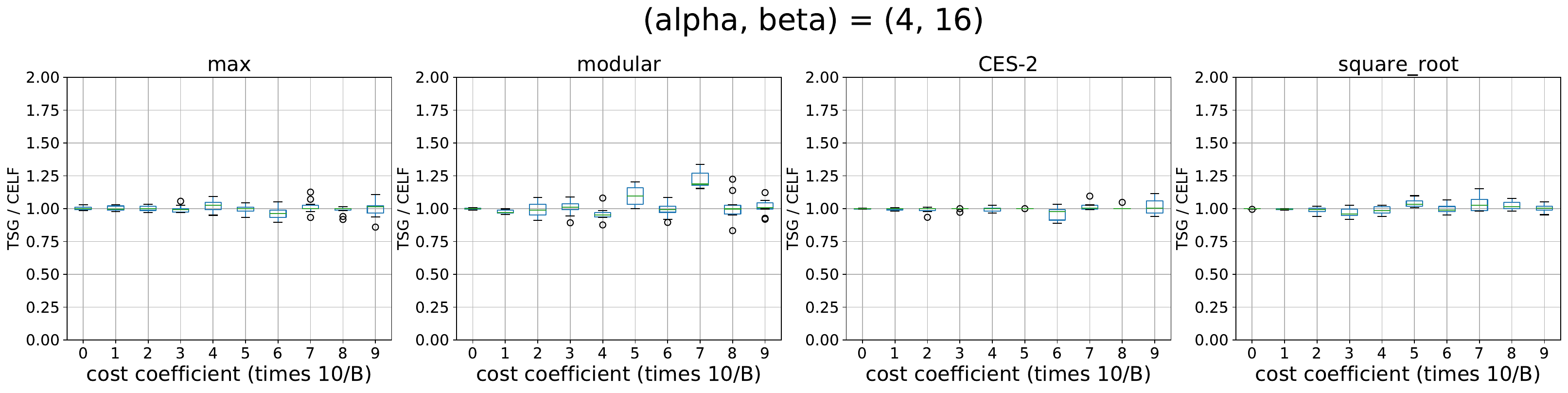}
		\caption{Comparing the output value of $\texttt{TSG}$ and that of $\texttt{CELF}$ for $(\alpha,\beta)=(10,10),(4,16)$ and for different values of $\lambda$ (the cost coefficient).}\label{fig:SE-ratios3}
	\end{center}
\end{figure}

\subsection{Supplementary plots for Figure~\ref{fig:SE-num-users}}\label{appendix:se-num-users}

Figure~\ref{fig:SE-num-users2} shows results for the comparison of the number of users selected by~$\texttt{TSG}$ and that by~$\texttt{CELF}$ when $(\alpha,\beta)=(5,5), (2,8)$.
\begin{figure}[h!]
	\begin{center}
		\includegraphics[width=6in]{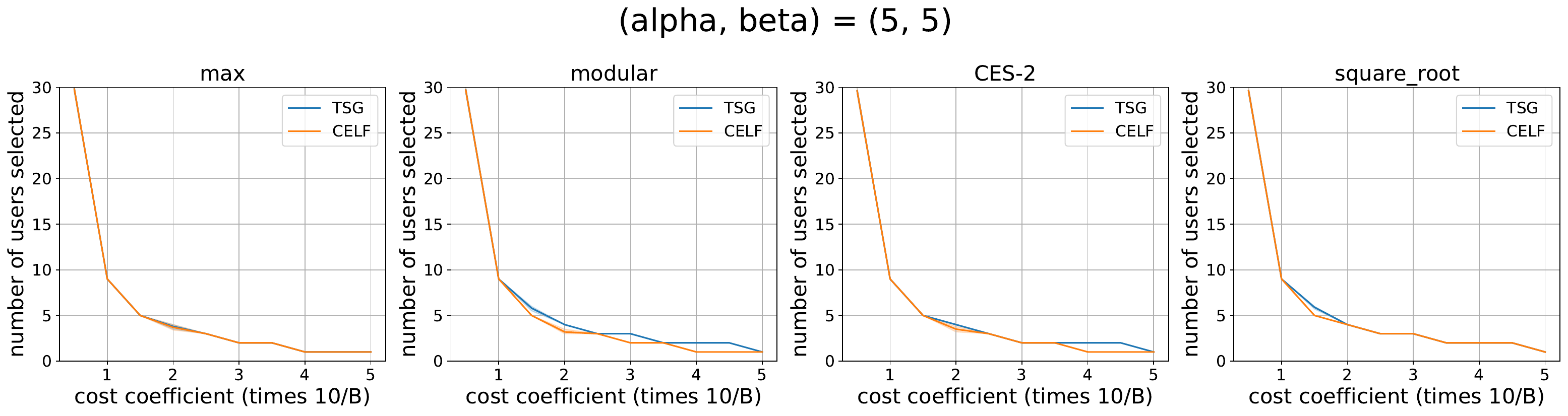}
		\includegraphics[width=6in]{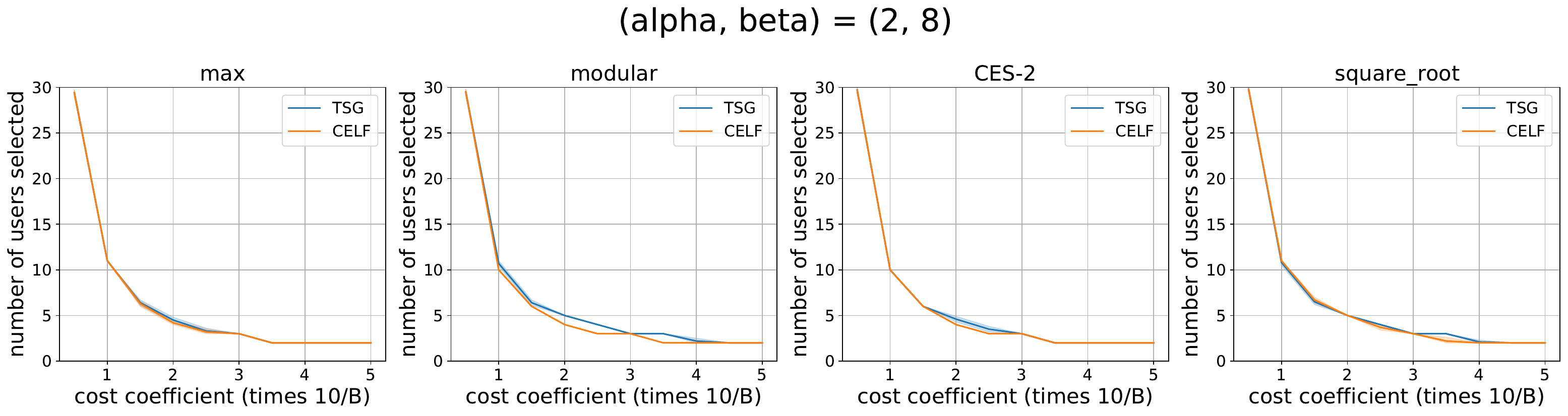}
		\caption{Comparing the number of users selected by $\texttt{TSG}$ and that by $\texttt{CELF}$. Showing the results for $(\alpha,\beta)=(5,5), (2,8)$ and for different values of $\lambda$ (the cost coefficient).}\label{fig:SE-num-users2}
	\end{center}
\end{figure}
Figure~\ref{fig:SE-num-users3} shows results for the comparison of  the number of users selected by~$\texttt{TSG}$ and that by~$\texttt{CELF}$ when $(\alpha,\beta)=(10,10), (4,16)$.
\begin{figure}[h!]
	\begin{center}
		\includegraphics[width=6in]{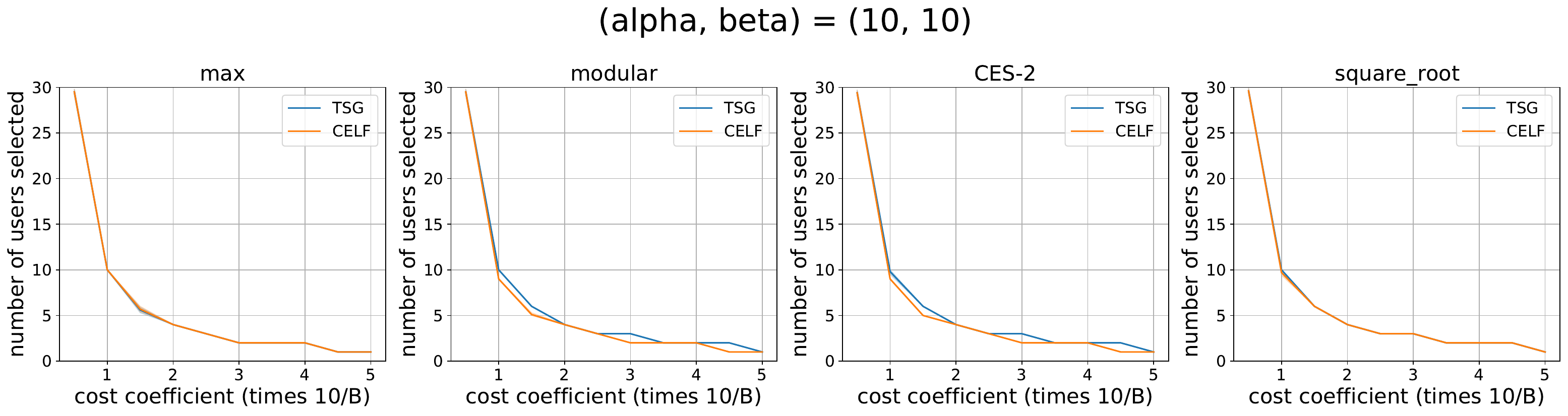}
		\includegraphics[width=6in]{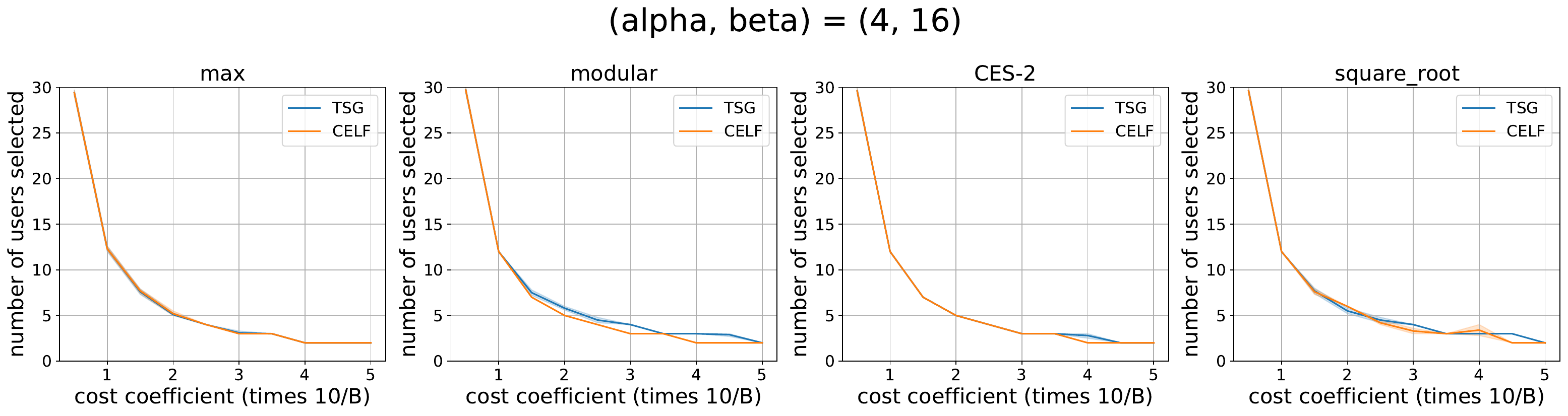}
		\caption{Comparing the number of users selected by $\texttt{TSG}$ and that by $\texttt{CELF}$. Showing the results for $(\alpha,\beta)=(10,10), (4,16)$ and for different values of $\lambda$ (the cost coefficient).}\label{fig:SE-num-users3}
	\end{center}
\end{figure}
\end{APPENDICES}

\end{document}